\documentclass[a4paper]{scrartcl}

\usepackage[english]{babel}

\usepackage{amssymb} 
\usepackage{amsbsy} 
\usepackage{amsgen} 
\usepackage{amsfonts} 
\usepackage[dvipsnames]{xcolor} 
\usepackage{siunitx}
\usepackage{multirow}
\usepackage{ulem} 
\usepackage{tikz}
\usepackage{amsmath}
\usepackage{upgreek}

\usepackage{booktabs}
\usepackage{threeparttable}
\usepackage{float,lscape}
\usepackage[section]{placeins}
\usepackage{verbatim}
\usepackage{lpic}

\usepackage{natbib}
\usepackage[margin=2.5cm
]{geometry}

\usepackage{authblk}

\title{Three-dimensional advective--diffusive boundary layers in open channels with parallel and inclined walls}

\author[1]{M. A. Etzold}
\author[2]{J. R. Landel}
\author[1]{S. B. Dalziel}

\affil[1]{Department of Applied Mathematics and Theoretical Physics, Centre for Mathematical Sciences, University of Cambridge, Wilberforce Road, Cambridge, CB3 0WA, UK}
\affil[2]{Department of Mathematics,  University of Manchester, Oxford Road, Manchester, M13 9PL, UK}

\date{\today}

\newcommand{\tavg}[1]{\overline{#1}}

\newcommand{\zavg}[1]{\overline{#1}}

\newcommand{\avg}[1]{\left<#1\right>}
\newcommand{\Peclet}{P\'eclet }
\newcommand{\Leveque}{L\'ev\^eque } 

\newcommand{\fd}[3]{\frac{\mathrm{d}^{#3}#1}{\mathrm{d}#2^{#3}}}
\newcommand{\fp}[3]{\frac{\partial^{#3}#1}{\partial#2^{#3}}}

\newcommand{\ImpPe}{\mathcal{P}e}
\newcommand{\Sh}{\mathcal{S}h}

\usepackage[implicit=false]{hyperref}

\begin{document}

\maketitle
\begin{abstract}
We study the steady laminar advective transport of a diffusive passive scalar released at the base of narrow three-dimensional longitudinal open channels with non-absorbing  side walls and rectangular or truncated-wedge-shaped  cross-sections. The scalar field in the advective--diffusive boundary layer  at the base of the channels is fundamentally  three-dimensional in the general case, owing to a three-dimensional velocity field and differing  boundary conditions at the side walls. 
We utilise three-dimensional numerical simulations and asymptotic analysis to understand how this inherent three-dimensionality influences the advective-diffusive transport as described by the normalised average flux, the Sherwood $\Sh$ or Nusselt numbers for mass or heat transfer, respectively. We show that $\Sh$ is well approximated by an appropriately formulated two-dimensional calculation, even when the boundary layer structure is itself far from two-dimensional. 
This is a key and novel results which can  significantly simplify the modelling of many laminar advection--diffusion scalar transfer problems.
The different transport regimes found depend on the channel geometry and a characteristic \Peclet number $\ImpPe$ based on the ratio of the cross-channel diffusion time and the longitudinal advection time.
We develop asymptotic expressions for $\Sh$ in the various limiting regimes, which mainly depend on the confinement of the boundary layer in the lateral and base-normal 
directions.
For $\ImpPe \gg 1$ we recover the classical \Leveque solution with a cross-channel-averaged shear rate $\overline{\gamma^{1/3}}$, $\mathcal{S}h\propto \overline{\gamma^{1/3}} \ImpPe^{1/3}$, for both geometries despite strongly curved boundary layers; for parallel walls a secondary regime with $\mathcal{S}h \propto \ImpPe^{1/2}$ is found for $\ImpPe \ll 1$. In the case of  truncated wedge channels, further regimes are identified owing to curvature effects, which we capture through a curvature-rescaled \Peclet number 
 $\ImpPe_{\beta}=\beta^2\ImpPe$, with $\beta$ the opening angle of the wedge. 
For $\ImpPe^{1/2}\ll \beta \ll 1$, the Sherwood number appears to follow $\Sh\sim \beta^{3/4}\ImpPe_{\beta}^{1/16}$. 
In all cases, we offer a comparison between our three-dimensional simulations, the asymptotic results and our two-dimensional simplifications, and can thus quantify the error in the flux from the simplified calculations.  Our findings are relevant to heat and mass transfer applications in confined U-shaped or V-shaped channels such as for the decontamination and cleaning of narrow gaps or transport processes in chemical or biological microfluidic devices.

\end{abstract}



\section{Introduction}\label{sec:intro}
%
%
The advective--diffusive  transfer of a scalar (e.g. mass or heat) at  solid--liquid boundaries in laminar channel flows is a  fundamental transport phenomenon found in numerous  applications. Mass transfer applications include: chemical \citep{zhang96,gervais06,kirtland09} and biological \citep{vijayendran03,squires08,hansen12} microfluidic reactors and sensors, porous microfluidic channels and membranes \citep{dejam19,kou19}, membrane extraction techniques \citep{jonsson00,marczak06}, micro-mixers \citep{Kamholz:1999,Ismagilov:2000,Kamholz:2001,Kamholz:2002,Stone:2004,Jiminez:2005,capretto11}, membraneless electrochemical fuel cells \citep{ferrigno02,cohen05,Braff:2013}, cross-flow membrane filtration \cite{porter72,bowen95,visvanathan00,herterich15}, crystal dissolution \citep{Bisschop2013}, aquifer remediation \citep{borden92,dejam14,kahler16}, and  cleaning \cite{Wilson:2005,fryer09,lelieveld14,pentsak19} and decontamination \cite{fitch03,settles06} in channels.
Heat transfer applications include:  film cooling \citep{acharya17}, heat exchangers  \citep{kakac02,ayub03}, and cooling and heating in micro-channels \citep{sobhan01,avelino04}.
Determining and predicting the advection-enhanced scalar flux at the transfer boundary  as a function of geometry, flow and scalar properties is highly desired in these problems. It allows assessment of the performance of the overall scalar transport. Also, scalar  transfer at the boundary is often a critical rate-limiting  step compared to other processes, particularly for mass transfer owing to low mass diffusivities compared to advection or reaction rates as commonly found in  applications \citep[e.g.][]{gervais06,squires08,kirtland09}.

Solving the  scalar transport problem in high-\Peclet number flows near boundaries was pioneered by the theoretical works of Graetz \citep{Graetz:1885}, Nusselt \citep{nusselt1916} and \Leveque \citep{leveque28} for  two-dimensional problems. They give analytical or scaling predictions for the scalar flux and the associated non-dimensional  transfer coefficient: the Sherwood  and Nusselt   numbers for mass and heat transfer, respectively.
Mass transfer problems have  benefited from progress in the understanding of heat transfer  \citep[e.g.][]{Bejan}, since heat and mass transfer problems are equivalent when both scalars are passive  or have the same properties. Henceforth, we  refer to the generic scalar non-dimensional  transfer coefficient as the Sherwood number, $\Sh$, for simplicity, as we assume a passive scalar in this study.  This assumption implies that the scalar transport equation and the governing equation for the flow are not fully coupled, such that the flow is independent of the tracer concentration, whereas the concentration field depends on the flow field. Thus, buoyancy or temperature changes that could affect the flow field are beyond the scope of this study. Nevertheless, our results  apply to analogous heat transfer problems provided that the temperature difference is sufficiently small. We will revisit these assumptions and their effect on the results in section \ref{sec:discussion}. 

Although  numerical simulations can now solve almost any  scalar transport problem with complex boundary conditions or geometries, the ease of use of simple theoretical predictions is still highly valuable for a broad range of applications.  
Theoretical models mostly rely on the key, widely-used simplifying assumption that the scalar transport problems modelled can be approximated by  two-dimensional problems. Transfer problems in steady axisymmetric channel flows with uniform lateral boundary conditions (e.g. Dirichlet or Neumann) can directly use  the two-dimensional axisymmetric theoretical results of Graetz: $\Sh \propto Re^\alpha Sc^\beta (\hat{D}/\hat{L})^\gamma$ \citep[e.g.][]{Bejan}, with $Re$  the Reynolds number, $Sc$  the Schmidt number, and $\hat{D}/\hat{L}$ the ratio of the  channel  diameter and the length of the  scalar transfer area. (Throughout this paper, hats denote dimensional quantities and dimensionless quantities remain undecorated.)   The positive exponents $\alpha$, $\beta$ and $\gamma$ vary depending on the flow profile (e.g. uniform or shear flow) and regime (laminar or turbulent), the wall roughness and whether the diffusive and momentum boundary layers are full developed or not. Non-axisymmetric three-dimensional problems, such as rectangular channel flows, also rely on empirical or asymptotic correlations based on Graetz' two-dimensional results by modifying the  Sherwood number such that: $\Sh \propto Re^\alpha Sc^\beta (\hat{D}_h/\hat{L})^\gamma$ \citep[e.g.][]{gekas87,bowen95}. The three-dimensional variations of the scalar field from a two-dimensional axisymmetric profile are thus captured  by the ratio $(\hat{D}_h/\hat{L})^\gamma$ in the relationship above, where the hydraulic diameter $\hat{D}_h$ accounts for non-circular channel cross-section. The key underlying assumption allowing non-axisymmetric three-dimensional problems to be modelled as two-dimensional axisymmetric problems is that the scalar boundary condition  is uniform, i.e. not mixed, at the side walls. 
This assumption has proven useful to many mass \citep[e.g. reviews][]{gekas87,bowen95} and heat \citep[e.g. reviews][]{sobhan01,ayub03,avelino04}  transfer problems.

Three-dimensional channel flows with mixed or differing scalar boundary conditions at the side walls can also be simplified to two-dimensional planar problems provided that the side walls with different boundary conditions  have a negligible effect on the overall transfer flux. This assumption is typically used when channel widths are larger than heights \citep[e.g.][]{squires08,Braff:2013}. Two-dimensional planar problems can then use advanced mathematical techniques such as potential flow and conformal mapping \citep{bazant04,choi05}, which provide analytical or semi-analytical solutions for any complex (planar) geometries. 

\begin{figure}
\centering
\includegraphics[keepaspectratio=true,width=0.6\textwidth]{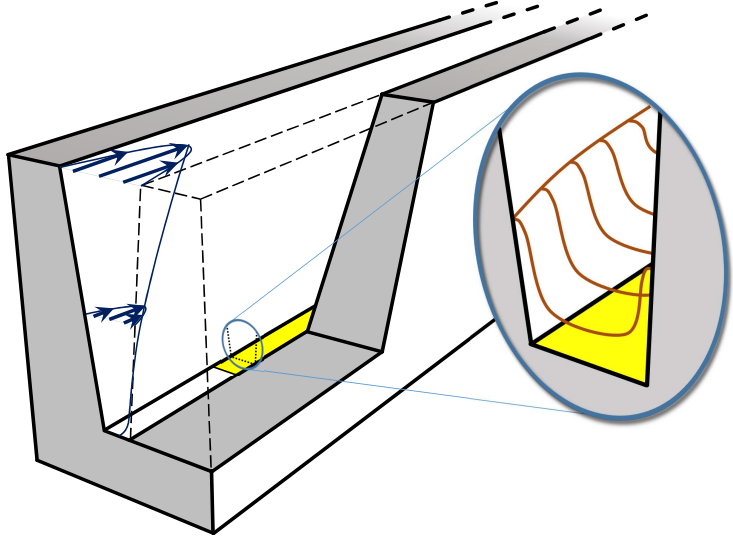}
\caption{
Schematic diagram of the scalar transfer problem in a narrow channel flow with a `truncated wedge' profile comprising a flat bottom and inclined walls. The shaded region at the base of the channel (yellow) represents the source of the scalar. The arrows at the left-hand end of the sketch show the three-dimensional profile of the velocity field and the enlargement of the section just beyond the start of the source of scalar shows the three-dimensional structure of the developing scalar  boundary layer. A real-life example would be the mass transfer from a flat viscous contaminant droplet trapped in a gap or crack \citep{Landel:2016}.}
\label{fig:conceptSketch}
\end{figure}

However, not all transport problems can be a priori reduced to simple axisymmetric or planar two-dimensional  problems. Many problems possess three-dimensional flow and scalar fields owing to three-dimensional geometries and differing lateral boundary conditions, thus rendering  analytical progress intractable. 
The main objective of this study, to predict the scalar flux and the Sherwood number as a function of the flow, scalar properties and geometry, requires us to analyse the impact of three-dimensional effects.
We focus on three-dimensional transport problems in laminar steady fully-developed longitudinal open channel flows with generic rectangular or {truncated} wedge geometries. As depicted in  figure~\ref{fig:conceptSketch}, we study the case where we have different scalar boundary conditions at the side walls with: fixed Dirichlet boundary condition  at the base of the channel, and no-flux boundary condition on all  other boundaries. Transport occurs at high \Peclet numbers such that a scalar boundary layer develops from the base of the channel. We define the channel aspect ratio as the ratio of the characteristic channel `height' $\hat{H}$, in the direction perpendicular to the base of the channel, to the characteristic channel width $\hat{w}$, in the lateral direction.
Three-dimensional effects are more significant when the channel aspect ratio is large and the  scalar boundary layer is narrowly confined in the lateral direction. We describe these geometries as `open channels' in the sense that when the channel has a finite height a free-slip boundary condition is assumed at the boundary opposite the base, and require this boundary to have a  width larger or equal to that of the base. As illustrated in figure~\ref{fig:conceptSketch}, the contour lines of the scalar field in  cross-sections of the channels can be strongly curved, whilst the profiles develop in the longitudinal direction. This  is due to the no-slip and no-flux boundary conditions {(for the velocity and scalar, respectively)} on the near side walls.

The problem considered here is a complex three-dimensional transport problem which has received little attention  in the literature for various reasons. For example, many engineering applications in heat and mass transfer seek to \textit{maximise} the interfacial transfer and thus tend to use geometrical design with aspect ratios corresponding to a ``thin layer'', such that the width of the channel is much larger than its height $\hat{w}\gg \hat{H}$. A large number of studies in the heat and mass transfer literature have thus focussed on enhancing the scalar transfer (i.e. the Sherwood number, $\Sh$, or the Nusselt number) in the small aspect ratio limit $\hat{H}/\hat{w}\ll 1$. However, the main novelty of our study is to focus on the opposite limit, $\hat{H}/\hat{w}\gg 1$ or the ``narrow channel'' limit, where $\Sh$ is naturally reduced due to confinement effects. This is less attractive for most engineering applications, which may explain why much less research has been done in the narrow channel limit. Importantly, in the narrow channel limit  traditional two-dimensional  approaches \citep[e.g.][]{Graetz:1885,nusselt1916,leveque28,gekas87,bowen95,sobhan01,ayub03,avelino04,squires08,Braff:2013,Bejan,dejam14,kou19,dejam19} that generally work in the thin layer limit cannot  be used \textit{a priori} since the flow and  scalar fields are both inherently three-dimensional. This is the central point that motivates our study and which should be of interest to  interfacial transfer problems where three-dimensional effects cannot be neglected.

The scenario shown in figure~\ref{fig:conceptSketch} closely models mass transfer applications in narrow spaces such as the cleaning and decontamination of  gaps, cracks and fractures. This kind of cleaning problems exist in most industrial activities and are of particular concern in the food \citep{Wilson:2005,fryer09,lelieveld14}, chemical \citep{pentsak19}, pharmaceutical and cosmetic industries, where purity, hygiene and cleanliness are essential. This scenario is also relevant to the decontamination of toxic liquid materials trapped in confined channels where the flow is laminar \citep{fitch03,settles06}. There are also potential applications to the pore-scale modelling of mass transfer phenomena in porous media, for instance in the context of aquifer remediation \citep{borden92,kahler16}, if the micropores have a  rectangular or truncated-wedge geometry.
Another application is for the transport of ions in membraneless electrochemical cells. In this last case, to obtain the ion flux and  deduce the current produced by the fuel cell, \cite{Braff:2013} assumed a two-dimensional plug flow between  electrodes in large  aspect ratio channels in order to simplify the ion transport problem. However, the laminar flow in this geometry is fundamentally three-dimensional, also resulting in a three-dimensional ion concentration field owing to differing boundary conditions at the side walls. Our study provides a posteriori justification for the two-dimensional assumption made by \cite{Braff:2013} and quantifies the associated error. 
The impact of three-dimensional effects has also been reported in microfluidic channels such as the T-sensor \citep{Kamholz:1999,Ismagilov:2000,Kamholz:2001,Kamholz:2002,Stone:2004}. \cite{Jiminez:2005} showed with numerical and asymptotic techniques that shear flows near the no-flux and no-slip solid boundaries at the side walls lead to wall boundary layers. His results confirmed the  power-laws found by \cite{Kamholz:2002} for the far-field region but not the initial square-root power-law. \cite{Jiminez:2005} also observed that, compared to the well-known case of longitudinal diffusion in a tube \citep[`Taylor dispersion';][]{taylor53}, the impact of the wall boundary layers on the effective mass transport is weak, the spreading rate changing by less than \SI{5}{\percent} between the near and far-field regions.

To achieve our objective of understanding mass transport, we use asymptotic analysis and numerical simulations to determine the main impact of three-dimensional effects. We seek to elucidate the different regimes that exist and what controls the transition between them, and to demonstrate that in each case an appropriate two-dimensional model can be developed that provides a good approximation to $Sh$.
These findings have important theoretical and practical implications. Theoretically, it could enable the use of more advanced two-dimensional mathematical techniques in the case of more complex longitudinal profiles of the channel geometry \citep{bazant04,choi05}. Practically, it enables computation of transfer fluxes in complex three-dimensional  applications  using simpler and faster techniques, whilst having clear estimates of the error made. This is particularly useful for end-users who may not have access to sophisticated computational tools or methods.

We begin by defining the problem  in \S\ref{sec:problemstatement}. In \S\ref{sec:flowfieldscale} we solve Stokes' equation to obtain an analytical solution for the  three-dimensional velocity field in
rectangular channels with parallel walls and truncated-wedge channels with angled walls. We introduce the three-dimensional scalar transport problem and a two-dimensional cross-channel averaged formulation in \S\ref{sec:scalartransport}. For channels with parallel walls, we use scaling arguments to obtain similarity solutions for the flux in cases where the diffusive boundary layer is much thinner (\S\ref{sec:straightwallsthin}) or much thicker (\S \ref{sec:straightwallsthick}) than the channel width. In \S\ref{sec:confinmentParallel}, vertical confinement effects are studied through a depth-averaged  advection--diffusion equation. In \S\ref{sec:straightwallstnumerics} and \S\ref{sec:ResultParallel}, three-dimensional numerical solutions of the transport problem demonstrate that two-dimensional  results give  accurate predictions for the Sherwood number across all \Peclet numbers, including those where asymptotic approaches are not valid.  In \S\ref{sec:angledwallsthin}, we study the thin boundary layer regime for the truncated wedge geometry and show asymptotically that the opening wedge geometry leads to a small increase in the flux compared to the parallel wall geometry. In \S\ref{sec:thickBLwedge}, thick boundary layers are studied for the  wedge geometry, revealing a much more complex behaviour due to the impact of the opening angle on diffusion through curvature effects and advection. In \S\ref{sec:confinmentwedge}, vertical confinement effects are studied for the truncated wedge geometry.  In \S\ref{sec:angledwallstransition} and \S\ref{sec:ResultParallelwedge}, three-dimensional numerical results for the truncated wedge geometry show  that appropriate two-dimensional  results give accurate predictions for the mass transfer in this geometry across all \Peclet numbers studied and for small opening angles. 
A more complex dependence with \Peclet number and geometry is found for the thick boundary layer regime. We also demonstrate the importance of a curvature-rescaled \Peclet number in this regime. In \S\ref{sec:discussion}, we discuss  implications of  our results for practical applications such as cleaning and decontamination    in confined channels.
In \S\ref{sec:conclusion} and table~\ref{tab:expcharacpara}, we summarize  all our scaling and asymptotic results for the Sherwood number in  the various regimes identified.

\begin{figure}%
\includegraphics[keepaspectratio=true,width=\textwidth]{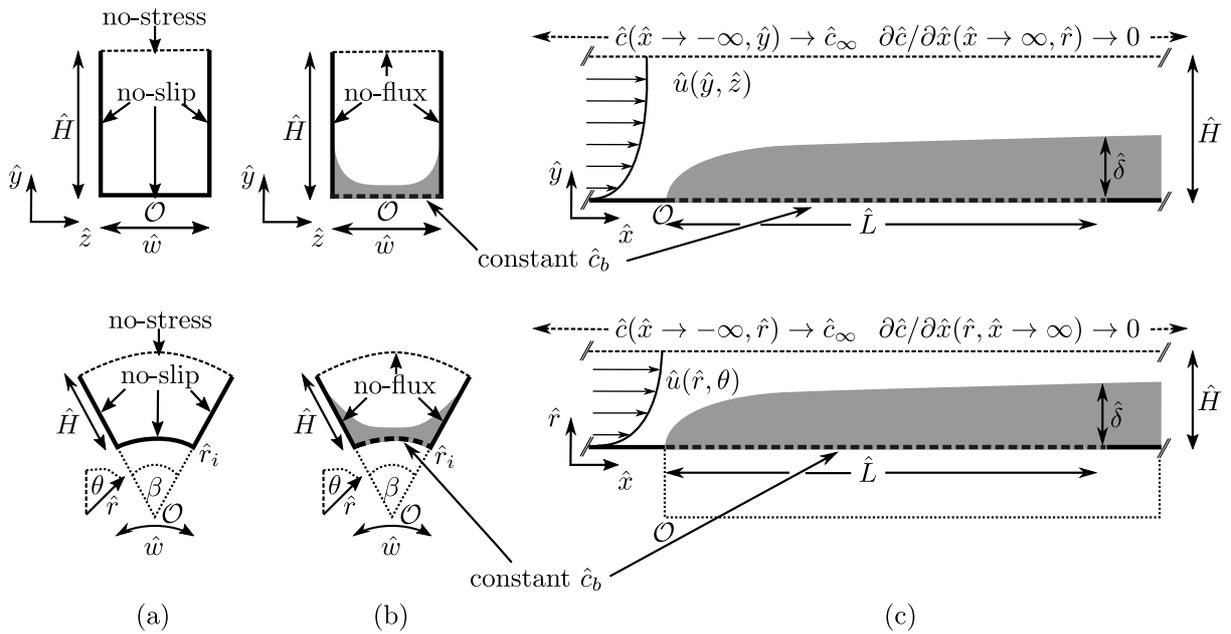}
\caption{
Schematic of  advection--diffusion  problem for a passive scalar of concentration $\hat{c}$. Top row:  rectangular channel geometry with parallel walls. Bottom row: truncated-wedge geometry with angled walls.  (\textit{a}) Cross sections with flow boundary conditions. (\textit{b}) Cross sections at $0<\hat{x}<\hat{L}$ with concentration boundary conditions. We impose $\hat{c}=\hat{c}_b$ at the channel base for $0< \hat{x} < \hat{L}$ (dashed lines); typical diffusive boundary layer of the concentration field (thickness $\hat{\delta}$) shown in light grey.  (\textit{c}) Side views at $\hat{z}=0$ (top) and $\theta=0$ (bottom) with boundary conditions; typical  velocity field $\hat{u}$ shown with arrows.}
\label{fig:geometry}
\end{figure}

\section{Model description}\label{sec:problemstatement}


We model the steady advective--diffusive transport of a passive scalar released from an area of length $\hat{L}$ in the flow direction and width $\hat{w}$. The release area, at the base of an infinitely long channel, is assumed to have zero thickness and have no effect on the velocity field. We study two generic three-dimensional geometries: a rectangular channel with parallel walls of arbitrary width $\hat{w}$ and  arbitrary height $\hat{H}$ (figure~\ref{fig:geometry}, top row); and a channel forming a truncated wedge with a base in the form of an arc of a circle and flat side walls (figure~\ref{fig:geometry}, bottom row). (Here we use the term `height' to represent the normal distance between the base and its opposite boundary or `top boundary' without reference to the direction of gravity.) The opening angle of the wedge is $\beta >0$ and the arc length at the base of the channel is $\hat{w}=\hat{r}_i \beta $, with $\hat{r}_i$ the truncation radius. In this study, we generally focus on the case of narrow channels, $\hat{w} \ll \hat{H}$. However, our problem formulation is sufficiently general so that  we are also able to discuss some results for $\hat{w} \sim \hat{H}$ and $\hat{w} \gg \hat{H}$.

For rectangular channels with parallel walls, we use Cartesian coordinates $(\hat{x},\hat{y},\hat{z})$, where $\hat{x}$ denotes the streamwise coordinate, $\hat{y}$ the direction normal to the channel base, and $\hat{z}$ the cross-channel direction. The origin $\mathcal{O}$ of the axes is placed at the intersection of the planes $\hat{x}=0$, the onset of the release area, $\hat{y}=0$, the base of the channel,  and $\hat{z}=0$, the channel mid-plane. We refer to this geometry as a parallel-wall channel hereafter.


For truncated wedges with angled walls, we use cylindrical coordinates $(\hat{x},\hat{r},\theta)$, where $\hat{x}$ denotes the streamwise coordinate, $\hat{r}$ the direction perpendicular to the base of the channel, $\theta$ the azimuthal direction. The origin $\mathcal{O}$ is placed at the intersection between the plane $\hat{x}=0$, the onset of the area of release, and the axis $\hat{r}=0$, the edge of the wedge before truncation. For small angles $\beta$, the curvature of the base could be neglected and the base of the channel  considered flat, thus approximating the channel sketched in figure~\ref{fig:conceptSketch}. We refer to this geometry as a truncated wedge hereafter.

The steady low-Reynolds-number open flow (see \S\ref{sec:flowfieldscale}) in either form of channel is taken as unidirectional and independent of $\hat{x}$. The cross-sectional structure is controlled by the combination of the no-slip boundary conditions (see figure~\ref{fig:geometry}(\textit{a})) on the side walls and base of the channel, and an assumed stress-free condition at the top located at $\hat{y}=\hat{H}$ or $\hat{r}=\hat{r}_i+\hat{H}$. 
The top boundary condition is an approximation for a liquid--gas interface, which could be curved due to surface tension effects. Surface tension and curvature effects at the top boundary are neglected in this study.

The passive scalar transport with concentration $\hat{c}$ is modelled using a steady advection--diffusion equation (see \S\ref{sec:scalartransport}). 
The area of release has a fixed concentration $\hat{c}_b>\hat{c}_{\infty}\geq 0$ (with $\hat{c}_{\infty}$ a fixed background concentration) over the region given by $0 < \hat{x} < \hat{L}$, $\hat{y}=0$ and $-\hat{w}/2< \hat{z}< \hat{w}/2$ for parallel-wall channels. Similarly, the area of release for truncated wedges is over $0 < \hat{x} < \hat{L}$, $\hat{r}=\hat{r}_i$ and $-\beta/2< \theta < \beta/2$. These regions are shown in figure~\ref{fig:geometry}(\textit{b,c}) for parallel-wall channels (top row) and  wedges (bottom row), respectively.  
All the channel walls  have a no-flux boundary condition, except for the area of release. In cases where we consider an infinite fluid layer thickness, we assume $\hat{c}\to 0$ at $\hat{y}\to +\infty$ or $\hat{r}\to +\infty$. Otherwise, for a finite fluid layer thickness, we impose a no-flux boundary condition at $\hat{y}=\hat{H}$ or $\hat{r}=\hat{r}_i+\hat{H}$. Upstream, we impose $\hat{c}\to \hat{c}_{\infty}$ for $\hat{x}\to -\infty$, and downstream, $\partial \hat{c}/\partial \hat{x} \to 0$ for $\hat{x}\to +\infty$.

\section{Flow field}\label{sec:flowfieldscale}

We assume an incompressible Stokes' flow.  Since the tracer is assumed passive, the governing equation for the fluid flow  is independent of the tracer concentration.  From the boundary conditions shown in figure \ref{fig:geometry}, by symmetry, the flow field has only a streamwise component $\hat{u}$, which depends  on $\hat{y}$ and $\hat{z}$ (respectively $\hat{r}$ and $\theta$ for  truncated wedges). The flow is driven by a constant streamwise gradient $\hat{G}=\partial\hat{P}/\partial\hat{x}<0$ in the non-hydrostatic component of the pressure $\hat{P}$, which could be created by gravity for instance. Thus, the flow is  three-dimensional in both geometries. Since we want to analyse three-dimensional effects on the scalar transport, it is  important to capture the  dependence of the flow with both coordinates.
We non-dimensionalise  spatial variables with the channel width or base arc length $\hat{w}$, the  length scale for the flow at the channel base,
\begin{equation}\label{eq:Stokesnormalisation}
 y =\frac{\hat{y}}{\hat{w}},\  z  = \frac{\hat{z}}{\hat{w}}, \  r =\frac{\hat{r}-\hat{r_i}}{\hat{w}}=\frac{\hat{r}}{\hat{w}}-\beta^{-1}, \ \textrm{and} \ H = \frac{\hat{H}}{\hat{w}},
\end{equation}
with $r$  the distance from the base of the truncated wedge, similar to $y$. (Since the flow is independent of $\hat{x}$, we defer its non-dimensionalisation  until \S \ref{sec:scalartransport}.) All velocities are non-dimensionalised with the characteristic velocity
$\hat{U}_0 = -\hat{G}  \hat{w}^2/(12 \hat{\mu})>0$,
with $\hat{\mu}$  the dynamic viscosity.
The factor of $1/12$ preserves the intuitive physical meaning of the cross-channel averaged velocity in channels with parallel walls far away from the base. 

\subsection{Flow field in channels with parallel walls}\label{sec:cartesianflowfield}

\begin{figure}
\centering
\includegraphics[width=\textwidth]{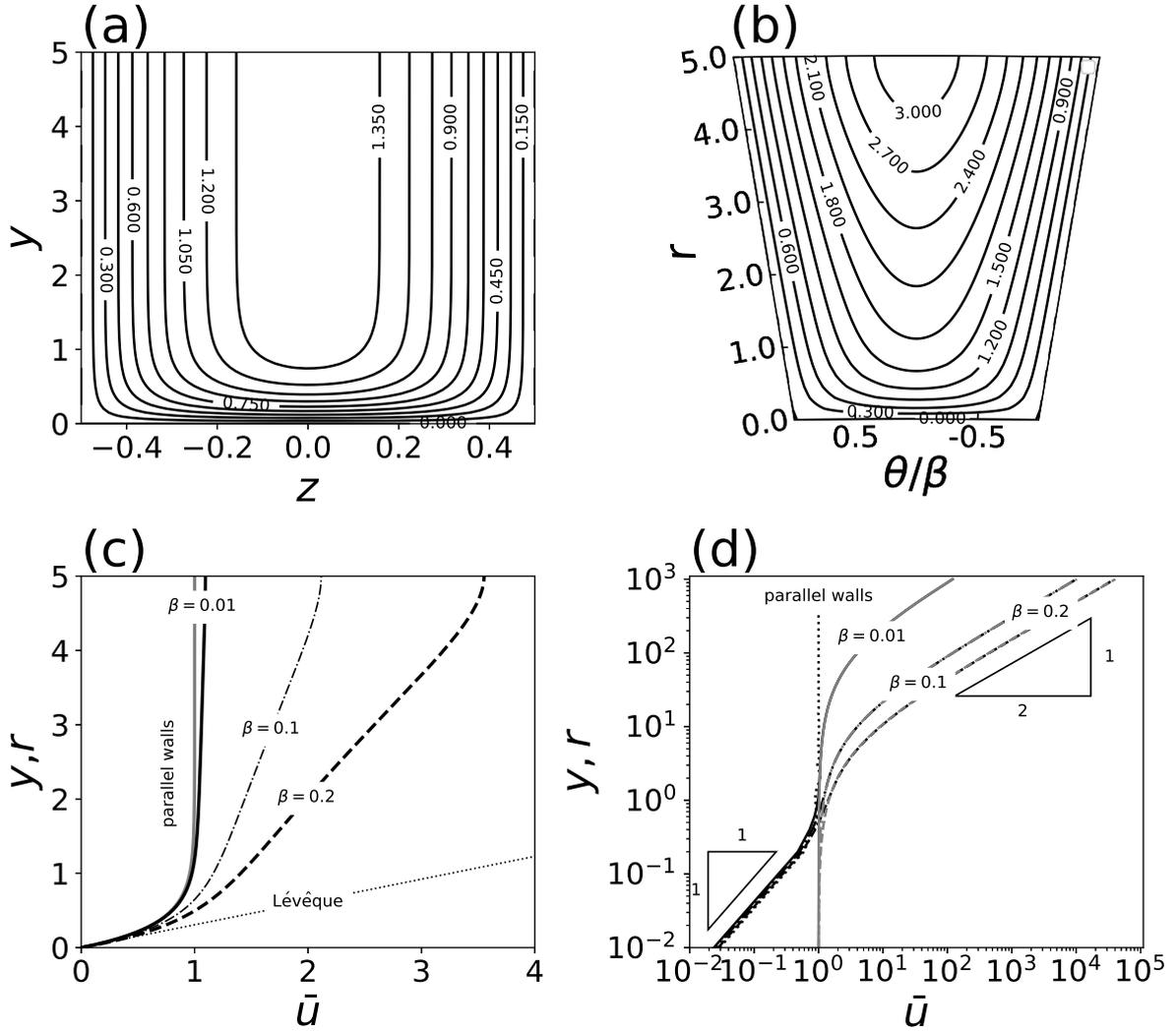}
\caption{Contour plots of the velocity $u$ in (\textit{a}) a parallel  channel ($H=5$) following (\ref{eq:straightflowfield}), and (\textit{b}) a   wedge ($\beta = 0.1$, $H=5$) following (\ref{eq:angledwallsflowfield}). (c) Vertical ($y$-) and radial ($r$-) profiles of the cross-channel averaged  velocity $\zavg{u}$ for both geometries with $H=5$. The \Leveque approximation $\overline{u} = \overline{\gamma}y$ (dotted line) uses (\ref{eq:shearUchannel}). 
(\textit{d}) Plot of $\zavg{u}$ in wedges, and in parallel channels (dotted line) for comparison. The far-field  velocity at small angles $u_W$  uses (\ref{eq:wedgefarfield}) (corresponding grey curves closely following the black curves for $r>1$). 
}
\label{fig:velocityfigure}
\end{figure}

The dimensionless Stokes equation for the flow in  channels with parallel walls is
\begin{equation}
\frac{\partial^2 u}{\partial y^2}+\frac{\partial^2 u}{\partial z^2} =-12,
\end{equation}
for $-1/2<z<1/2$, $0<y<H$, with boundary conditions (figure~\ref{fig:geometry}, top row) 
\begin{equation}\label{eq:bcwithparallelwallsForAppendix}
u( y =0, z ) = 0,\ \frac{\partial u}{\partial  y }( y =H, z ) = 0,\ u( y ,  z = \pm 1/2) = 0.
\end{equation} 
The solution of this inhomogeneous problem is described by the infinite series
\begin{equation}\label{eq:straightflowfield}
u( y , z ) = 12H y -6 y ^2-\sum_{n=0}^{+\infty} C_n\sin(\lambda_n  y )\cosh(\lambda_n  z ),
\end{equation}
where the eigenvalues $\lambda_n$ and  coefficients $C_n$ are, for all integers $n\geq 0$,
\begin{equation}\label{eq:lambdaCn}
\lambda_n = \frac{2n+1}{2H}\pi,\ C_n = \frac{192H^2}{\pi^3(2n+1)^3 \cosh \left(\lambda_n/2\right)}.
\end{equation}
The velocity  (\ref{eq:straightflowfield}) is shown in figure \ref{fig:velocityfigure}(\textit{a}) for $H=5$, truncated after 1000 terms. The flow is clearly three-dimensional near the base of the channel owing to the influence  of the solid boundaries on three sides.
However, for $1\ll  y  \leq H$, the influence of the solid base decreases and the velocity field tends to a two-dimensional Poiseuille profile  
\begin{equation}\label{eq:straightwallsfarfield}
u_P( z ) = \frac{3}{2}\left(1 - 4 z ^2\right),
\end{equation}
valid only for $H\gg 1$. 
For $ y  \ll 1$, the flow  is influenced by the base and $u\approx \gamma y$, where $\gamma = \hat{\gamma}/(\hat{U}_0/\hat{w})$ is the  dimensionless shear rate at $ y =0$. In general, the shear rate is 
\begin{equation}\label{eq:shearUchannel}
\gamma( z )= \left. \frac{\partial u}{\partial  y }\right|_{ y =0} = 12 H-\sum_{n=0}^{+\infty} C_n \lambda_n \cosh(\lambda_n  z ).
\end{equation}
The dependence of $\gamma$ with $z$ is important   for $H \gtrsim 1$. For $H\ll 1$, $\gamma$ is uniform and approaches the semi-parabolic Nusselt film limit in the interior of the channel, owing to vertical confinement effect, with a dependence with $z$ limited to the corners, $|z|\to 1/2$.
Although (\ref{eq:shearUchannel}) contains $H$ as a parameter, for  $H>1$  the cross-channel average of the shear rate appears to be independent of $H$ and approaches $\zavg{\gamma}\approx 3.26$ asymptotically  rapidly (see figure \ref{fig:GammaVSHandBeta}(\textit{a}), appendix~\ref{apdx:AddMat}). This is  related to the fact that we impose a constant streamwise pressure gradient to drive the flow in the channel.

We also plot the cross-channel averaged velocity $\zavg{u}$, with $\zavg{{\boldsymbol\cdot}}=\int_{-1/2}^{1/2} {\boldsymbol\cdot} \, \mathrm{d}z$, in figure \ref{fig:velocityfigure}(\textit{c}) (solid grey line), along with the two asymptotic limits: $\zavg{u}\sim \gamma y$ for $y  \ll 1$ and $\zavg{u}\sim 1$ for $y  \gg 1$. When analysing  the scalar transport in the next sections, we will decompose the velocity  such that $u=\zavg{u}+u'$,  where $\zavg{u}=\zavg{u}(y)$ and $u'=u'(y,z)$. Thus,  three-dimensional effects related to the flow are contained in the cross-channel variation velocity $u'$.

\subsection{Flow field in  a truncated wedge channel }\label{sec:polarflowfield}

The dimensionless Stokes equation for the flow in truncated wedge channels is
\begin{equation}\label{eq:floweqangled}
\left( \frac{\partial^2 u}{\partial  r ^2}+\frac{1}{ r +\beta^{-1}} \frac{\partial u}{\partial  r } + \frac{1}{( r +\beta^{-1})^2} \frac{\partial^2 u}{\partial \theta^2}\right) = -12,
\end{equation}
for $0 <  r < H$, $-\beta/2< \theta < \beta/2$,
with boundary conditions (figure~\ref{fig:geometry}, bottom row)
\begin{equation}
u( r =0,\theta) = 0,\ \frac{\partial u}{\partial  r }(r=H,\theta) = 0,\ u( r ,\theta = \pm \beta/2) = 0.
\end{equation}
Similar to  parallel channels (see \S\ref{sec:cartesianflowfield}), the solution for the velocity  is three-dimensional, 
\begin{equation}\label{eq:angledwallsflowfield}
u( r ,\theta)  =  6 (H+\beta^{-1})^2 \ln (1+\beta r )-3  r  ( r +2\beta^{-1})
-\sum_{n=0}^{+\infty} S_n \sin \left(\chi_n \ln(1+\beta  r ) \right)\cosh(\chi_n \theta),
\end{equation}
where the eigenvalues $\chi_n$ and the coefficients $S_n$ are, for all integers $n\geq 0$,
\begin{eqnarray}
\chi_n &=& \frac{(2n+1)}{2\ln(1+\beta H)}\pi,\\
S_n &=& \frac{192\beta^{-2} \left((2n+1)\pi \ln ^2(1+\beta H) + 4 (-1)^n (1 + \beta H)^2 \ln ^3(1+\beta H) \right)}
{(2n+1)^2\pi^2 \left((2n+1)^2\pi^2 + 16 \ln^2(1+\beta H)\right)  \cosh(\chi_n\beta /2) }.
\end{eqnarray}

In figure~\ref{fig:velocityfigure}(\textit{b}) we show contour plots of the velocity  (\ref{eq:angledwallsflowfield}) in a channel with  $\beta=0.1$, $H=5$ (see  table~\ref{tab:numdet} in appendix~\ref{apdx:NumSimDet} for the number of eigenvalues used).  For small opening angles, the flow field is similar to parallel channels (figure~\ref{fig:velocityfigure}\textit{a}). Far away from the top and  base boundaries but closer to the side walls, for  $(\beta  r +1) \ll  r  \ll H$, (\ref{eq:floweqangled}) simplifies to $\partial^2u/\partial\theta^2=-12( r +\beta^{-1})^2$, which gives, at leading order,
\begin{equation}\label{eq:wedgefarfield}
u_W( r ,\theta) = \frac{3}{2} (\beta   r +1)^2 \left(1- 4\frac{\theta^2}{\beta^2}\right).
\end{equation}
In contrast with the far-field velocity in parallel channels (see  (\ref{eq:straightwallsfarfield})), the far-field velocity $u_W$ in truncated wedges remains three-dimensional, except in the limit $\beta\ll 1/r\ll 1$. 

We  plot in figure \ref{fig:velocityfigure}(\textit{c},\textit{d}) the cross-channel averaged  velocity $\tavg{u}$, with $\tavg{{\boldsymbol\cdot}}=\beta^{-1} \int_{-\beta/2}^{\beta/2} {\boldsymbol\cdot} \, \mathrm{d}\theta$,
for opening angles $\beta=0.01$ (black solid line), $\beta=0.1$ (black dash-dotted line), and $\beta=0.2$ (black dashed line)  for $H=5$ (\textit{c}) and $H=1000$ (\textit{d}). Note that the noticeable change in slope for $\tavg{u}$ near $r=5$ for $\beta=0.1$ and 0.2   is due to the no-stress boundary condition at the top. 
Near the base, for $ r \ll 1$, $\tavg{u}\approx \gamma(\theta) r $, similar to  parallel  channels, whilst in the far field $\tavg{u}\sim  (\beta r)^2$, characteristic of a far-field flow in a narrow wedge.
The shear rate at the base of the channel is given by 
\begin{equation}\label{eq:wedgegamma}
\gamma(\theta) = \left. \frac{\partial u}{\partial  r }\right|_{ r =0} = 12H + 6 \beta H^2 - \sum_{n=0}^{+\infty} S_n \chi_n\beta \cosh(\chi_n \theta).
\end{equation}
The dependence on $\theta$ vanishes in the interior of the channel for $H\ll 1$, owing to radial confinement effects, where it is limited to the corners, $|\theta|\to \beta/2$. The cross-channel average $\tavg{\gamma}$ depends on $\beta$ and $H$. For $\beta \rightarrow 0$, $H\gg 1$, $\tavg{\gamma}$ rapidly approaches the value for parallel channels: $\tavg{\gamma}\approx 3.26$. For larger $\beta$, $H\leq 100$,  $\zavg{\gamma}$  is approximately: 3.47 for $\beta=0.1$, 4.00 
for $\beta=0.3$, and 8.68 for $\beta=1$, see also appendix~\ref{apdx:AddMat} and figure~\ref{fig:GammaVSHandBeta}(\textit{b}).

We will use the decomposition $u=\tavg{u}+u'$, with $\tavg{u} = \tavg{u}(r)$ and $u'=u'(r,\theta)$), in the next sections to study the impact of the three-dimensional  cross-channel azimuthal variations $u'$ on scalar transport in wedges.


\section{Scalar transport}\label{sec:scalartransport}

As noted in \S \ref{sec:intro}, the objective of this work is to determine the impact of three-dimensional effects on the flux of a passive scalar released from the base of a channel flow in the two geometries described in figure~\ref{fig:geometry}. The steady transport of a passive scalar is governed by the general advection--diffusion equation, assuming Fick's law for molecular diffusion.
We focus on the case where the scalar concentration field forms a slender diffusive boundary layer that develops in the $\hat{x}$ direction such that $\hat{\delta}/\hat{L}=Pe_L^{-1/2}=(\hat{U}_{\delta}\hat{L}/\hat{D})^{-1/2} \ll 1$, with $\hat{\delta}$ a characteristic  diffusive boundary layer thickness,  $\hat{U}_{\delta}$ a characteristic streamwise velocity at $\hat{y}\sim \hat{\delta}$,  and $\hat{D}$ the scalar diffusivity. This implies that  streamwise diffusion is negligible \citep{Bejan}. 
As in (\ref{eq:Stokesnormalisation}), we use $\hat{w}$ and $\hat{U}_0$ as non-dimensionalising quantities. We also use the following non-dimensionalisation
\begin{equation}
x = \frac{\hat{x}}{\hat{w} Pe_w},\  L  = \frac{\hat{L}}{\hat{w}},\  \delta =\frac{\hat{\delta}}{\hat{w}},\ \mathrm{and}\  c =\frac{\hat{c}-\hat{c}_\infty}{\hat{c}_b-\hat{c}_\infty},
\end{equation}
where $\hat{x}$ has been rescaled with the \Peclet number $Pe_w = \hat{U}_0\hat{w}/\hat{D}$. We  choose $\hat{w}$ as the characteristic length scale for the  transport problem since the ratio between the diffusive boundary layer thickness $\hat{\delta}$ and the gap width $\hat{w}$ is key to describe the different regimes for the scalar transport and resulting flux. The advection--diffusion equation for parallel channels is then
\begin{equation}\label{eq:cartesianfullproblem_a}
u\frac{\partial c}{\partial x} =  \frac{\partial^2 c}{\partial y^2}  +\frac{\partial^2 c}{\partial z^2}, 
\end{equation}
for $0<x<L/Pe_w$, $0<y<H$,  $\left|z\right| < 1/2$, with boundary conditions (figure~\ref{fig:geometry})
\refstepcounter{equation}\label{eq:cartesianfullproblem_aBC}
\begin{align}
& c(x=0,y,z)=0, \tag{\theequation \textit{a}}\label{eq:cartesianfullproblem_aBCa}\\
& c(x,y=0,z)=1,\ c(x,y\to+\infty,z)\to 0 \ \textrm{or}\ \fp{c}{y}{}(x,y=H,z)=0, \tag{\theequation \textit{b--d}}\label{eq:cartesianfullproblem_aBCb-d}\\
& \fp{c}{z}{}(x,y,z=\pm 1/2)=0. \tag{\theequation \textit{e,f}}\label{eq:cartesianfullproblem_aBCe-f}
\end{align}
For truncated wedge channels, the governing advection--diffusion equation is
\begin{equation}\label{eq:angledwallblequation}
u\frac{\partial c}{\partial x} =  \frac{\partial^2 c}{\partial r^2}+\frac{1}{(r+\beta^{-1})}\frac{\partial c}{\partial r} + \frac{1}{(r+\beta^{-1})^2} \frac{\partial^2 c}{\partial \theta^2},
\end{equation}
for $0<x<L/Pe_w$, $0<r<H$, $ |\theta| < \beta/2$, with boundary conditions (figure~\ref{fig:geometry})
\refstepcounter{equation}\label{eq:angledwallblequationBC}
\begin{align}
& c(x=0,r,\theta)=0, \tag{\theequation \textit{a}}\label{eq:angledwallblequationBCa}\\
& c(x,r=0,\theta)=1,\ c(x,r\to+\infty,\theta)\to 0 \ \textrm{or}\ \fp{c}{r}{}(x,r=H,\theta)=0, \tag{\theequation \textit{b--d}}\label{eq:angledwallblequationBCb-d}\\
& \fp{c}{\theta}{}(x,r,\theta=\pm\beta/2)=0. \tag{\theequation \textit{e,f}}\label{eq:angledwallblequationBCe-f}
\end{align}
The concentration field $c$ and resulting flux can be fully determined by solving (\ref{eq:cartesianfullproblem_a}) and (\ref{eq:angledwallblequation}) for $0<x<L/Pe_w$ using the velocity  $u$ defined in (\ref{eq:straightflowfield}) and (\ref{eq:angledwallsflowfield}), respectively.

In regimes dominated by  cross-channel diffusion, we  use the cross-channel average of  (\ref{eq:cartesianfullproblem_a}) and (\ref{eq:angledwallblequation}) to determine the cross-channel averaged concentration  and the  flux. As introduced previously, we use $u=\zavg{u}+ u'$ and $c=\zavg{c}+ c'$, where overbars denote  cross-channel averages (along the $z$-direction for parallel channels and along the $\theta$-direction for wedges), and primes indicate  cross-channel variations. We obtain for parallel channels
\begin{equation}\label{eq:crosschannelavg1}
\zavg{u}\frac{\partial \zavg{c}}{\partial x}+\frac{\partial }{\partial x}\zavg{u' c'} = \frac{\partial^2 \zavg{c}}{\partial y^2},
\end{equation}
for $0<x<L/Pe_w$, $0<y<H$,  with boundary conditions
\refstepcounter{equation}\label{eq:crosschannelavg1BC}
\begin{equation}
\zavg{c}(x=0,y)=0,\ \zavg{c}(x,y=0)=1,\ \zavg{c}(x,y\to+\infty)\to 0 \ \textrm{or}\ \fp{\zavg{c}}{y}{}(x,y=H)=0.
 \tag{\theequation \textit{a--d}}\label{eq:crosschannelavg1BCa--d}
\end{equation}
For truncated wedge channels we obtain
\begin{equation}\label{eq:crosschannelavg2}
\tavg{u}\frac{\partial \tavg{c}}{\partial x}+\frac{\partial }{\partial x} \left(\tavg{u'c'}\right) =  \frac{\partial^2 \tavg{c}}{\partial r^2}+\frac{1}{(r+\beta^{-1})}\frac{\partial \tavg{c}}{\partial r},
\end{equation}
for $0<x<L/Pe_w$, $0<r<H$, with boundary conditions
\refstepcounter{equation}\label{eq:crosschannelavg2BC}
\begin{equation}
\tavg{c}(x=0,r)=0,\ \tavg{c}(x,r=0)=1,\ \tavg{c}(x,r\to+\infty)\to 0 \ \textrm{or}\ \fp{\tavg{c}}{r}{}(x,r=H)=0.
 \tag{\theequation \textit{a--d}}\label{eq:crosschannelavg2BCa--d}
\end{equation}

In both  geometries, concentration iso-surfaces are in general three-dimensional. Owing to the boundary conditions, concentration profiles at a given $0<x<L/Pe_w$ are curved upwards. The effect of curved concentration profiles, combined with  curved velocity profiles (as shown in figure~\ref{fig:velocityfigure}), is captured by the fluctuation flux $\zavg{u' c'}$  in (\ref{eq:crosschannelavg1}) and (\ref{eq:crosschannelavg2}). If  $c'$ or $u'$ are small, this term may be negligible  and the equations become two-dimensional. Otherwise, this term can either enhance or reduce the overall transport and flux. 
We investigate the effect of the three-dimensional fluctuation flux in detail in the next sections by considering the different limits for the ratio $\delta=\hat{\delta}/\hat{w}$. 


\section{Channels with parallel walls}\label{sec:straightwalls}

\subsection{Thin boundary layer regime, $\hat{\delta} \ll \hat{w}$} \label{sec:straightwallsthin}

If $\delta\ll 1$, we can use the \Leveque approximation \citep{leveque28} $u= \gamma y +O(\delta^2)$ in the diffusive boundary layer, for $y = O(\delta)$ \citep[for a discussion in English of some of L\'{e}v\^{e}que's main results see ][]{glasgow10}. The base shear rate $\gamma=O(1)$ is a function of $z$, with a small dependence on $H$ (see (\ref{eq:shearUchannel})). The advection--diffusion equation (\ref{eq:cartesianfullproblem_a}) becomes
\begin{equation}\label{eq:adv-diff-thinparallel}
 \left(\gamma y+O(\delta^2) \right)  \frac{\partial  c }{\partial x} =  \frac{\partial^2  c }{\partial  y ^2}  +\frac{\partial^2  c }{\partial  z ^2}.
\end{equation}
The different terms in (\ref{eq:adv-diff-thinparallel}) scale such that
\begin{equation}\label{eq:scalingthinBLparallel}
 \delta  \frac{1}{ L/Pe_w } \sim \frac{1}{ \delta ^2}\vee\ 1,
\end{equation}
where $a\vee b$ selects whichever of $a$ and $b$ is dominant. 
The dominant balance is $\delta ^3 \sim  L/Pe_w $ in the diffusive boundary layer, resulting in the well-known \Leveque problem \citep{leveque28} at leading order, 
\begin{equation}\label{eq:straightwalllevequepde}
 \gamma y \frac{\partial  c }{\partial x} = \frac{\partial^2  c }{\partial  y ^2},
\end{equation}
where the boundary conditions (\ref{eq:cartesianfullproblem_aBC}\textit{a--c}) apply. Although $\partial^2 c/\partial z^2\ll 1$, the problem remains three-dimensional as for each `slice'  $\gamma$ depends parametrically on $z$. We designate this modified \Leveque  problem  as the `slice-wise problem' hereafter.
The scaling (\ref{eq:scalingthinBLparallel}) also suggests that the characteristic \Peclet number in this problem is
\begin{equation}\label{eq:characteristicPeclet}
\ImpPe = \frac{Pe_w}{L} = \frac{\hat{U}_0 \hat{w}^2}{\hat{L}\hat{D}}.
\end{equation}
 The rescaled \Peclet number $\ImpPe$ compares the diffusion time across the channel width $\hat{w}$ with the advection time along the length of release area $\hat{L}$. Thus, the diffusive boundary layer thickness is $\delta \sim \ImpPe^{-1/3}$ in the \Leveque regime, which is valid for $\ImpPe^{1/3}\gg 1$.

A similarity solution for (\ref{eq:straightwalllevequepde}) exists with similarity variable $y /x^{1/3}$ \citep{Bejan}
\begin{equation}\label{eq:levequestraightslicewiseconcentrationfield}
 c (x, y , z ) = \frac{\Upgamma (1/3, \gamma(z) y^3/(9x))}{\Upgamma\left(1/3\right)},
\end{equation}
where $\Upgamma({\boldsymbol\cdot},{\boldsymbol\cdot})$ denotes the upper incomplete Gamma function and $\Upgamma({\boldsymbol\cdot})=\Upgamma({\boldsymbol\cdot},0)$ the Gamma function. By construction, our slice-wise solution (\ref{eq:levequestraightslicewiseconcentrationfield}) satisfies only the boundary conditions (\ref{eq:cartesianfullproblem_aBC}\textit{a--c}) in the $x$- and $y$- directions, but not the no-flux boundary conditions (\ref{eq:cartesianfullproblem_aBC}\textit{e,f}) at the side walls since $\partial c/\partial z$ diverges as $|z|\to 1/2$ when $\gamma\to 0$. In fact, a lateral diffusive boundary layer exists at the side walls of characteristic thickness $\delta_{wall} \sim \delta \sim \ImpPe^{-1/3}$, across which cross-channel ($z$) diffusion is not negligible. In their two-dimensional channel geometry, \cite{Jiminez:2005} resolved a similar wall boundary layer using a matched asymptotic solution, requiring the numerical resolution of an elliptic problem. The correction to the mean flux was small and higher order terms had to be found numerically. Since our problem is   inherently three-dimensional near the corners at $|z|= 1/2$ for both the velocity and concentration fields, we choose to compute the small correction to the flux due to the wall boundary layers using three-dimensional numerical calculations of the governing equations.We will  discuss this further in \S\ref{sec:straightwallstnumerics}.

We define the dimensionless  flux per unit area as \citep{Landel:2016}
\begin{equation}\label{eq:dimlessmassfluxnum}
j = \frac{\hat{j} \hat{w}}{\hat{D}(\hat{c}_b-\hat{c}_\infty)}=-\left. \fp{ c }{ y }{} \right|_{ y =0},
\end{equation}
where $\hat{j}$ is the (dimensional) diffusive flux per unit area, with  $j>0$ for a positive flux into the channel. 
We can then obtain the dimensionless average flux or Sherwood number for the slice-wise modified \Leveque limit from the concentration field 
\begin{equation}\label{eq:slicewiseflux}
\Sh= \avg{j} = \frac{3^{4/3} \zavg{\gamma^{1/3}} }{2\Upgamma(1/3)}  \ImpPe^{1/3},
\end{equation}
where $\avg{{\boldsymbol\cdot}}=( L/Pe_w )^{-1}\int_{0}^{ L/Pe_w }\int_{-1/2}^{1/2}{\boldsymbol\cdot}\, \mathrm{d} z \, \mathrm{d}x$ represents the average over the area of release. The cross-channel variations of the velocity, which varies as $\cosh(z)$ according to (\ref{eq:straightflowfield}), are captured in the term $\zavg{\gamma(z)^{1/3}}$ in our result (\ref{eq:slicewiseflux}). 

As a further simplification of the slice-wise \Leveque problem, we consider a two-dimensional solution based on approximating the velocity near the base as $u_b(y) = \zavg{\gamma}y$ instead of $u_b(y,z) = \gamma(z)y$ in  (\ref{eq:straightwalllevequepde}), where boundary conditions (\ref{eq:cartesianfullproblem_aBC}\textit{a--c})  apply. We designate this problem   hereafter as the `two-dimensional' problem. 
The two-dimensional solution $\zavg{c}$ is obtained by replacing $\gamma(z)$ in (\ref{eq:levequestraightslicewiseconcentrationfield}) by $\zavg{\gamma}$. The corresponding two-dimensional Sherwood number depends on $(\zavg{\gamma})^{1/3}$ instead of $\zavg{\gamma^{1/3}}$ in (\ref{eq:slicewiseflux}).

For $H\gg 1$, the two-dimensional Sherwood number deviates from the slice-wise Sherwood number (\ref{eq:slicewiseflux}) by $(\zavg{\gamma^{1/3}}-(\zavg{\gamma})^{1/3})/\zavg{\gamma^{1/3}}\approx\SI{-2.39}{\percent}$ (computed for $H=5$ and using $n=1000$ eigenvalues in (\ref{eq:straightflowfield})). This small deviation is close to the maximum asymptotic deviation found for $H\gg 1$, since $\gamma$ becomes independent of $H$ in this limit. The deviation decreases with decreasing $H$ as the velocity  (\ref{eq:straightflowfield}) converges towards the two-dimensional semi-parabolic Nusselt film solution for $H\ll 1$. However, for $H\ll 1$, the top boundary condition for $c$ (\ref{eq:cartesianfullproblem_aBC}\textit{c}) is not valid anymore and should be replaced with the no-flux boundary condition (\ref{eq:cartesianfullproblem_aBC}\textit{d}). This vertical confinement effect modifies the solution for $c$, as we will discuss in \S\ref{sec:confinmentParallel}. Therefore, our slice-wise solutions (\ref{eq:levequestraightslicewiseconcentrationfield}) for $c$ and  (\ref{eq:slicewiseflux}) for $\Sh$, and the corresponding two-dimensional solutions, are only valid for  $H\gg 1$. 

\subsection{Thick boundary layer regime, $\hat{\delta} \gg \hat{w}$}\label{sec:straightwallsthick}


If  $\delta \gg 1$, the concentration still follows (\ref{eq:cartesianfullproblem_a}). In this limit, $u$ in the diffusive boundary layer is independent of the $y$-coordinate and  parabolic in the $z$-direction, with $\zavg{u}= 1 + O(\delta^{-2})$ (see (\ref{eq:straightwallsfarfield})) and $u'=O(1)$. A scaling analysis of (\ref{eq:cartesianfullproblem_a}), using $u\sim U_\delta\sim 1$, $x\sim L/Pe_w=\ImpPe^{-1}$, $y\sim \delta\gg 1$ and $z\sim 1$, shows that $c$ follows $\partial^2 c/\partial z^2 = 0$ at leading order to satisfy all the boundary conditions (\ref{eq:cartesianfullproblem_aBC}). Hence, $c=\zavg{c}$ at leading order owing to the no-flux boundary conditions at the side walls.
The dependence of $\zavg{c}$ with $x$ and $y$ can be obtained using the cross-channel averaged advection--diffusion equation (\ref{eq:crosschannelavg1}), where  $\zavg{u'c'}$ is negligible since $c'/\zavg{c}=O(\delta^{-2})\ll 1$ from the above scaling analysis. Thus, 
\begin{equation}\label{eq:2dparallelwallthickbl}
\frac{\partial \zavg{c}}{\partial x}=\frac{\partial^2 \zavg{c}}{\partial  y ^2},
\end{equation}
for $0< x < L/Pe_w$,  $0< y < H$, is valid for $\ImpPe^{1/2}\ll 1$ since $\delta \sim \ImpPe^{-1/2}$. It is physically intuitive that $c$ is nearly uniform across the channel since we expect cross-channel diffusion to dominate for thick diffusive boundary layers and small \Peclet numbers.

First, we solve  (\ref{eq:2dparallelwallthickbl}) for a finite domain height with $1\ll \delta \lesssim H<\infty$,  under the boundary conditions (\ref{eq:crosschannelavg1BC}\textit{a,b,d}).
Using separation of variables, we find
\begin{equation}
\zavg{ c }(x, y ) = 1-\sum_{n=0}^{+\infty}  \frac{2}{H\sigma_n} \exp\left(-\sigma_n^2 x\right)\sin\left(\sigma_n  y \right), \label{eq:parallelwallfiniteheightcfield}
\end{equation}
with $\sigma_n = \pi (2n+1)/(2H)$. (Note that the eigenvalue here is the same as for the velocity field in (\ref{eq:lambdaCn}).)
The Sherwood number, computed using (\ref{eq:dimlessmassfluxnum}), is
\begin{equation}\label{eq:parallelwallfiniteheightflux}
\Sh=\ImpPe\sum_{n=0}^{+\infty}\frac{2}{H\sigma_n^2} \left(1-\exp\left(-\sigma_n^2\ImpPe^{-1} \right) \right).
\end{equation}
In the limit  $H^2 \ImpPe\to 0$, corresponding to  $ \delta  \rightarrow H$, our result (\ref{eq:parallelwallfiniteheightcfield}) shows that $c$ becomes uniform across the channel, as expected intuitively, with $\zavg{ c }\to 1$ everywhere since $x\sim \ImpPe^{-1}$. In addition, (\ref{eq:parallelwallfiniteheightflux}) predicts that, for $H^2 \ImpPe\to 1$, the Sherwood number behaves as  
\begin{equation}\label{eq:limlowPe}
\Sh\sim H \ImpPe ,
\end{equation} 
confirming that the flux vanishes in this limit.

Second, if $1\ll \delta  \ll H$, we can assume a semi-infinite domain in $ y $. We solve  (\ref{eq:2dparallelwallthickbl}) for $0< x< L/Pe_w$, $0< y$ under  (\ref{eq:crosschannelavg1BC}\textit{a,b,c}).
A similarity solution exists  \citep{Bejan}
\begin{equation}\label{eq:straightwallthickblconcentration}
\zavg{ c }(x,y) = \textrm{Erfc}\left(\frac{y}{2x^{1/2}} \right),
\end{equation}
where $\textrm{Erfc}({\boldsymbol\cdot})$ is the complementary error function. We find  the Sherwood number
\begin{equation}\label{eq:straightwallfarfieldflux}
\Sh =\frac{2}{\sqrt{\pi}}\ImpPe^{1/2}.
\end{equation}
Thus, we see that without vertical confinement, the Sherwood number increases at a faster rate in the limit of small $\ImpPe$, as $\Sh \sim \ImpPe^{1/2}$ in (\ref{eq:straightwallfarfieldflux}) instead of $\sim \ImpPe$ in (\ref{eq:limlowPe}).


\subsection{Vertical confinement, $\hat{\delta}\sim \hat{H}$}\label{sec:confinmentParallel}

To study the impact of vertical confinement, $\delta\sim H$, on $\Sh$ we  use the cross-channel averaged advection--diffusion equation (\ref{eq:crosschannelavg1}), under the no-flux top boundary condition (\ref{eq:crosschannelavg1BC}\textit{d}). Integrating (\ref{eq:crosschannelavg1}) in the streamwise direction from $0$ to $L/Pe_w$, we obtain
\begin{equation} \label{eq:differentialflux}
-\frac{\partial \avg{j_y}}{\partial y}= \fp{\avg{c}}{y}{2}= \ImpPe   \left. \zavg{u} \, \zavg{c}\right|_{x=L/Pe_w} + \ImpPe   \left.\zavg{u'c'}\right|_{x=L/Pe_w} = q_m + q'
\end{equation}
with $j_y=-\partial c/\partial y$ the vertical flux at a given $y$ coordinate. The quantity $q_m$ represents the vertical ($y$-) profile of the contribution to the flux from the cross-channel averaged concentration field at the end of the area of release, $x=L/Pe_w$. The quantity $q'$ represents the vertical ($y$-) profile of the contribution to the flux from the cross-channel fluctuations of the concentration field at $x=L/Pe_w$. We refer to $q_m$ and $q'$ as the local mean flux and  local fluctuation flux, respectively. Thus, the vertical variation of the  vertical average flux $-\partial \avg{j_y}/\partial y$  depends on the contributions of both $q_m$ and $q'$. Integrating again in the vertical direction from $0$ to $H$, we obtain
\begin{equation}\label{eq:totalcomposedflux}
\Sh = \ImpPe \int_0^H \left. \zavg{u} \, \zavg{c}\right|_{x=L/Pe_w} \mathrm{d}y + \ImpPe \int_0^H \left.\zavg{u'c'}\right|_{x=L/Pe_w} \mathrm{d}y= \avg{j_m}+\avg{j'},
\end{equation}
where $\avg{j_m}$ and $\avg{j'}$ are the total contributions from the mean and fluctuation fluxes to $\Sh$.
We now assume that $q'$ is either negligible compared to $q_m$ or scales in a similar fashion to $q_m$.  We will discuss this assumption in detail in \S\ref{sec:ResultParallel}, but we note that in the thick boundary layer regime we have already shown that $q'\ll q_m$ (see \S\ref{sec:straightwallsthick}). In the limit $\delta\sim H$, we must have $\zavg{c}(x=L/Pe_w) \sim 1$, therefore the Sherwood number scales as
\begin{equation}\label{eq:jFluxVerticalConfined}
\Sh \sim Q \ImpPe = U H\ImpPe,
\end{equation}
with $Q=\int_0^H \zavg{u} \mathrm{d}y$ (and $u$ from (\ref{eq:straightflowfield})) the channel volume flow rate and $U\sim U_{\delta}$ the mean channel velocity. In the limits of small or large channel heights, we find that the vertically confined Sherwood number is: $\Sh \sim  H^3\ImpPe$ for $H\ll 1$, since $U\sim H^2$ in the $y$ direction; and $\Sh \sim  H\ImpPe $ for $H\gg 1$ since $U\sim 1$, as also found in our theoretical result (\ref{eq:limlowPe}).


\subsection{Transition regime, $\hat{\delta}\sim \hat{w}$, and numerical formulations for three- and two-dimensional problems} \label{sec:straightwallstnumerics}

For $\ImpPe\sim 1$, or $\delta \sim 1$, the streamwise ($x$) advection,  vertical ($y$) and cross-channel ($z$) diffusion are all of similar order of magnitude in the advection--diffusion equation (\ref{eq:cartesianfullproblem_a}). 
Thus, $c$ is strongly three-dimensional in the transition regime. To analyse the impact on the flux or $\Sh$, we solve (\ref{eq:cartesianfullproblem_a}) numerically under (\ref{eq:cartesianfullproblem_aBC}\textit{a,b,d--f}),  using our three-dimensional result (\ref{eq:straightflowfield}) for $u$. We   vary $\ImpPe$ to compare the numerical results with our asymptotic results  in the thin (\S\ref{sec:straightwallsthin}), thick (\S\ref{sec:straightwallsthick}) and vertically confined (\S\ref{sec:confinmentParallel}) regimes. 
We formulate the problem for a finite channel height. This Graetz-type problem can be solved using separation of variables \citep{Graetz:1885,Bejan}.  Hence, 
\begin{equation}\label{eq:numconcfield}
 c (x, y , z ) = 1-\sum_{n=1}^{+\infty}{k_n}\exp(-\nu_n x) A_n( y , z ).
\end{equation}
The eigenpairs $A_n$ and $\nu_n$ are solutions of the homogeneous eigenvalue problem
\begin{equation}\label{eq:straightevproblem}
-u \nu_n A_n = \frac{\partial^2 A_n}{\partial  y ^2} + \frac{\partial^2 A_n}{\partial  z ^2},
\end{equation}
for all integers $n\geq 1$, $0<x<L/Pe_w$, $0< y <H$, $\left| z \right| < 1/2$,  with boundary conditions
\begin{equation}\label{eq:straightevproblemBC}
A_n( y =0, z ) = 0,\ \frac{\partial A_n}{\partial  y}  ( y =H, z )=0,\  \frac{\partial A_n}{\partial  z}  ( y , z =\pm 1/2)=0.
\end{equation}
%
Since the velocity  (\ref{eq:straightflowfield}) involves an infinite sum, which is impractical for analytical progress, we solve a second-order finite difference formulation of  (\ref{eq:straightevproblem}) using the SLEPc implementation \citep{SLEPc} of the LAPACK library (Linear Algebra Package, \cite{LAPACK}). We verified our numerical scheme against known solutions as documented in  \ref{app:verification}. The agreement between the numerical solutions and asymptotic solutions obtained here provides further verification.
We then compute the amplitudes  $\left|A_n\right|$ in (\ref{eq:numconcfield})  using the upstream boundary condition $c(x=0 ,y,z)=0$ and the orthogonality of the eigenfunctions. 

Once $A_n$ and $\nu_n$  are calculated, we  compute the Sherwood number following (\ref{eq:dimlessmassfluxnum}),
\begin{equation}\label{eq:cartesianfullflux}
\Sh= \ImpPe \sum_{n=0}^{+\infty} \frac{1}{\nu_n} \left(1-\exp\left(-\nu_n \ImpPe^{-1}\right)\right)\int_{-1/2}^{1/2} \left. \frac{\partial A_n}{\partial  y }\right|_{y=0} \mathrm{d} z.
\end{equation}
The relevant dimensionless group is again $\ImpPe$. Due to the decreasing exponential functions in  (\ref{eq:numconcfield}) and (\ref{eq:cartesianfullflux}), $c$ at the  end of the area of release is mainly described by small eigenvalues. The numerical solution suggests that  the significant $\left|A_n\right|$ decrease approximately hyperbolically with $n$ (not shown), whilst the eigenvalues $\nu_n$ increase monotonically with $n$. Thus, for a given $x< L/Pe_w$, only a small number of  eigenvalues  is required to compute the solution accurately, representing the local behaviour of the boundary layer solution, as will be shown  in the next section.

For comparison, we also solve a two-dimensional  formulation of this problem based on the cross-channel averaged advection--diffusion equation (\ref{eq:crosschannelavg1}), neglecting $\zavg{u'c'}$:
\begin{equation}\label{eq:parallel2davgproblem}
\zavg{u}\frac{\partial \zavg{ c }}{\partial x} = \frac{\partial^2 \zavg{ c }}{\partial  y ^2} 
\end{equation}
for $0<x< L/Pe_w$, $0< y <H$,
under (\ref{eq:crosschannelavg1BC}\textit{a,b,d}).
The boundary conditions can also be homogenised to obtain a one-dimensional eigenvalue problem, which we solve using a shooting method \citep{Berry:1952} to obtain $\zavg{ c }$ and a two-dimensional $\Sh$.
This simpler two-dimensional formulation of the advection--diffusion problem  allows us to assess \textit{a posteriori} the error on $Sh$ when neglecting the three-dimensional flux $\zavg{u'c'}$.

More details about the three-dimensional and two-dimensional numerical calculations, and  the numerical results shown in this paper  can be found in appendix~\ref{apdx:NumSimDet} and table~\ref{tab:numdet}.

\subsection{Results in parallel wall channels}\label{sec:ResultParallel}

In this section, we compare our asymptotic predictions for $\delta$ and  $\Sh$ in the parallel channels with three-dimensional and two-dimensional  numerical calculations of the advection--diffusion equation. The aim here is to assess whether three-dimensional effects related to the corners at the base of the channel or due to confinement have a strong impact on $\delta$ and $\Sh$ in the different regimes identified previously.
We study the influence of $\ImpPe$,  lateral and vertical confinement effects. We also analyze the relative magnitude of the three-dimensional fluctuation flux $\zavg{u'c'}$ and whether it can be neglected in  (\ref{eq:crosschannelavg1}).

\subsubsection{Concentration field}

\begin{figure}
\includegraphics[width=\textwidth]{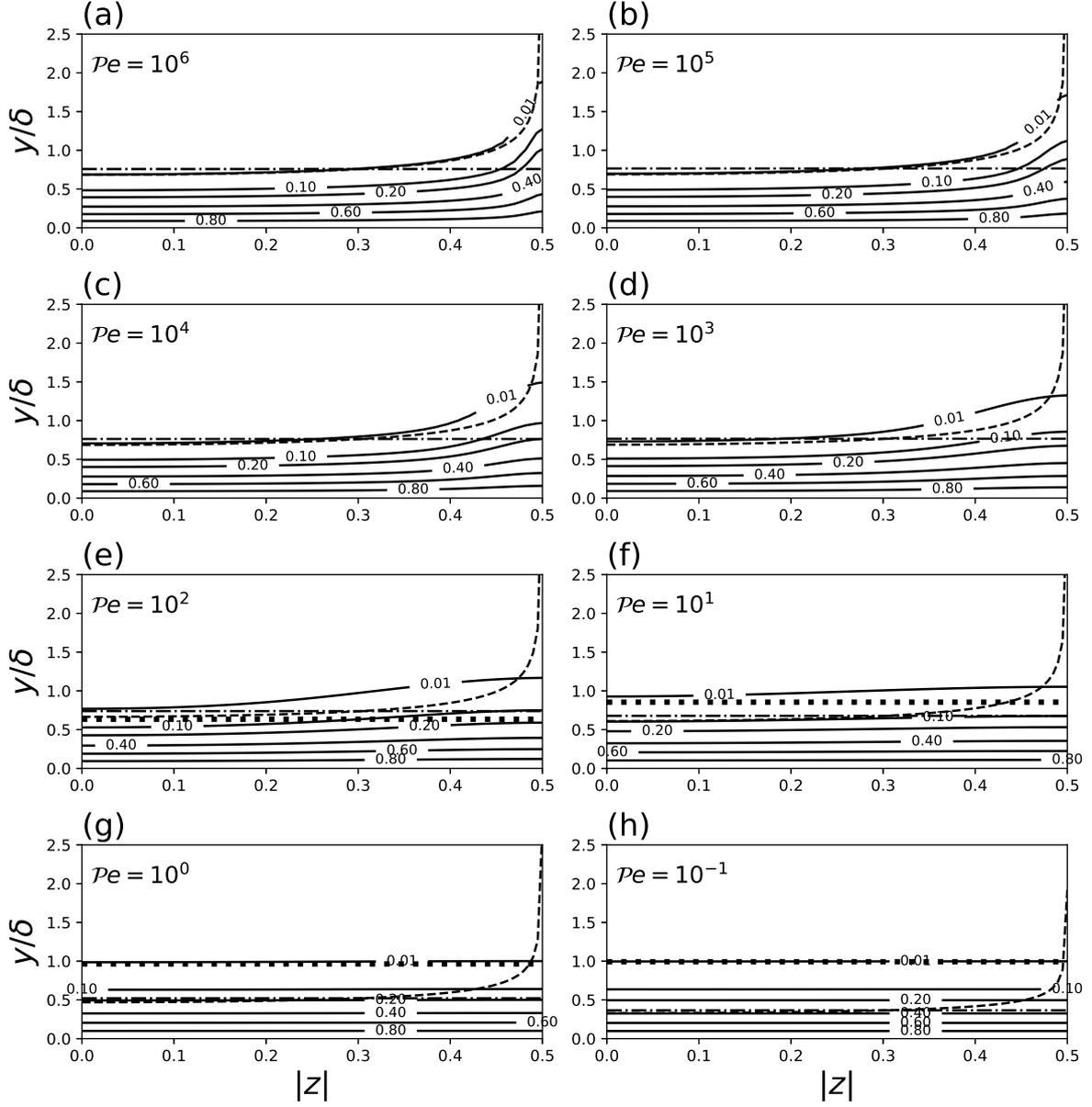}
\caption{
Contour plots of the three-dimensional concentration field  computed numerically (solid lines) using  (\ref{eq:numconcfield})--(\ref{eq:straightevproblemBC}) (see details in table~\ref{tab:numdet}, appendix~\ref{apdx:NumSimDet}),  at $x=L/Pe_w$, for various $\ImpPe$. In (\textit{a--h}),  dashed lines show the slice-wise thin boundary predictions (\ref{eq:levequestraightslicewiseconcentrationfield}) for $\delta$ ($\ImpPe^{1/3}\gg 1$); dash-dotted lines show the two-dimensional predictions for $\delta$ based on (\ref{eq:levequestraightslicewiseconcentrationfield}). In (\textit{e--h}),  dotted lines show the thick boundary layer predictions (\ref{eq:straightwallthickblconcentration}) for $\delta$ ($H^{-1}\ll\ImpPe^{1/2}\ll 1$). }\label{fig:contourcart}
\end{figure}

In figure~\ref{fig:contourcart} we show contour plots of $c$ for $0\leq z\leq 1/2$ (note the symmetry with $z=0$) at the end of the area of release, $x=L/Pe_w$, for various \Peclet numbers: from $\ImpPe=10^6$ (figure \ref{fig:contourcart}\textit{a}) to $\ImpPe=10^{-1}$ (figure~\ref{fig:contourcart}\textit{h}). Solid lines show the   numerical solution of the three-dimensional formulation (\ref{eq:numconcfield})--(\ref{eq:straightevproblemBC}) using the three-dimensional velocity field (\ref{eq:straightflowfield}).  To ensure an accurate resolution of the boundary layer, we imposed $H\geq 2\delta$. 
%
%
We normalise the $y$-axis by $\delta$, computed as $\delta=\zavg{y_{\delta}}$ with $c(L/Pe_w,y_{\delta},z)=0.01$. All the theoretical predictions shown in figure~\ref{fig:contourcart} for the contour representing $\delta$ are  referenced to the same value. The dashed lines are plotted using the asymptotic concentration  (\ref{eq:levequestraightslicewiseconcentrationfield}) in the slice-wise thin boundary layer regime, which used $\gamma(z)$ but assumed no cross-channel diffusion. The dash-dotted lines are plotted using (\ref{eq:levequestraightslicewiseconcentrationfield}) assuming a two-dimensional velocity profile (i.e. replacing $\gamma(z)$ by $\zavg{\gamma}$). These two predictions, corresponding to $\delta \ll 1$ or $\ImpPe^{1/3} \gg 1$, are shown in all graphs in figure~\ref{fig:contourcart}. The dotted lines, only shown in figures~\ref{fig:contourcart}(\textit{e--h})  where $\ImpPe=10^2$--$10^{-1}$, respectively, are plotted using the solution (\ref{eq:straightwallthickblconcentration}) for $\zavg{c}$ and correspond to the thick boundary layer regime: $1\ll \delta \ll H$ or $H^{-1}\ll \ImpPe^{1/2} \ll 1$.


For $\ImpPe \geq 100$ (figures~\ref{fig:contourcart}\textit{a--e}), the two-dimensional predictions for $\delta$ in the thin boundary layer regime (dash-dotted lines) are in  agreement with the three-dimensional  numerical results in the interior of the channel $|z|< 0.4$. Near the side walls ($1/2-|z|\lessapprox 0.1$), the two-dimensional predictions underestimate the numerical three-dimensional results ($c=0.01$ contour plotted with a solid line) owing to the (basal) diffusive boundary layer at the wall. The diffusive boundary layer is better captured by the slice-wise thin boundary layer  prediction (\ref{eq:levequestraightslicewiseconcentrationfield}) (dashed lines). The agreement improves as $\ImpPe$ increases (see figures~\ref{fig:contourcart}\textit{a,b}), since the influence of the three-dimensional wall boundary layers, not captured by (\ref{eq:levequestraightslicewiseconcentrationfield}), reduces. At lower values of $\ImpPe$, we can see in figures~\ref{fig:contourcart}(\textit{e,f}) ($\ImpPe= 100$ and $10$, respectively) that the characteristic wall boundary layer thickness (in the $z$-direction) increases inwards and $\delta_{wall}\sim 1$ is not small anymore. The thin boundary layer  predictions are not valid anymore and increasingly underestimate $\delta$ with decreasing $\ImpPe$. As $\ImpPe\approx 10$ to $100$,  the thick boundary layer predictions for $\delta$ based on (\ref{eq:straightwallthickblconcentration}) (dotted lines) are in qualitative agreement. The agreement improves significantly when $\ImpPe$ decreases, confirming the change of regime to the thick boundary layer regime, valid for $\ImpPe^{1/2}\ll 1$, as shown in figures~\ref{fig:contourcart}(\textit{g,h}) where $\ImpPe=1$, $0.1$, respectively. The dotted lines and the contour line $c=0.01$ almost overlap in figures~\ref{fig:contourcart}(\textit{g,h}). The concentration profile becomes uniform across the channel width as we predicted in \S\ref{sec:straightwallsthick}.

\subsubsection{Three-dimensional fluxes}

\begin{figure}
\includegraphics[keepaspectratio=true,width=\textwidth]{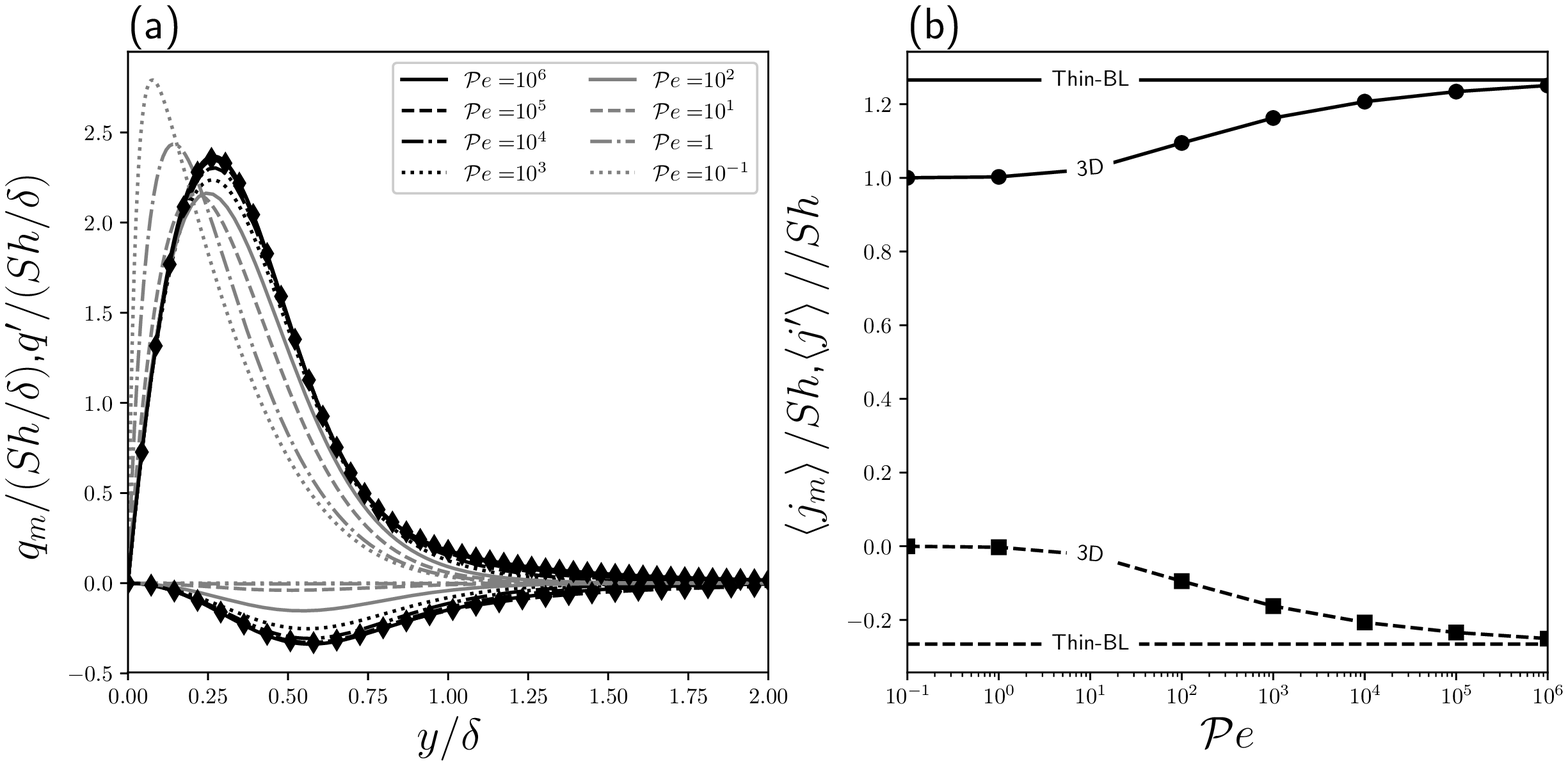}
\caption{
(\textit{a}) Vertical profiles of the local mean flux $q_m\geq 0$  and local fluctuation flux $q'\leq 0$ at $(x=L/Pe_w,y)$ computed following (\ref{eq:differentialflux}) using the three-dimensional  numerical simulations  for different $\ImpPe$ (see details in table~\ref{tab:numdet}, appendix~\ref{apdx:NumSimDet}).
The curves range $\ImpPe=10^6$ to $10^{-1}$. 
The   thin boundary layer predictions (solid lines with lozenges) follow (\ref{eq:levequestraightslicewiseconcentrationfield}).
(\textit{b}) Variations of the normalised total  mean flux $\avg{j_m}\geq 0$ (large dots) and normalised total fluctuation flux $\avg{j'}\leq 0$ (squares) with $\ImpPe$ (the straight lines joining the symbols are for visual aid),  computed  numerically following (\ref{eq:totalcomposedflux}).  The thin boundary layer predictions for $\avg{j_m}$ (solid line) and $\avg{j'}$ (dashed line) follow (\ref{eq:levequestraightslicewiseconcentrationfield}).
}
\label{fig:fluxcontributions}
\end{figure}

To analyze the impact of the three-dimensional fluctuation flux $\zavg{u'c'}$ on the total flux or Sherwood number, we plot  in figure~\ref{fig:fluxcontributions} $q_m$, $q'$, $\avg{j_m}$ and $\avg{j'}$ from (\ref{eq:differentialflux}) and (\ref{eq:totalcomposedflux}), computed  numerical using  (\ref{eq:numconcfield}--\ref{eq:straightevproblemBC}) (see  table~\ref{tab:numdet}, appendix~\ref{apdx:NumSimDet}, for more details). We also show the asymptotic predictions for $\delta\ll 1$ (lines with lozenges) computed using  (\ref{eq:levequestraightslicewiseconcentrationfield}).

The results indicate that the effect of the mean flux $\zavg{u}\, \zavg{c}$ is much stronger  than the  effect of the fluctuation flux $\zavg{u'c'}$ since $|q'|\ll q_m$ for most $y$ (figure~\ref{fig:fluxcontributions}\textit{a})  and $|\avg{j'}|\ll |\avg{j_m}|$  (figure~\ref{fig:fluxcontributions}\textit{b}) across all regimes:  the thin boundary layer regime, $\ImpPe\gg 1$;  the transition regime, $\ImpPe\sim 1$; and the thick boundary layer regime for $\ImpPe\ll 1$. We also note that $\zavg{u'c'}$ tends to reduce the flux and Sherwood number since  $q'$ and $\avg{j'}<0$. 
The fluctuation flux, which has the strongest effect at large $\ImpPe$, is primarily due to the negative effect of the wall boundary layers that develop for both  $u$ and  $c$. Close to the wall, $u$ decreases and $u'<0$ (figure~\ref{fig:velocityfigure}\textit{a}), whilst $c$ increases and $c'>0$ (figure~\ref{fig:contourcart}\textit{a--d}), thus producing a negative fluctuation flux in average. It is also interesting to note that the maximum of the fluctuation flux $\avg{\zavg{u'c'}}$ occurs at mid-depth in the diffusive boundary layer across all regimes. This is due to the contribution being from the product of an increasing function of $y$, the velocity fluctuation $u'$, and a decreasing function of $y$, the concentration fluctuations $c'$. Overall,  the average fluctuation flux $|\avg{j'}|$ does not exceed more that \SI{25}{\percent} of the mean flux $\Sh$ for all $\ImpPe$, and $|\avg{j'}|/\avg{j_m}\leq \SI{20}{\percent}$, which strongly suggests that it can be neglected at leading order. In particular, $\avg{j'}$ vanishes in the thick boundary layer regime, confirming \textit{a posteriori} our assumption to neglect $\zavg{u'c'}$ when $\ImpPe^{1/2}\ll 1$ (\S\ref{sec:straightwallsthick}).

\subsubsection{Sherwood number}

\begin{figure}
\centering
\includegraphics[width=0.8\textwidth]{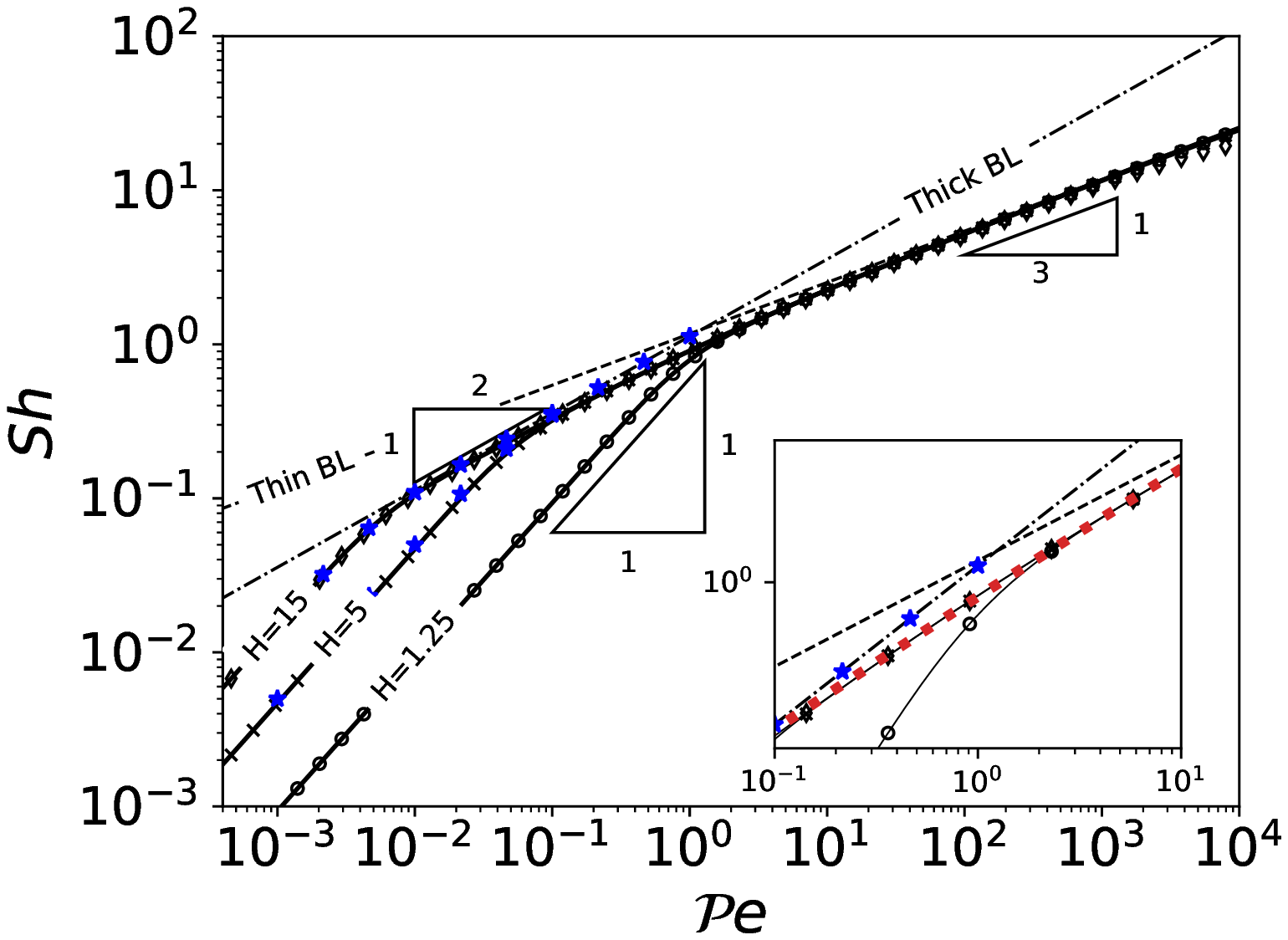}
\caption{
Sherwood number versus   \Peclet number in parallel channels.  Three-dimensional  numerical results (black symbols) follow (\ref{eq:cartesianfullflux})  for three channel heights (see details in table~\ref{tab:numdet}, appendix~\ref{apdx:NumSimDet}).
Two-dimensional  numerical results using (\ref{eq:parallel2davgproblem}) (neglecting the three-dimensional flux $\zavg{u'c'}$) are plotted with solid lines closely following the symbols. The slice-wise  prediction (\ref{eq:slicewiseflux}) in the thin boundary layer regime  (large $\ImpPe$)  is plotted with a dashed line. The  prediction (\ref{eq:straightwallfarfieldflux}) in the thick boundary layer regime  (small $\ImpPe$) and for $1\ll \delta\ll H$ is plotted with a dash-dotted line.
The  prediction (\ref{eq:parallelwallfiniteheightflux}) in the thick boundary layer regime and for $1\ll \delta\leq H$ is plotted with  blue stars for $H=5$ and $15$. 
As $\ImpPe \to 0$, the scaling $\Sh\sim H\ImpPe$ due to the impact of vertical confinement is predicted by (\ref{eq:jFluxVerticalConfined}). For $\ImpPe\sim 1$,  $\Sh_{approx}= 1.96\ImpPe^{1/2}/(1+1.18\ImpPe^{1/6})$ is shown with a red dotted line in the inset. 
}\label{fig:cartfull}
\end{figure}

In figure~\ref{fig:cartfull}, we plot the three-dimensional  numerical results for  $\Sh$, designated as $\Sh_3$, computed  using  (\ref{eq:cartesianfullflux}) as a function of $\ImpPe$, with different open black symbols for different domain heights: $H=1.25$ (circles), $H=5$ (crosses) and $H=15$ (lozenges). 
%
For $10^{-3}\leq \ImpPe\leq 10^{4}$, the two-dimensional  numerical results (solid lines closely following the symbols), designated as $\Sh_2$, based on (\ref{eq:parallel2davgproblem}) and neglecting $\zavg{u'c'}$  are in good agreement with $\Sh_3$, for all three $H$. 
In the transition region $10^{-1}\leq\ImpPe \leq 10$ (see inset in figure~\ref{fig:cartfull}) where the distribution for both $u$ and $c$ are inherently three-dimensional, the  numerical  two-dimensional results are close to the  numerical  three-dimensional results. 
We find a relative deviation, $|\Sh_{3}-\Sh_{2}|/\Sh_{3}$, less than \SI{5}{\percent}  for $\ImpPe\leq 740$, and less than \SI{20}{\percent} for $740\leq \ImpPe\leq 10^4$ for all $H$. We note that for $H=1.25$ and \num{5} the deviation remains less than \SI{5}{\percent} over the whole range shown. 
Part of this deviation is due to numerical limitations (numerical resolution and truncation in the number of eigenpairs), particularly at large $\ImpPe$. 
For all $H$, the deviation increases monotonically with increasing $\ImpPe$, in agreement with the results  in figure~\ref{fig:fluxcontributions}, which show that the contribution of $\zavg{u'c'}$ increases  at large $\ImpPe$. 
At large $\ImpPe$, the  deviation  $(\Sh_{3}-\Sh_{2})/\Sh_{3}$ should converge to the theoretical deviation between the slice-wise asymptotic $\Sh$ and the two-dimensional asymptotic $\Sh$: $(\Sh_{3}-\Sh_{2})/\Sh_{3}\to (\zavg{\gamma^{1/3}}-\zavg{\gamma}^{1/3})/\zavg{\gamma^{1/3}}\approx \SI{-2.4}{\percent}$. Indeed we have shown in \S\ref{sec:straightwallsthin} that as $\ImpPe\to \infty$, $\Sh_3$ converges to the slice-wise prediction (\ref{eq:slicewiseflux}), whilst $\Sh_3$ converges to the two-dimensional prediction, which replaces $\zavg{\gamma^{1/3}}$ by $\zavg{\gamma}^{1/3}$ in (\ref{eq:slicewiseflux}). Our numerical results appear to confirm this  prediction. For $H=1.25$ in figure~\ref{fig:cartfull}, we find $(\Sh_{3}-\Sh_{2})/\Sh_{3}\approx \SI{-2.4}{\percent}$ as $\ImpPe\rightarrow \num{e4}$. For larger $H$, we find that the magnitude of the deviation is smaller than \SI{2.4}{\percent} for $\ImpPe\leq 756$ ($H$=5) and $\ImpPe\leq 92$ ($H=15$). Computation of additional eigenpairs  for $\Sh_{3}$ would extend these ranges to larger $\ImpPe$. 
Therefore, the results in figure~\ref{fig:cartfull} strongly suggest \textit{a posteriori} that the three-dimensional flux $\zavg{u' c'}$ contributes to a small portion of $\Sh$ for all $\ImpPe$ and all $H$. 

An important implication for practical applications where high accuracy is not critical is that $\zavg{u' c'}$ can be neglected  to solve the simpler two-dimensional problem (\ref{eq:parallel2davgproblem}), thus reducing computational burden. For a given resolution $\delta x$ in all directions, a three-dimensional solution requires more memory for the storage of the grid by a factor of at least $\delta x/w$ compared with a  two-dimensional solution. For matrix-based solvers such as LAPACK \citep{LAPACK},  computational time increases by a factor of approximately $(\delta x/w)^3$ in the three-dimensional case. Therefore, though not fully optimised, the shooting method used to solve the two-dimensional case is memory efficient and could be run on portable platforms with  limited memory, such as mobile phones. 

At large $\ImpPe$, the slice-wise thin boundary layer prediction (\ref{eq:slicewiseflux}) for $\Sh$ (dashed line in figure~\ref{fig:cartfull}) is in agreement with $\Sh_3$. At $\ImpPe = 10^4$, the  deviation between them is $\leq \SI{1.5}{\percent}$  for $H=1.25$,  $\leq \SI{3.3}{\percent}$ for $H=5$, and $\leq \SI{18.8}{\percent}$ for  $H=15$. The increase of the deviation with increasing  $H$ is due to the numerical limitations mentioned above: a combination of the truncation error from taking a finite number of terms in (\ref{eq:cartesianfullflux}) and a reduced resolution since the number of grid points is fixed for all our computational domains (see also table \ref{tab:numdet},  appendix~\ref{apdx:NumSimDet}). This is a common  problem when solving eigenvalue problems using finite-difference methods \citep{Pryce1993}.
The effects of truncation error and reduced resolution are noticeable at large $\ImpPe$ for the results  in figure~\ref{fig:cartfull} for $\ImpPe > \num{e6}$ ($H=1.25$,  not shown), $\ImpPe > \num{2.6e4}$ ($H=5$, not shown), $\ImpPe > \num{2e3}$ ($H=15$). 
 This emphasises the importance of our  asymptotic solutions providing accurate predictions in regimes where numerical results are computationally expensive and prone to numerical errors. 

At small $\ImpPe$,  the thick boundary layer prediction (\ref{eq:straightwallfarfieldflux}) $\Sh\sim \ImpPe^{1/2}$ (dash-dotted line) follows  the  numerical results  as long as $\delta\ll H$. As $\delta \sim H$, the Sherwood number follows a different regime: $\Sh\sim \ImpPe$, as predicted by (\ref{eq:parallelwallfiniteheightflux}) (filled blue stars). The transition between the confined regime ($\delta\sim H$) and the unconfined regime ($\delta\ll H$) can be estimated at low \Peclet numbers using $\delta\sim \ImpPe^{-1/2}\sim H$. We find for $H=1.25$ (circles), $H=5$ (crosses) and $H=15$ (lozenges) that the transition occurs for $\ImpPe\sim$ 0.6, 0.04 and \num{4e-3}, respectively, which agrees with the results shown in figure~\ref{fig:cartfull}.   In the confined regime we  also find that $\Sh$ increases approximately linearly with $H$ at a sufficiently small and fixed $\ImpPe$, as predicted by the asymptotic scaling $\Sh\sim H \ImpPe$ in (\ref{eq:limlowPe}).

In the transition region for $\ImpPe\sim 1$ (inset in figure~\ref{fig:cartfull}) the maximum error between the asymptotic theoretical predictions and the three-dimensional  numerical calculations, found at the intersection of $\Sh\sim \ImpPe^{1/3}$ (dashed line) and $\Sh\sim \ImpPe^{1/2}$ (dash-dotted line), is always less than approximately $\SI{30}{\percent}$. Since we expect the transition to be smooth, at least for Stokes flow, we propose a Pad\'{e} approximant combining both asymptotic limits: 
\begin{equation}\label{eq:Shfit}
\Sh_{approx} = \frac{1.96 \ImpPe^{1/2}}{1+1.18 \ImpPe^{1/6}}
\end{equation}
(red dotted line in the inset), where the two numerical coefficients have been computed using a  least-squares fit. The approximant agrees with the three-dimensional numerical results to better than $1$\%  for $0.3\leq\ImpPe\leq 10$, and to better than 7\% for $0.06\leq\ImpPe\leq 50$. Therefore, in practical applications requiring slightly less accuracy, the asymptotic predictions and the combined fit (\ref{eq:Shfit}) can provide  instantaneous quantitative predictions of the Sherwood number as long as $\delta\ll H$. The asymptotic scaling (\ref{eq:limlowPe}) also provides qualitative predictions of $\Sh$ in the confined regime $\delta\sim \ImpPe^{-1/2}\sim H$.


\section{Channels with a truncated wedge geometry}


\subsection{Thin boundary layer regime, $\hat{\delta}\ll \hat{w}$}\label{sec:angledwallsthin}

In truncated wedges (figure~\ref{fig:geometry}), for  $\delta \ll 1$ we can  use the \Leveque approximation $u =  \gamma r +O(\delta^2)$   in the diffusive boundary layer, similar to parallel channels (\S\ref{sec:straightwallsthin}). The shear rate $\gamma=O(1)$ depends on $\theta$ following (\ref{eq:wedgegamma}). In this regime, the four terms in the advection--diffusion equation (\ref{eq:angledwallblequation}) (in cylindrical coordinates)  scale such that
\begin{equation}\label{eq:scalingthinBLwedge}
\delta \frac{1}{L/Pe_w}\sim \frac{1}{\delta^2} \vee \frac{\beta}{\delta} \vee 1,
\end{equation}
which suggests that $\delta\sim \ImpPe^{-1/3}$, as  found for parallel channels. The cross-channel diffusion term $(r+\beta^{-1})^{-2}\partial^2 c/\partial \theta^2$ is negligible since $c'= O(\delta^2)$ or smaller. The curvature term (second term on the right hand side of (\ref{eq:angledwallblequation})), not present in parallel  channels, is also negligible at leading order, and of order $O(\beta \delta)$ compared with the  $O(1)$ radial diffusion term and axial advection term. We note that $\beta$ can be $\sim 1$ or $\ll 1$. At leading order,  (\ref{eq:angledwallblequation}) reduces to the slice-wise modified \Leveque problem: $\gamma r \partial c/\partial x = \partial^2 c/\partial r^2$, where $\gamma$ depends parametrically on $\theta$, making the problem three-dimensional.  This is the same equation as in parallel channels (see (\ref{eq:straightwalllevequepde})). Hence, the slice-wise Sherwood number  is 
\begin{equation}\label{eq:truncatedwegeleveque}
\Sh=\frac{3^{4/3} \tavg{\gamma^{1/3}}}{2\Upgamma(1/3)} \ImpPe^{1/3},
\end{equation}
for $\ImpPe^{1/3}\gg 1$.
The  diffusive boundary layers along the side walls, where  $(r+\beta^{-1})^{-2}\partial^2 c/\partial \theta^2$ is not negligible, are very thin. Their thickness, in the cross-channel ($r$-) direction, is of the order $\delta_{wall}\sim \delta$.
Their contribution to the flux $j$ can therefore be neglected at leading order. Similar to parallel channel, for $\beta \rightarrow 0$ the small deviation between our slice-wise solution (which assumes a three-dimensional velocity  and use $\overline{\gamma(z)^{1/3}}$ in (\ref{eq:truncatedwegeleveque})) and the two-dimensional solution (which assumes a uniform velocity and use $\overline{\gamma}^{1/3}$ instead) is $(\overline{\gamma}^{1/3}-\overline{\gamma^{1/3}})/\overline{\gamma^{1/3}} \approx \SI{-2.38}{\percent}$ (for $\beta=\num{1e-6}$). The deviation $(\overline{\gamma}^{1/3}-\overline{\gamma^{1/3}})/\overline{\gamma^{1/3}}$  increases slightly with the opening angle. For $\beta = 0.5$, $1.0$ and $\pi/2$, we find: \SI{-2.80}{\percent}, \SI{-3.38}{\percent} and  \SI{-4.11}{\percent} (with $n=5000$ eigenpairs), respectively (see  figure \ref{fig:gammabarrationwedge}(\textit{a}), appendix~\ref{apdx:AddMat}). 

We now consider the influence of the higher order curvature term, neglected above. We still assume  $u= \gamma r$, i.e. the next terms in $O(\delta^2)$ are neglected. We also assume $\delta \ll \beta$ so that the curvature term in (\ref{eq:angledwallblequation}) is much larger than the cross-channel diffusion term.
The advection--diffusion equation (\ref{eq:angledwallblequation}) becomes
\begin{equation}\label{eq:polarlevequebase}
 \gamma r  \frac{\partial  c }{\partial x} = \frac{\partial^2  c }{\partial  r ^2} +\frac{1}{ r +\beta^{-1}}\frac{\partial  c }{\partial r }. 
\end{equation}
We change the variables from $(x, r )$ to $(\xi,\eta)$, with $\xi = x^{1/3}/\beta^{-1}$, which represents the ratio of $ \delta\sim x^{1/3} $ and $r_i=\beta^{-1}$, and $\eta =  r /x^{1/3}$ the similarity variable for the advection--diffusion equation at leading order. 
 After substituting a Poincar\'e expansion: $ c (\xi,\eta)= c _0(\eta) +\xi  c _1(\eta)  + \ldots$, we find that the next term at order $\xi^1$ (see appendix \ref{sec:appthinboundarylayerregime} for further details), is
\begin{equation}\label{eq:1stordercorrectioncthinBLwedge}
c_1(x,r,\theta) = -\frac{r}{2x^{1/3}} \frac{\Upgamma (1/3, \gamma(\theta) r^3/(9x))}{\Upgamma\left(1/3\right)}.
\end{equation}
Hence, we obtain the slice-wise Sherwood number, with the first order correction $c_1$,
\begin{equation}\label{eq:polarthinexpansionflux}
\Sh  = \frac{3^{4/3} \tavg{\gamma^{1/3}}}{2\Upgamma(1/3)} \ImpPe^{1/3}+\frac{\beta}{2},
\end{equation}
for $\delta\ll 1$ or $\ImpPe^{1/3}\gg 1$, and $\beta\ll 1$. If  $O(\delta^2)$ terms are included in $u$ in (\ref{eq:polarlevequebase}), we find a similar correction  for $\Sh$ with $\beta/2$ in (\ref{eq:polarthinexpansionflux})  replaced by $f(\gamma,\beta)\beta$  where the $O(1)$ function $f(\gamma,\beta)$ must be computed numerically. We  note that this expansion, at first order in $\xi$, is valid only if $\delta \ll \beta$. If $\delta \sim \beta $ or $\gg \beta$, the scaling analysis (\ref{eq:scalingthinBLwedge}) shows that the cross-channel diffusion term, neglected  in  (\ref{eq:polarlevequebase}), is of the same order or larger than the curvature term. Thus, cross-channel diffusion would  need to be included in  (\ref{eq:polarlevequebase}). This is intuitively  expected as the wedge  approaches the parallel  channel as $\beta\to 0$.


\subsection{Thick boundary layer regime}\label{sec:thickBLwedge}

The terms in the governing advection--diffusion equation (\ref{eq:angledwallblequation}) for $c$ scale such that
\begin{equation}\label{eq:scalingthickBLwedge}
\frac{1}{\delta^2} \sim \frac{1}{\delta^2} \vee \frac{\beta}{\delta(1+\beta\delta)} \vee \frac{1}{(1+\beta\delta)^2},
\end{equation}
where we used $u\partial c/\partial x \sim U_{\delta}/(L/Pe_w)=1/\delta^2$ and $Pe_L=\hat{U}_{\delta}\hat{L}/\hat{D}= L^2/\delta^2$ in the diffusive boundary layer. 
In this regime, the  boundary layer thickness is much larger than the local width of the channel: $ \delta \gg (1+ \delta \beta)$, which implies  strong cross-channel diffusion (last term in (\ref{eq:scalingthickBLwedge})) compared with streamwise advection, radial diffusion and the curvature--diffusion term (first, second and third terms in (\ref{eq:scalingthickBLwedge}), respectively). Thus, we need to examine the influence of two small independent parameters: a physical parameter $1/\delta \ll 1$; and a geometrical parameter $\beta\ll 1$, the opening angle, which shows that the curvature term is also negligible compared with cross-channel diffusion.
Therefore, similar to  parallel channels (see \S\ref{sec:straightwallsthick}), cross-channel diffusion dominates in (\ref{eq:angledwallblequation}) and we have $1/(r+\beta^{-1})^2 \partial^2 c/\partial \theta^2 = 0$ at leading order. This implies $c=\tavg{c}+O(\delta^{-2},\beta/\delta,\beta^2)$ 
is independent of $\theta$ at leading order, owing to the no-flux boundary condition at the walls. 

To analyse the two-dimensional dependence of $c$ on $x$ and $r$, we use the cross-channel averaged advection--diffusion equation (\ref{eq:crosschannelavg2}), where  $\tavg{u'c'}=O(\delta^{-2},\beta/\delta,\beta^2)$ is negligible compared with  $\tavg{u}\,\tavg{c}=O(1)$. Equation (\ref{eq:crosschannelavg2}) becomes, for $0< x < L/Pe_w$ and  $0< r < H$,
\begin{equation}\label{eq:crosschannelavg3}
\zavg{u}\frac{\partial \tavg{c}}{\partial x} =  \frac{\partial^2 \tavg{c}}{\partial r^2}+\frac{1}{(r+\beta^{-1})}\frac{\partial \tavg{c}}{\partial r}.
\end{equation}
The terms in (\ref{eq:crosschannelavg3}) scale as the first three terms in (\ref{eq:scalingthickBLwedge}), which shows that different  balances can arise depending on the ratio of the two small parameters $1/\delta$ and $\beta$, i.e. $\beta \delta$. We examine three sub-regimes: if $\beta\delta \ll 1$, sub-regime (i), the dominant balance is between streamwise advection and radial diffusion; if $\beta\delta \sim 1$, sub-regime (ii), or $\beta\delta \gg 1$, sub-regime (iii) the curvature term is also important and all three terms need to be taken into account at leading order to determine $\tavg{c}$ and eventually the Sherwood number $\Sh$. 

(i) For $\beta \delta \ll 1$, the wedge velocity is $\tavg{u}=1+O \left(\delta^{-2},\beta,(\beta\delta)^2 \right)$.  Then, substituting  $\eta=r/x^{1/2}$ and $\epsilon=1/x^{1/2}$ in  (\ref{eq:crosschannelavg3}) and
using a two-parameter  expansion: $\tavg{c}(\eta,\epsilon)=\tavg{c}_0(\eta)+\epsilon\tavg{c}_{11}(\eta)+(\beta/\epsilon)\tavg{c}_{12}(\eta)+O\left(\delta^{-2},\beta,(\beta\delta)^2 \right)$, we find at leading order  $\tavg{c}_0=\textrm{Erfc}(\eta/2)$  (see appendix \ref{sec:appthickboundarylayerregime} for further details), similar to (\ref{eq:straightwallthickblconcentration}) in  parallel channels as expected intuitively. At the next order in $O(\epsilon)$, we find $\tavg{c}_{11} = 0$. At order $O(\beta/\epsilon)$, we find 
\begin{equation}\label{eq:thickBLwedge1stordercorrectioni}
\tavg{c}_{12} = -\frac{\eta}{2} \textrm{Erfc}\left(\frac{\eta}{2}\right).
\end{equation}
The Sherwood number including the corrections at order $O(\epsilon,\beta/\epsilon)$, is
\begin{equation}\label{eq:thickBLwedgei}
\Sh = \frac{2}{\sqrt{\pi}} \ImpPe^{1/2}  + \frac{\beta}{2},
\end{equation}
for $\beta\ll 1/\delta \ll 1$ with $\delta\sim \ImpPe^{-1/2}$ and $\delta\ll H$. To compute higher-order corrections for $\Sh$, the velocity field must also be expanded at the next order in $O\left(\delta^{-2},\beta,(\beta\delta)^2 \right)$.

(ii) For $\beta \delta \sim 1$, we effectively have only one small parameter $\beta\ll 1$. The velocity  is $\tavg{u}=(1+\beta r)^2+O(\beta^2)$. All three terms in (\ref{eq:crosschannelavg3}) are important, and the resulting equation
\begin{equation}
\left((1+\beta r)^2 +O(\beta^2)\right) \frac{\partial \tavg{c}}{\partial x} =  \frac{\partial^2 \tavg{c}}{\partial r^2}+\frac{1}{(r+\beta^{-1})}\frac{\partial \tavg{c}}{\partial r}
\end{equation}
is not amenable for  asymptotic expansions. Thus, we compute $\tavg{c}$ and  $\Sh$  numerically in this sub-regime in \S\ref{sec:ResultParallelwedge}. However, we expect  that $\delta \sim \beta^{-1} \sim \ImpPe^{-1/2}$, for $\delta\ll H$. Then, we intuitively expect  $\Sh$ to be a function of $\beta$ and $\ImpPe^{1/2}$ at leading order, with $\beta\sim\ImpPe^{1/2}$.

(iii) For $\beta \delta \gg 1$ we have two small parameters: $\beta\ll 1$ and $1/(\beta\delta)\ll 1$, and $\tavg{u}=(\beta r)^2+ 2\beta r + O(\delta^2\beta^4,1,\delta\beta^2)$. Similar to (ii), all three terms in (\ref{eq:crosschannelavg3}) are important and 
\begin{equation}
\left(1 + \frac{2}{\beta r} + O(\beta^2,(\delta\beta)^{-2},\delta^{-1})\right) \beta^2 r^2 \frac{\partial \tavg{c}}{\partial x} =  \frac{\partial^2 \tavg{c}}{\partial r^2}+\frac{1}{(r+\beta^{-1})}\frac{\partial \tavg{c}}{\partial r}
\end{equation}
is  not amenable for asymptotic expansions. We also compute $\tavg{c}$ and  $\Sh$  numerically in \S\ref{sec:ResultParallelwedge} in this sub-regime. Nevertheless, we can expect  that $\delta \sim \beta^{-1/2}\ImpPe^{-1/4}$, for $\delta\ll H$. We  also expect $\Sh$ to be  a function of $\beta^{1/2}$ and  $\ImpPe^{1/4}$ at leading order, following the results found in other regimes. We will show in \S\ref{sec:ResultParallelwedge} that $\delta \sim \beta^{-1/2}\ImpPe^{-1/4}$ is indeed the correct scaling, whilst the Sherwood number varies slightly from the expected scaling.

It is also worth noting that in sub-regime (iii), $\beta\ll 1$ and $\ImpPe\ll \beta^2$, curvature effects have a direct impact on $\delta$ and  $\Sh$ through a curvature-rescaled \Peclet number $\ImpPe_{\beta}=\beta^2\ImpPe$. This rescaling is due to the opening geometry of the wedge allowing the velocity to increase as $U_{\delta}\sim (\beta \delta)^2$. Hence, we have $\delta \sim \ImpPe_{\beta}^{-1/4}$. The curvature-rescaled \Peclet number $\ImpPe_{\beta}$ is somewhat analogous to the Dean number, $De = Re\sqrt{D/(2R_c)}$ (with $Re$ the characteristic pipe flow Reynolds number, $D$ the pipe diameter and $R_c$ a characteristic radius of curvature of the pipe flow), which accounts for  secondary recirculation flows due to curvature effects in slightly bent pipe flows \citep[e.g.][]{berger83}. 

In summary, the two-dimensional thick boundary layer regime exists for  wedge flows provided  both $\beta\ll 1$ and $\delta\gg 1$. Sub-regime (i) only exists for small enough opening angle: $\beta\ll 1/\delta\ll 1$, which is effectively possible for $\beta \lesssim 0.01$. Sub-regime (iii) only exists for thick enough diffusive boundary layers: $\delta \gg 1/\beta\gg 1$, which is only possible for $\delta \gtrsim 100$. If either $\beta\sim 1$ or $\delta \sim 1$, the diffusive boundary layer is not thick compared with the local width of the gap and the   thick boundary layer regime does not apply. Terms in the governing equation (\ref{eq:crosschannelavg2}), which have been neglected or considered small  in this regime, can become important. In \S\ref{sec:ResultParallelwedge}, we explore using numerical calculations whether the two-dimensional thick boundary layer regime holds beyond its theoretical range of validity or whether three-dimensional effects become important.


\subsection{Radial confinement, $\hat{\delta}\sim \hat{H}$}\label{sec:confinmentwedge}

Similar to \S\ref{sec:confinmentParallel}, we study the impact of radial confinement $\delta\sim H$ on $\Sh$ using the cross-channel averaged advection--diffusion equation (\ref{eq:crosschannelavg2}) under the free-slip and no-flux top boundary condition (\ref{eq:crosschannelavg2BC}\textit{d}). Integrating (\ref{eq:crosschannelavg2}) in the streamwise direction from $0$ to $L/Pe_w$, we obtain
\begin{align} \label{eq:differentialfluxwedge}
-\frac{\partial \avg{j_r}}{\partial r} - \frac{\avg{j_r}}{(r+\beta^{-1})} & = \fp{\avg{c}}{r}{2} + \frac{1}{(r+\beta^{-1})} \fp{\avg{c}}{r}{} \nonumber\\
& = \ImpPe   \left. \zavg{u} \, \zavg{c}\right|_{x=L/Pe_w} + \ImpPe   \left.\zavg{u'c'}\right|_{x=L/Pe_w} = q_m + q',
\end{align}
with $j_r=-\partial c/\partial r$ the radial flux at a particular $r$ coordinate. A new term exists compared to parallel channels and  (\ref{eq:differentialflux}): the second term on the left-hand side is due to curvature. Integrating again in the radial direction from $0$ to $H$, we obtain
\begin{align}\label{eq:totalcomposedfluxwedge}
\Sh &+ \left[ \frac{\avg{c}}{(r+\beta^{-1})} \right]^H_0 + \int_0^H  \frac{\avg{c}}{(r+\beta^{-1})^2} \mathrm{d}r \nonumber\\
&= \ImpPe \int_0^H \left. \zavg{u} \, \zavg{c}\right|_{x=L/Pe_w} \mathrm{d}r + \ImpPe \int_0^H \left.\zavg{u'c'}\right|_{x=L/Pe_w} \mathrm{d}r= \avg{j_m}+\avg{j'},
\end{align}
where the curvature term has been integrated by parts.
Similar to \S\ref{sec:confinmentParallel}, we assume that $q'$ is either negligible compared to $q_m$ or scales in a similar fashion. We will discuss this assumption in detail in \S\ref{sec:ResultParallelwedge}, but we note that in the thick boundary layer regime (see \S\ref{sec:thickBLwedge}) we showed that $q'\ll q_m$. For $\delta\sim H$, we must have $\zavg{c}(x=L/Pe_w) \sim 1$. Hence,
\begin{equation}\label{eq:confinedfluxwedge}
\Sh \sim Q \ImpPe = U H\ImpPe,
\end{equation}
with $Q=\int_0^H \zavg{u} \mathrm{d}y$ (and $u(r,\theta)$ from (\ref{eq:angledwallsflowfield})) the wedge volume flow rate and $U$ the mean channel velocity. We have neglected the weak dependence of  $\avg{c}$ with $r$  in the integral on the left hand side of (\ref{eq:totalcomposedfluxwedge}).
In the limit of small or large channel heights, we find that the radially confined Sherwood number is: $\Sh \sim  H^3\ImpPe$ for $H\ll 1$, since $U\sim H^2$ in the $r$ direction; $\Sh \sim  H\ImpPe $ for $H\gg 1$ and $\beta H\ll 1$ or $\sim 1$, since $U$ is nearly uniform in the $r$ direction at leading order for small enough opening angles; and $\Sh \sim  H^3 \ImpPe_{\beta}$ for $H\gg 1$ and  $\beta H\gg 1$, where the curvature-rescaled \Peclet number $\ImpPe_{\beta}= \beta^2\ImpPe$ appears again, as in sub-regime (iii) of the thick boundary layer regime (see \S\ref{sec:thickBLwedge}).


\subsection{Transition regime, $\hat{\delta}\sim \hat{w}$ or $\beta\sim 1$, and numerical formulations for three- and two-dimensional problems}\label{sec:angledwallstransition}

In wedge flows, for $\ImpPe\sim 1$ or $\delta\sim 1$, or for $\beta\sim 1$, and in sub-regimes (ii) and (iii) of the thick boundary layer regime (see \S\ref{sec:thickBLwedge}), $c$ is three-dimensional. We study the impact of three-dimensional effects on $\Sh$ by solving (\ref{eq:angledwallblequation}) numerically under (\ref{eq:angledwallblequationBC}\textit{a,b,d--f}) and using our three-dimensional result (\ref{eq:angledwallsflowfield}) for $u$. Using the same method as in \S \ref{sec:straightwallstnumerics}, homogenisation of the boundary conditions, followed by separation of variables, leads to 
\begin{equation}\label{eq:numconcfieldwedge}
 c (x, r , \theta ) = 1-\sum_{n=1}^{\infty}\exp(-\rho_n x) B_n( r, \theta ).
\end{equation}
The eigenpairs $B_n$ and $\rho_n$ are solutions of the homogeneous eigenvalue problem
\begin{equation}\label{eq:wedgeevproblem}
-u \rho_n B_n = \frac{\partial^2 B_n}{\partial  r ^2} + \frac{1}{r+\beta^{-1}}\frac{\partial B_n}{\partial r}+\frac{1}{(r+\beta^{-1})^2}\frac{\partial^2 B_n}{\partial  \theta ^2},
\end{equation}
for all integers $n\geq 1$, $0<x<L/Pe_w$, $0< r <H$, $\left| \beta \right| < \theta/2$,  with boundary conditions
\begin{equation}\label{eq:wedgeevproblemBC}
B_n( r =0, \theta ) = 0,\ \frac{\partial B_n}{\partial  r}  ( r =H, \theta )=0,\  B_n( r , \theta =\pm 1/2)=0.
\end{equation}
We compute $\left|B_n\right|$ in (\ref{eq:numconcfieldwedge}) using ${c(x=0 ,r,\theta)=0}$ and the orthogonality of the eigenfunctions. As in  parallel  channels, we solve a second-order finite difference formulation of  (\ref{eq:wedgeevproblem}) using LAPACK \citep{LAPACK} (see more detail  in appendix \ref{apdx:NumSimDet}).

For comparison, we also solve a two-dimensional  formulation of this problem  based on the cross-channel averaged equation  (\ref{eq:crosschannelavg2}), neglecting the three-dimensional flux $\tavg{u'c'}$:
\begin{equation}\label{eq:2dpolarproblem}
\tavg{u}\frac{\partial \tavg{ c }}{\partial x} = \frac{\partial^2 \tavg{ c }}{\partial  r ^2}+ \frac{1}{ r +\beta^{-1}} \frac{\partial \tavg{ c }}{\partial  r },
\end{equation}
for $0 <  r  < H$, $0 < x < \infty$, under (\ref{eq:crosschannelavg2BC}\textit{a,b,d}).
Homogenisation of the boundary conditions leads to a one-dimensional eigenvalue problem, which we solve using a shooting method \citep{Berry:1952} to obtain $\tavg{ c }$ and a two-dimensional $\Sh$. This simpler two-dimensional formulation of the transport problem in wedges allows us to assess \textit{a posteriori} the error on $\Sh$ when neglecting the three-dimensional flux $\overline{u'c'}$.


\subsection{Results in truncated wedges}\label{sec:ResultParallelwedge}

In this section, we compare our asymptotic predictions for $\delta$ and  $\Sh$ in the wedge geometry with  three- and two-dimensional  numerical calculations of the advection--diffusion equation.  Similar to \S\ref{sec:ResultParallel}, the aim here is to assess whether three-dimensional effects related to the corners  or due to confinement have a strong impact on $\delta$ and $\Sh$ in the different regimes identified previously.
We study the influence of $\ImpPe$, $\beta$, which controls the importance of  curvature effects, not present in  parallel channels, and lateral and radial confinement effects. We also analyze the relative magnitude of the three-dimensional fluctuation flux $\zavg{u'c'}$ and whether it can be neglected in  (\ref{eq:crosschannelavg2}).

\subsubsection{Concentration field}

\begin{figure}
\includegraphics[keepaspectratio=true,width=0.9\textwidth]{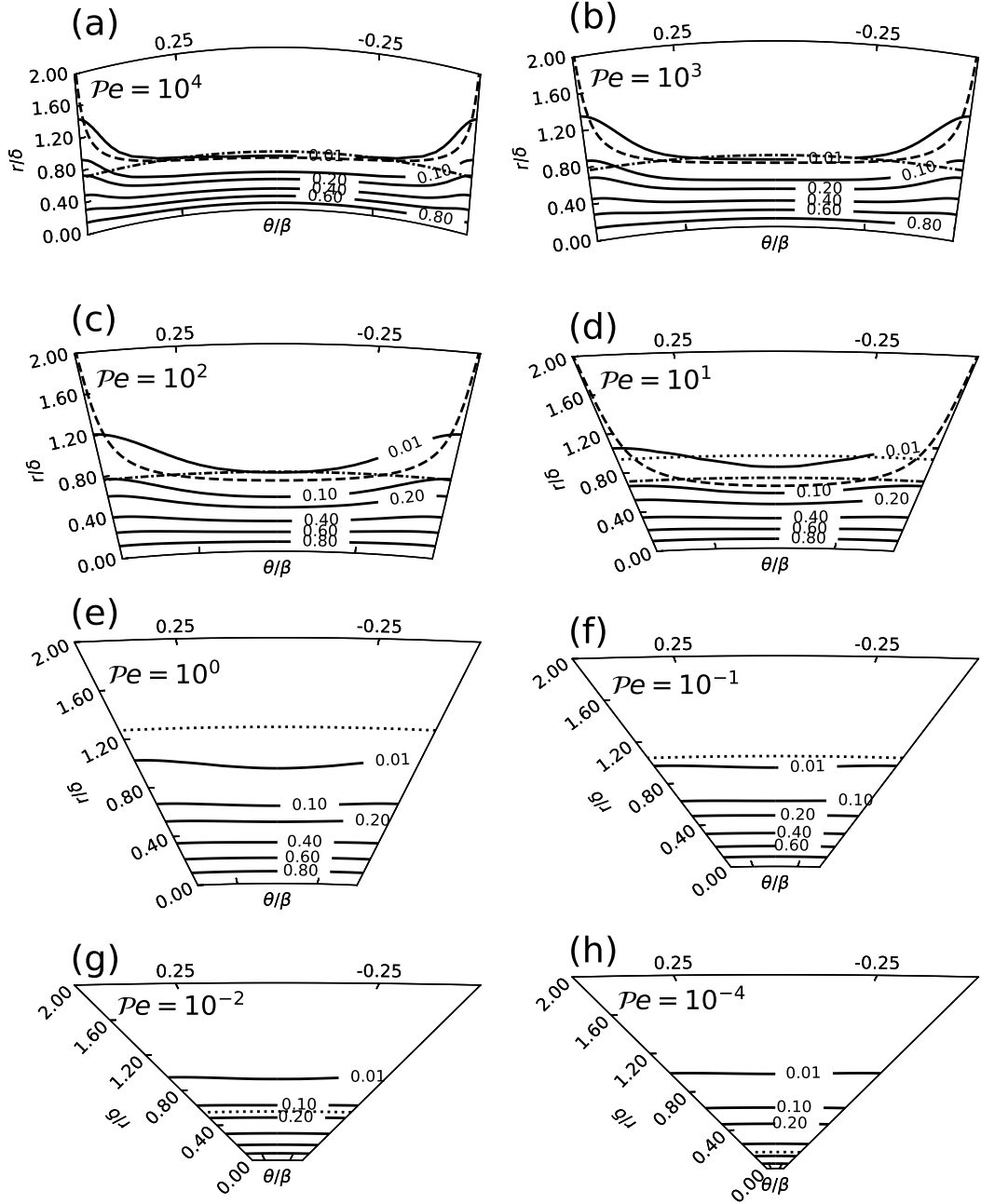}
\caption{
Contour plots of the three-dimensional concentration field computed  numerically (solid lines) using (\ref{eq:numconcfieldwedge}--\ref{eq:wedgeevproblemBC}) for $\beta = 0.3$, at $x=L/Pe_w$, for various $\ImpPe$ (see details in table~\ref{tab:numdet}, appendix~\ref{apdx:NumSimDet}).  In (\textit{a--d}),  dashed lines show the slice-wise thin boundary layer predictions (\ref{eq:levequestraightslicewiseconcentrationfield}) (substituting $(y,z)$ by $(r,\theta)$) for  $\delta$  with the first order curvature correction (\ref{eq:1stordercorrectioncthinBLwedge}) ($\ImpPe^{1/3}\gg 1$). Dash-dotted lines show the two-dimensional predictions  for $\delta$ based on (\ref{eq:levequestraightslicewiseconcentrationfield}). In (\textit{d--h}),  dotted lines show the thick boundary layer predictions  for $\delta$ in sub-regime (i) (\ref{eq:straightwallthickblconcentration}) (substituting $(y,z)$ by $(r,\theta)$) with the first order curvature correction (\ref{eq:thickBLwedge1stordercorrectioni}) ($\beta \ll \beta\delta \ll 1\ll \delta\ll H$ or $\beta \ll \ImpPe^{1/2} \ll 1$). Although all panels have the same $\beta=0.3$, the scaling distorts the region so that as $\ImpPe$ is decreased the range of $r$ increases.
}
\label{fig:polarcontourplots}
\end{figure}

In  figure~\ref{fig:polarcontourplots}, we show contour plots in polar coordinates $(0\leq r\leq 2\delta,-\beta/2\leq \theta\leq \beta/2)$ of $c$ at the end of the area of release, $x=L/Pe_w$, for various \Peclet numbers: from $\ImpPe=10^4$ (figure \ref{fig:polarcontourplots}\textit{a}) to $10^{-4}$ (\ref{fig:polarcontourplots}\textit{h}). For conciseness, we only show results for $\beta=0.3$. At smaller angles $\beta$, the concentration  converges towards the parallel  geometry, while curvature effects are increasingly important at larger $\beta$. Solid lines show the three-dimensional numerical results computed using (\ref{eq:numconcfieldwedge}--\ref{eq:wedgeevproblemBC}). 
We normalise the $r$-axis by $\delta$, computed as $\delta=\zavg{r_{\delta}}$ with $c(L/Pe_w,r_{\delta},\theta)=0.01$. As can be seen in figure~\ref{fig:polarcontourplots}, this leads to a distortion of the region being viewed, with lower $\ImpPe$ cases having a much greater range of $r$. The dashed lines, shown in figures~\ref{fig:polarcontourplots}(\textit{a--d})  where $\ImpPe\geq 10$, are plotted using the thin boundary layer predictions (\ref{eq:levequestraightslicewiseconcentrationfield}) (substituting $(y,z)$ by $(r,\theta)$) with the first-order curvature correction (\ref{eq:1stordercorrectioncthinBLwedge}), which used $\gamma(\theta)$, from (\ref{eq:wedgegamma}), but assumed no cross-channel diffusion. The dash-dotted lines are plotted using (\ref{eq:levequestraightslicewiseconcentrationfield}) and (\ref{eq:1stordercorrectioncthinBLwedge}) assuming a two-dimensional velocity profile, i.e. replacing $\gamma(\theta)$ by $\zavg{\gamma}$. These two predictions correspond to $\delta \ll 1$ or $\ImpPe^{1/3} \gg 1$. The dotted lines, shown in figures~\ref{fig:polarcontourplots}(\textit{d--h})  where $\ImpPe\leq 10$, are plotted using  (\ref{eq:straightwallthickblconcentration}) (substituting $(y,z)$ by $(r,\theta)$) for $\tavg{c}$ with the first-order curvature correction (\ref{eq:thickBLwedge1stordercorrectioni}) in $\beta/\epsilon=x^{1/2}\beta$. These lines show the asymptotic predictions in the thick boundary layer regime, sub-regime (i), for $\beta \ll \beta\delta \ll 1\ll \delta\ll H$ or $\beta \ll \ImpPe^{1/2} \ll 1$. 

Similar to the parallel geometry, for $\ImpPe \gtrsim 100$ (figures~\ref{fig:polarcontourplots}\textit{a--c}) the two-dimensional thin boundary layer  predictions (dash-dotted lines) are in reasonable agreement (within \SI{12}{\percent} deviation) with the three-dimensional numerical results in the interior of the channel $|\theta/\beta|< 0.25$. The diffusive boundary layer is better captured by the slice-wise thin boundary layer predictions  (dashed lines) at large $\ImpPe$ since the influence of the wall boundary layer reduces (see figures~\ref{fig:polarcontourplots}\textit{a,b}) for $0.3<|\theta/\beta|< 0.5$). The main distinction between this and the parallel  geometry is that the transition between the thin and thick boundary layer regimes can occur at lower $\ImpPe$ in the  wedge and over a wider range: approximately $10^{-4} \lesssim \ImpPe \lesssim 10$ (see figures~\ref{fig:polarcontourplots}\textit{d--h}). The transition occurred for $1 \lesssim \ImpPe \lesssim 100$ in parallel channels (see figure~\ref{fig:contourcart}). This is due to curvature effects when $\beta$ is not very small, such as here with $\beta=0.3$. Then, as $\ImpPe$ decreases, the concentration contours flatten owing to cross-channel diffusion, which becomes the dominant effect at low $\ImpPe$. 
We can also notice that the thick boundary layer prediction for $\delta$ in  sub-regime (i)  (dotted line in figures~\ref{fig:polarcontourplots}\textit{d--h})  only  has approximate agreement with the  numerical results (see concentration contour $c=0.01$) in a sub-range of the transition:  for $10^{-1} \lesssim \ImpPe \lesssim 10^{1}$. 
At lower  $\ImpPe$, the prediction in sub-regime (i) consistently underestimates $\delta$, with increasing  deviation from the numerical results as $\ImpPe$ decreases. This is due to the fact that $\beta=0.3$ is too large for sub-regime (i) because this sub-regime is theoretically valid for $\beta \lesssim 0.01$ (\S\ref{sec:thickBLwedge}). Nevertheless,  the contour plots reveal that the asymptotic results from sub-regime (i)  still provide qualitative prediction at  angles an order of magnitude larger than its theoretical range of validity. 
For $\beta=0.3$, sub-regimes (ii) and (iii) are valid for $\ImpPe\sim \SI{9e-2}{}$  and $\ImpPe\ll \SI{9e-2}{}$, as shown in table~\ref{tab:regimeswedgeresults}. In these two sub-regimes, curvature effects become more important, enhancing radial diffusion and leading to thicker boundary layers, comparatively with sub-regime (i) or parallel channels. 

\begin{table}
\centering
\scalebox{0.95}{
\begin{tabular}{lcccc}
\hline
$\beta=$ & $0.01$ & $0.1$ & $0.2$ & $0.3$ \\ \hline
\textit{Thin boundary layer regime} & $\ImpPe\gtrsim \SI{e3}{}$ & $\ImpPe\gtrsim \SI{e3}{}$ & $\ImpPe\gtrsim \SI{e3}{}$ & $\ImpPe\gtrsim \SI{e3}{}$ \\
\textit{Thick boundary layer regime} &  &  &  & \\
Sub-regime (i) & $\ImpPe\sim \SI{e-2}{}$ & -- & -- & -- \\
Sub-regime (ii) & $\ImpPe\sim \SI{e-4}{}$ & $\ImpPe\sim \SI{e-2}{}$ & $\ImpPe\sim \SI{4e-2}{}$ & $\ImpPe\sim \SI{9e-2}{}$ \\
Sub-regime (iii) & $\ImpPe\ll \SI{e-4}{}$ & $\ImpPe\ll \SI{e-2}{}$ & $\ImpPe\ll \SI{4e-2}{}$ & $\ImpPe\ll \SI{9e-2}{}$ \\
$\delta\sim H$ (radial confinement) & $\ImpPe\lesssim \SI{4e-3}{}$ & $\ImpPe\lesssim \SI{2e-3}{}$ & $\ImpPe\lesssim \SI{5e-4}{}$ & $\ImpPe\lesssim \SI{2e-4}{}$ \\
Confinement regime & (i) & (iii) & (iii) & (iii) \\ \hline
$\beta=$ & $0.5$\textdagger & $1$\textdagger & $\pi/2$\textdagger &  \\ \hline
\textit{Thin boundary layer regime} & $\ImpPe\gtrsim \SI{e3}{}$ & $\ImpPe\gtrsim \SI{e3}{}$ & $\ImpPe\gtrsim \SI{e3}{}$ &  \\
\textit{Thick boundary layer regime} &  &  &  & \\
Sub-regime (i) & -- & -- & -- &  \\
Sub-regime (ii) & -- & -- & -- &   \\
Sub-regime (iii) & -- & -- & -- &  \\
$\delta\sim H$ (radial confinement) & $\ImpPe\lesssim \SI{8e-5}{}$ & $\ImpPe\lesssim \SI{2e-5}{}$ & $\ImpPe\lesssim \SI{8e-6}{}$ &  \\
Confinement sub-regime & (iii) & (iii) & (iii) &  
\end{tabular}
}
\caption{Range of \Peclet numbers for the  asymptotic regimes found in figures~\ref{fig:polarcontourplots}--\ref{fig:polar1}. \textdagger For $\beta=0.5$, $1$ and $\pi/2$, $\beta$ is theoretically not small enough for the thick boundary layer  regime to exist. Nevertheless, we have computed the transition to the radially confined regime assuming that $\delta$  follows sub-regime (iii): $\delta \sim \ImpPe_{\beta}^{-1/4}=\beta^{-1/2}\ImpPe^{-1/4}$. The transition \Peclet number to the radially confined regime is shown with a black star in figure~\ref{fig:polar1} for  each $\beta$.
}
\label{tab:regimeswedgeresults}
\noindent\rule{\textwidth}{0.4pt}
\end{table}

\subsubsection{Three-dimensional fluxes}

\begin{figure}
\centering
\includegraphics[keepaspectratio=true,width=0.7\textwidth]{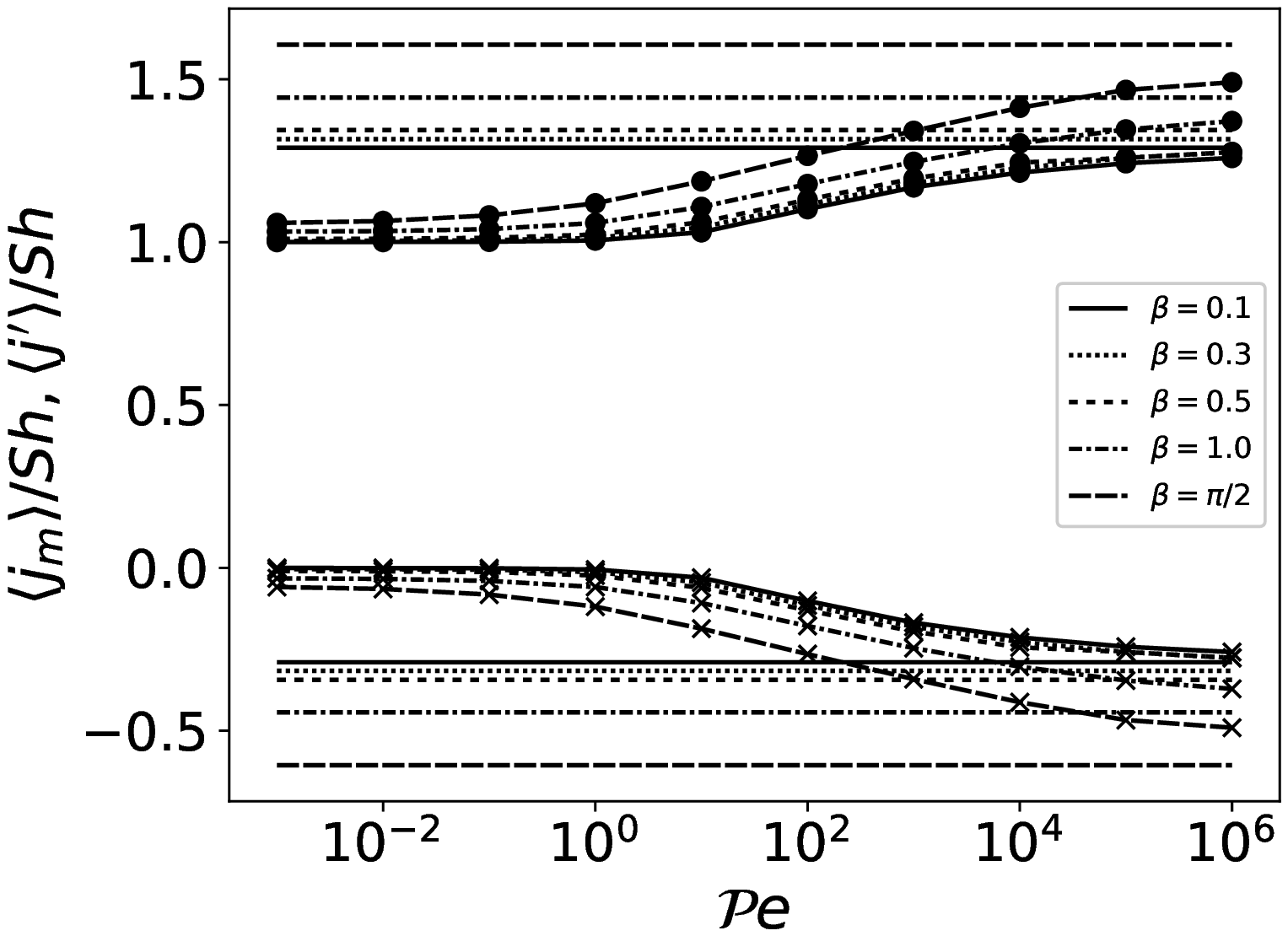}
\caption{ 
Variations of the normalised total  mean flux $\avg{j_m}>0$ (dots) and  total fluctuation flux $\avg{j'}\leq 0$ (crosses) with $\ImpPe$ in  wedges for $0.1\leq \beta \leq \pi/2$. 
The fluxes $\avg{j_m}$ and $\avg{j'}$ are computed  numerically using  (\ref{eq:totalcomposedfluxwedge}) (see details in table~\ref{tab:numdet}, appendix~\ref{apdx:NumSimDet}).  The  thin boundary layer predictions  (\ref{eq:levequestraightslicewiseconcentrationfield}) are the horizontal lines with a line style matching the numerical calculations for each $\beta$.
}
\label{fig:fluxcontributionswedge}
\end{figure}

To analyze the impact of the three-dimensional fluctuation flux $\zavg{u'c'}$ on the total flux or Sherwood number, we plot  in figure~\ref{fig:fluxcontributionswedge} $\avg{j_m}$ and $\avg{j'}$ from  (\ref{eq:totalcomposedfluxwedge}), computed  numerically using  (\ref{eq:numconcfieldwedge}--\ref{eq:wedgeevproblemBC}) for $\beta=0.1$ (symbols and solid line), $\beta = 0.3$ (symbols and dotted line), $\beta=0.5$ (symbols and dashed line), $\beta = 1$ (symbols and dash-dotted line) and $\beta = \pi/2$ (symbols and long-dashed line) (see  table~\ref{tab:numdet}, appendix~\ref{apdx:NumSimDet}, for more details). We  show the asymptotic predictions for $\delta\ll 1$  computed  numerically  using  (\ref{eq:levequestraightslicewiseconcentrationfield}) (substituting $(y,z)$ by $(r,\theta)$)  and  the first-order curvature correction (\ref{eq:1stordercorrectioncthinBLwedge}). They correspond to the horizontal lines plotted with a line style matching the  numerical results for each $\beta$.


Figure~\ref{fig:fluxcontributionswedge} shows that in the thick boundary layer regime (low $\ImpPe$), the negative contribution of the total fluctuation flux $\avg{j'}$  vanishes for all $\beta$. For $\ImpPe\sim 1$, larger values of $\beta$ lead to a faster increase in the contribution of $\avg{j'}$, thus extending the range of the transition between thin and thick boundary layer regimes for $\beta\sim 1$.
In the  thin boundary layer regime (large $\ImpPe$),  $\avg{j'}$  reduces the contribution from that evaluated just on the mean flux $\avg{j_m}$ by between approximately \SI{25}{\percent} ($\beta \ll 1$) and \SI{50}{\percent} ($\beta =\pi/2$).  The  numerical calculations (symbols) converge asymptotically towards the predictions in  the thin boundary layer regime at large $\ImpPe$. The thin boundary layer  predictions  capture the increasing trend in the contribution of $\avg{j'}$ with increasing $\beta$.
For $\beta=0.1$, the results are similar to those obtained in parallel  channels (figure~\ref{fig:fluxcontributions}\textit{b}).
This suggests that  a two-dimensional description of the flux in  wedges is also appropriate at leading order  for the full range of \Peclet numbers studied, provided $\beta\ll 1$. At this stage, it is uncertain whether  a two-dimensional description remains  accurate at large $\ImpPe$ and for $\beta\sim 1$ or whether three-dimensional effects must be included.  We discuss this further below. 

\subsubsection{Sherwood number}

\begin{figure}
\includegraphics[width=\textwidth,keepaspectratio=true]{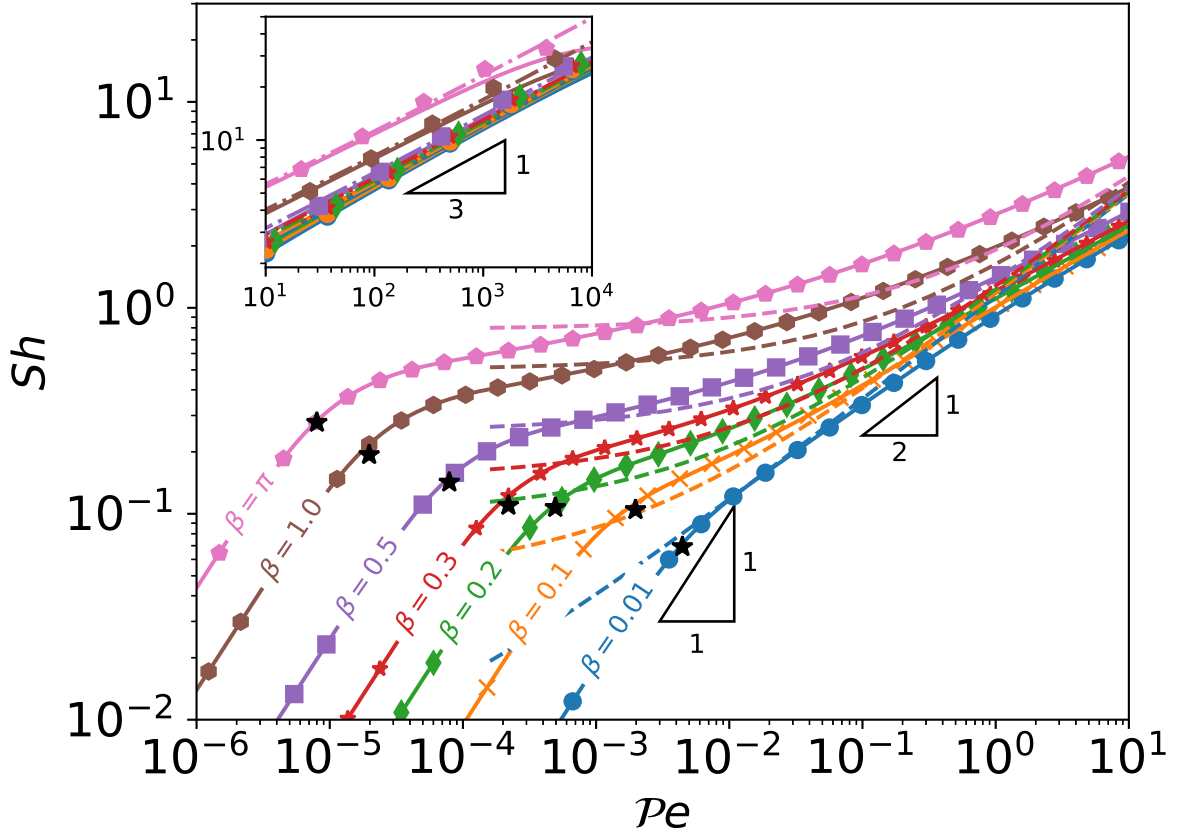}
\caption{
Sherwood number versus  \Peclet number in wedges. Three-dimensional numerical results based on (\ref{eq:numconcfieldwedge}--\ref{eq:wedgeevproblemBC}) for $0.01\leq \beta \leq \pi$ are shown as symbols. 
Corresponding two-dimensional numerical results based on (\ref{eq:2dpolarproblem}) (neglecting the three-dimensional flux $\tavg{u'c'}$) are plotted with solid lines closely following the symbols.
Thick boundary layer predictions (\ref{eq:thickBLwedgei}), sub-regime (i), are plotted  using dashed lines in matching colours. 
As $\ImpPe \to 0$, the transition to the radially confined regime $\Sh\sim H\ImpPe$ is marked by black stars (see table~\ref{tab:regimeswedgeresults}).
Inset:  thin boundary layer  predictions (\ref{eq:polarthinexpansionflux})  plotted with dash-dotted lines in matching colours. 
}
\label{fig:polar1}
\end{figure}

Figure~\ref{fig:polar1} shows $\Sh$ computed from the three-dimensional  numerical calculation of  (\ref{eq:numconcfieldwedge}--\ref{eq:wedgeevproblemBC})  versus $\ImpPe$, for $H=15$, and for: $\beta = 0.01$ (blue circles), $\beta=0.1$ (orange crosses), $\beta=0.2$ (green lozenges), $\beta=0.3$ (red stars), $\beta=0.5$ (violet squares), $\beta=1.0$ (brown hexagons), $\beta=\pi/2$ (pink pentagons). The solid lines closely following the symbols  correspond to the two-dimensional  numerical results based on (\ref{eq:2dpolarproblem}), neglecting the three-dimensional  flux $\zavg{u'c'}$ (see table~\ref{tab:numdet},  appendix \ref{apdx:NumSimDet}, for  details about the numerical computations). 
For $\SI{e-6}{}\leq \ImpPe\leq \SI{e4}{}$, the two-dimensional  numerical results are mostly in  agreement with the three-dimensional  numerical results. For $\beta=\pi/2$, we find that the deviation between the two-dimensional and the three-dimensional results is within  less than \SI{5}{\percent}  for $\ImpPe \leq10^2$, and  \SI{10}{\percent} for $10^2<\ImpPe\leq  \num{3e2}$. For $\beta \leq 0.3$, the deviation is within less than \SI{5}{\percent}  for  $\ImpPe \leq \num{4.2e2}$, and \SI{10}{\percent} for $\num{4.2e2}<\ImpPe \leq \num{1.5e3}$. The  deviation for  intermediate $\beta$ are within the same bounds. 
The increased deviation observed for  $\beta\sim 1$ and $\ImpPe\geq 100$  (brown and pink curves, inset of figure~\ref{fig:polar1}) is due to a combination of truncation error and reduced resolution in the  calculation of $\Sh$, which is performed using different methods between the two-dimensional and  three-dimensional numerical calculations. 
In general, we find that increasing the resolution and the number of eigenpairs for the calculation of $\Sh$ reduces the deviation  at large $\ImpPe$ and for $\beta$ up to $\pi/2$ (see  appendix~\ref{sec:appconv}, figure~\ref{fig:gammabarrationwedge}(\textit{b})). 
This also improves the agreement between the numerical results and the slice-wise thin boundary layer   predictions (\ref{eq:polarthinexpansionflux}) at large $\ImpPe$ (dash-dotted lines  using matching colours for each $\beta$, inset only). We can also notice that  the curves do not collapse at large $\ImpPe$. This is due to the fact that $\Sh$ depends on $\beta$ (see (\ref{eq:polarthinexpansionflux})).

The results in figure~\ref{fig:polar1} clearly demonstrate that for applications not requiring a high accuracy for $\Sh$, the three-dimensional fluctuation flux $\zavg{u'c'}$ can be neglected and the  two-dimensional formulation (\ref{eq:2dpolarproblem}) can be used for all $\ImpPe$ and  $\beta$ up to at least $\pi/2$. As mentioned in \S\ref{sec:ResultParallel}, the two-dimensional formulation significantly  reduces  computational burden  whilst preserving reasonable accuracy.
In addition, our slice-wise  thin boundary layer predictions (\ref{eq:polarthinexpansionflux})  provide fast and accurate complementary estimates of $\Sh$ in the computationally challenging regime at $\ImpPe^{1/3}\gg 1$  and for all $\beta$.

As $\ImpPe$ decreases, a more complex behaviour emerges due to the increased effect of curvature  for non-negligible opening angles. For $\beta=0.01$ (blue symbols and curves in figure~\ref{fig:polar1}), curvature effects are negligible and, as long as $\delta\ll H$ ($H\ImpPe^{1/2}\gg 1$), the two-dimensional thick boundary layer predictions (\ref{eq:thickBLwedgei}), sub-regime (i), $\Sh \sim \ImpPe^{1/2}+\beta/2$, $\beta\ll \beta\delta \ll 1\ll \delta \ll H$ or $\beta \ll \ImpPe^{1/2}\ll 1$ (dashed lines using matching colours for each $\beta$ in main graph), agree with the numerical computations in the range predicted in table~\ref{tab:regimeswedgeresults}. Then, as $\beta$ increases,  $\Sh$ increases at fixed $\ImpPe$, departing from this prediction (see all colours other than blue). This is due to the fact that $\beta\delta$ increases and sub-regime (i) is not valid any more. As shown in table~\ref{tab:regimeswedgeresults}, for $0.01\ll\beta\ll 1$ sub-regime (i) disappears and the diffusive boundary layer can be in sub-regimes (ii) or (iii) of the thick boundary layer regime, where the curvature term in the advection--diffusion equation (\ref{eq:crosschannelavg3}) becomes non-negligible and no asymptotic predictions exist for $\Sh$ in sub-regimes (ii) and (iii). Table~\ref{tab:regimeswedgeresults}  presents the range of $\ImpPe$ where sub-regimes (ii) and (iii) are valid, provided $\beta\ll 1$. For $\beta\geq 0.5$, the results shown in violet, brown and pink cannot be considered in the thick boundary layer regime since $\delta\sim 1+\beta\delta$ (see scaling analysis in \S\ref{sec:thickBLwedge}). 

Then, if $\beta\sim 1$, all the terms in the governing advection--diffusion equation (\ref{eq:angledwallblequation}) are important, making the  problem even more three-dimensional and requiring full numerical calculation of (\ref{eq:angledwallblequation}). As can be seen in figure~\ref{fig:polar1}, an increase in $\beta$ leads to an increase in $\Sh$, which appears to tend towards a plateau, reducing its dependence with $\ImpPe$. 

At very low $\ImpPe$, radial confinement  becomes important and the curves follow another regime: $\Sh\sim  H\ImpPe$ (see (\ref{eq:confinedfluxwedge})) similar to parallel channels (see  figure~\ref{fig:cartfull}). The $\ImpPe$ at which the the radially confined regime occurs depends on the regime that the boundary layer would be without confinement effect. The corresponding transitional $\ImpPe$ and the associated regime are indicated in the rows ``$\delta\sim H$'' and ``Confinement regime'' in table~\ref{tab:regimeswedgeresults}. The predictions for the transitional $\ImpPe$ (black stars in figure~\ref{fig:polar1}) agree with the numerical results. For $\beta\geq 0.5$, since the  thick boundary layer regime is not theoretically valid, as discussed previously,  the predictions given in table~\ref{tab:regimeswedgeresults} assume that the diffusive boundary layer is in sub-regime (iii) of the thick boundary layer regime. As shown in figure~\ref{fig:polar1}, the estimated transitional $\ImpPe$  are still accurate even for $\beta\geq 0.5$, at least up to $\pi/2$ (see black stars for the violet squares, brown hexagons and pink pentagons). We note that the locus of the confinement transition is not a simple curve. This is  partly due to the fact that the confinement transition occurs in different sub-regimes, but also that for $\beta=0.5$, 1 and $\pi/2$ the transition does not occur in an asymptotic regime, as stated in table~\ref{tab:regimeswedgeresults}.

\begin{figure}
\includegraphics[width=\textwidth,keepaspectratio=true]{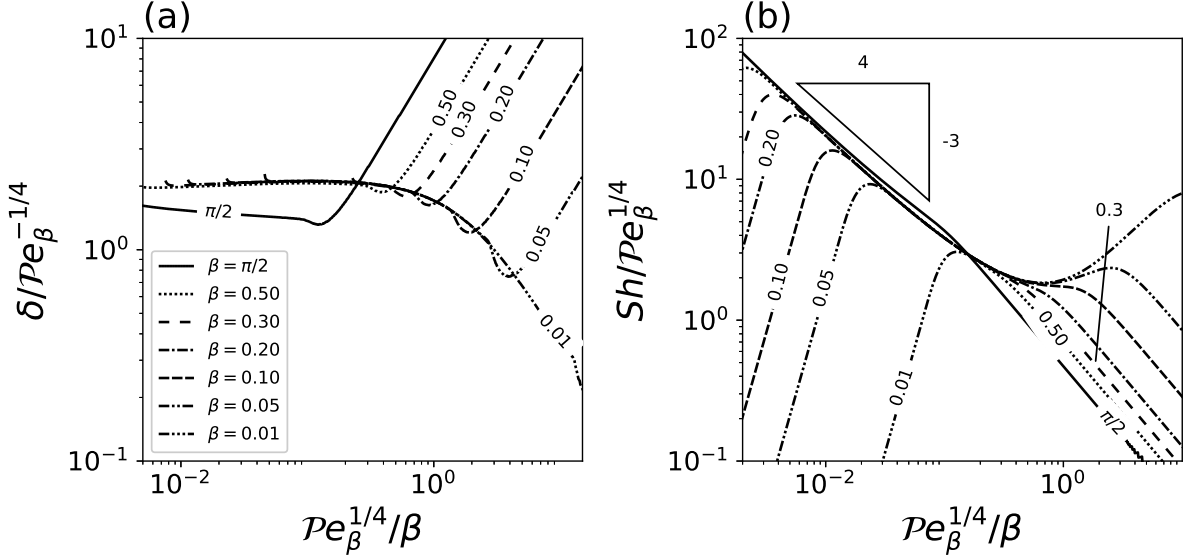}
\caption{
Study of the thick boundary layer regime, sub-regime (iii), (see \S\ref{sec:thickBLwedge}) where $\beta \ll 1 \ll \beta\delta \ll \delta\ll H$ or $\ImpPe_{\beta}^{1/4}\ll \beta\ll 1$ and the dependence with the curvature-rescaled \Peclet number $\ImpPe_{\beta}=\beta^2\ImpPe$. (\textit{a}) Variation of the normalised $\delta$ with $\ImpPe_{\beta}^{1/4}/\beta$  for various  $\beta$. 
(\textit{b}) Variation of $\Sh/\ImpPe_{\beta}^{1/4}$ with $\ImpPe_{\beta}^{1/4}/\beta$ and for the same $\beta$ as in (\textit{a}). 
\label{fig:polar2}
}
\end{figure}

Sub-regime (iii) of the thick boundary layer regime occurs only at  very low \Peclet numbers: $\ImpPe^{1/2}\ll \beta \ll 1$. As seen in table~\ref{tab:regimeswedgeresults}, this regime may only appear in figure~\ref{fig:polar1}   for a very limited range of $\ImpPe$ and  for $\beta=0.2$ and 0.3 only, as  radial confinement effects also become important at similar $\ImpPe$. In sub-regime (iii), we noted in   \S\ref{sec:thickBLwedge} the importance of a curvature-rescaled \Peclet numbers $\ImpPe_{\beta}=\beta^2\ImpPe$ since $U_{\delta}\sim (\beta\delta)^2$. In general, we must have $\ImpPe_{\beta}^{-1/4}\ll H$ when $\ImpPe_{\beta}^{1/4}/\beta \ll 1$ for sub-regime (iii) to exist without  being affected by radial confinement effects. To show sub-regime (iii) more clearly, we plot in figure~\ref{fig:polar2}(\textit{a})  $\delta/\ImpPe_\beta^{-1/4}$ for various $0.01\leq \beta\leq 0.1$ as a function of  $\SI{e-2}{} \leq \ImpPe_{\beta}^{1/4}/\beta\leq 10$, effectively ranging $\SI{e-12}{}\leq \ImpPe \leq 100$. 
All the results shown in figure~\ref{fig:polar2}(\textit{a}) and (\textit{b}) were computed numerically using the two-dimensional formulation (\ref{eq:2dpolarproblem}), for $H=1000$ and $n=5000$. We decided to use the two-dimensional formulation, instead of the exact three-dimensional formulation, due to computational difficulties in reaching sufficiently low $\ImpPe$. We expect  the results to remain accurate since, as we have shown previously, the error made using the two-dimensional formulation remains small, particularly at low $\beta$ and low $\ImpPe$.
We can see that for $\ImpPe_{\beta}^{1/4}/\beta\ll 1$, the predicted transition for sub-regime (iii), all the curves collapse and $\delta\sim \ImpPe_{\beta}^{-1/4}$, as suggested by our scaling analysis in \S\ref{sec:thickBLwedge}. 

In contrast with $\delta$, we find that $\Sh$ (figure~\ref{fig:polar2}(\textit{b})) does not follow the intuitive scaling  $\Sh\sim \ImpPe_\beta^{1/4}$
. 
Instead, the  collapse of the curves suggests a different trend in sub-regime (iii):  $Sh/\ImpPe_\beta^{1/4}\sim (\ImpPe_\beta^{1/4}/\beta)^{-3/4}$, or equivalently $\Sh\sim \beta^{3/4} \ImpPe_\beta^{1/16}$. However, we have not been able to confirm this result analytically. We find that this empirical collapse occurs for $\ImpPe_{\beta}^{1/4}/\beta \ll 1$, as long as radial confinement effects are not important. Radial confinement effects occur when $\delta\sim H$, or $\ImpPe_{\beta}^{-1/4}\sim H$, as shown by the radical change of regime at lower $\ImpPe_{\beta}^{1/4}/\beta$ in figure~\ref{fig:polar2}(\textit{b}). 

The  weaker dependence of $\Sh$ on $\ImpPe$  in the limit of vanishing $\ImpPe$ and $\ImpPe_{\beta}^{1/4}/\beta \ll 1$, assuming no radial confinement effects, shows that multiple effects become important in addition to the streamwise advection--radial diffusion balance. This has been observed both in the asymptotic sub-regime (iii) in figure~\ref{fig:polar2}(\textit{b}) and the regime $\beta\sim 1$ in figure~\ref{fig:polar1}. Curvature effects become  important and the velocity field increases such that $U_\delta\sim (\beta \delta)^2$ due to the opening of the wedge channel. 
These combined effects are the cause of the observed plateau  in the Sherwood number in figure~\ref{fig:polar1}, as $\ImpPe\to 0$ and $\beta\sim 1$, which implies an enhanced mass transfer compared with parallel-wall channels. This is intuitively expected as lateral confinement effects vanish  with increasing  opening angle.


 \section{Discussion and implications for practical applications}\label{sec:discussion}

As mentioned in \S\ref{sec:intro}, this study applies to  the removal of contaminant 
trapped in sub-surface features such as gaps, cracks or folds. We assumed  that  the area of release is constant and flat, and does not impact the velocity field.  In practice, the contaminant may take the shape of a  droplet which can perturb the flow in various ways: for instance through changes in the height of the channel, particularly when $\delta\sim H$,  causing a change in velocity  and modifying $\ImpPe$, thereby affecting the mass transfer.
We also assumed that the scalar released is passive. It is likely that this assumption is justified for slowly dissolving or low solubility substances such as found in many cleaning and decontamination scenarios. 
 Otherwise, changes to the  density or viscosity of the cleaning agent  need to be accounted for.
 For example, if transport of the scalar leads to a significant change in fluid density, buoyancy effects should be considered. 
 Significant changes of the fluid viscosity with tracer concentration could change the shear profile in ways which could affect  the Sherwood number. This may be of particular importance for materials such as highly soluble liquids with high viscosity. 

 Another potential limitation of this study is the geometrical simplification of the bottom of the channel, particularly in the case of mass transfer. A dissolving droplet or solid at the bottom of the channel may not have a flat surface, as assumed in the parallel-wall geometry, or a convex circular surface, as assumed in the truncated wedge geometry; or its shape may be affected by the dissolution process itself. We did not consider this level of detail in order to obtain simple analytical expressions and deduce key physical insight, which might have been lost in a full numerical treatment.

Applications with slow changes in time of the source concentration can also exploit our results under the assumption of a quasi-steady diffusive boundary layer. 
The concentration profile and mass transfer in the diffusive boundary layer can be considered to adjust  instantaneously to the changes in the source concentration \citep[see][]{Landel:2016}.

The key and most intuitive implication of our findings to decontamination  and cleaning applications is that increasing the \Peclet number $\ImpPe$  improves the flux, which then allows for better neutralisation of the substance through reactions in the bulk. We find that increasing the width of the channel $\hat{w}$ has the strongest impact on increasing $\ImpPe$. Indeed, we have $\ImpPe\sim \hat{w}^4$ since the characteristic channel velocity $\hat{U}_0$ increases quadratically with $\hat{w}$. However, changes of the channel width are only possible through alterations of the material. Such techniques may not be favoured due to their  destructive potential for  substrates, but could be considered at the designer stage for some applications. 

The main physical parameter generally controlled in cleaning and decontamination applications, and which can increase $\ImpPe$ in a less destructive way, is the flow velocity since  $\ImpPe\propto \hat{U}_0$. 
The local velocity in the channel is controlled by pressure forces, gravity, viscosity and capillary forces. 
Therefore, reducing the viscosity of the  cleansing flow, through through the formulation or an increase of temperature for instance, or increasing the pressure gradient, could lead to increasing $\ImpPe$.  
Depending on the geometry and the regime, different gains in the flux can be obtained. For example, in the case of parallel channels, the highest gain is obtained when the flow is confined vertically: doubling the speed $\hat{U}_0$ will also double the Sherwood number $\Sh$ and thus the overall flux. If the boundary layer is unconfined vertically but confined in the lateral direction (thick boundary layer regime), doubling $\hat{U}_0$ yields an increase by $2^{1/2}\approx 1.41$ in the flux. In the thin boundary layer regime, where the boundary layer is  unconfined, doubling $\hat{U}_0$ yields an increase of only $2^{1/3}\approx 1.26$ in the flux. However, these results are valid provided that the  boundary layer does not change regime. As  $\hat{U}_0$ increases, $\delta$ decreases,  reducing confinement effects and potentially leading to a change in regime. 
Consequently, while there are still gains in the flux, the gains may be smaller.
Increasing $\hat{U}_0$ inside sub-surface channels can be challenging. Most decontamination and cleaning techniques involve surface washing  which has a limited effect on  the velocity in sub-surface features, which may be driven purely by gravitational draining.

Mass transfer from the area of  release through a diffusive boundary layer is  a first but key step towards complete removal in the context of cleaning and decontamination. For very long channels, scalar transport beyond the area of release, i.e. for $\hat{L}<\hat{x}<\infty$, becomes a  Taylor--Aris problem \citep{taylor53,aris56} with non-uniform inlet scalar profile (see figures \ref{fig:contourcart} and \ref{fig:polarcontourplots} for concentration profiles at the downstream end of the area of release). \cite{Giona:2009} study the dispersion of a scalar transported in laminar channel flows with various smooth and non-smooth cross-sectional geometries. They  consider the case of  impulse feeding with no-flux boundary condition on all the channel walls. Their results describe the evolution of the scalar distribution beyond the area of release but at a finite distance, thus complementing the works of \cite{taylor53} and \cite{aris56} who looked at the far field distribution. Similar to our $\ImpPe$, their effective \Peclet number compares the axial convective time scale to the transverse or cross channel diffusive time scale. In the limit of vanishing effective \Peclet number they recover the Taylor--Aris regime, whilst for effective \Peclet numbers of more than 10 they find an \textit{advection dominated dispersion} regime which is characterised by wall boundary layers with slow advective transport. Their advection-dominated dispersion regime has parallels with our thick boundary layer regime, thus making the Taylor--Aris regime analogous with our thick boundary layer regime. 


The existence of the advection dominated dispersion regime has practical implications for decontamination problems. The scalar can be trapped in boundary layers close to the walls \citep{Adrover:2009}. This can increase its dwelling time in the channel and could potentially enable ingress into  absorbing channel walls, thus, dispersing  contaminants further.

Our results are also relevant to turbulent flows, provided the diffusive boundary layer is thinner than the viscous sub-layer 
of the turbulent wall boundary layer if the flow. In the case of shallow cracks and gaps on a  substrate, or for rough substrates, the thin film flow washing the surface can be turbulent above these features. In general the viscous sub-layer develops faster than the diffusive boundary layer, due to the high Schmidt numbers involved in typical cleaning and decontamination problems, of the order of $10^3$ to $10^4$.

 \section{Summary and conclusion}\label{sec:conclusion}

We have studied in this paper the convective  transport of a passive scalar released at the base of generic rectangular channels with parallel walls and channels with a truncated wedge cross-section with angled walls. Our main objective was to predict the flux or Sherwood number $\Sh$ as a function of the flow, scalar properties, and the geometry. Due to the  lateral and vertical or radial confinement, the resulting diffusive boundary layer for the scalar is three-dimensional. This makes the problem too complex to solve analytically in the  general case.
Using a combination of asymptotic analysis and numerical calculations, we have found that different regimes  exist for $\Sh$ depending mainly on the ratio of the diffusive boundary layer thickness and the gap width $\delta=\Hat{\delta}/\Hat{w}$.  We have  also shown  that  $\delta$ is a function of  a characteristic \Peclet number, $\ImpPe = (\hat{w}^2/\hat{D})/(\hat{L}/\hat{U}_0)$, and the opening angle $\beta$ for wedges, depending on the regime. An important and unexpected conclusion is that in  all the regimes identified,  two-dimensional approximate models can provide accurate quantitative predictions for $\Sh$ across all the parameters and geometries explored, despite the problem begin fundamentally three-dimensional.  We summarize in table~\ref{tab:expcharacpara} the different predictions for $\Sh$, $\delta$ and $U_\delta$ for each regime and channel geometry. The main remarks of our study are:

\begin{landscape}
\begin{table}
  \begin{minipage}{\textwidth}
    \centering
    \resizebox*{!}{0.7\textheight}{
    \scalebox{0.96}{
    \begin{tabular}{@{}lcccc@{}}
\hline
Channel geometry  & Parameter &  &  &  \\ 
and regime & range & $U_\delta$ & $\delta$ & $\Sh$ \\ \hline \vspace{0.1cm}
\textit{Parallel wall} &   &  &  & \\
Thin boundary layer regime &   &  &  & \\
$\delta \ll 1$ (slice-wise \Leveque regime) & $\ImpPe^{1/3}\gg 1$  & $\gamma(z) y+O(\delta^2)$ & $ \sim \ImpPe^{-1/3}$  & $\frac{3^{4/3} \zavg{\gamma^{1/3}} }{2\Upgamma(1/3)}  \ImpPe^{1/3}$\\
$\delta \sim H \ll 1$ (confined) & $\ImpPe^{1/3}\gg 1$  & $\sim H^2$ & $ \sim H$  & $\sim H^3  \ImpPe$\\
Thick boundary layer regime &   &  &  & \\
$1 \ll \delta \ll H$ & $\ImpPe^{1/2}\ll 1$  & $\frac{3}{2}(1-4z^2) + O(\delta^{-2})$ & $\sim \ImpPe^{-1/2}$  & $\frac{2}{\sqrt{\pi}}  \ImpPe^{1/2}$\\
$1 \ll \delta \sim H$ (confined) & $\ImpPe^{1/2}\ll 1$  & $\sim 1$ & $ \sim H$  & (\ref{eq:parallelwallfiniteheightflux}) $\sim H  \ImpPe$\\\textbf{}
Transition &   &  &  & \\
$ \delta \sim 1 \ll H$ & $\ImpPe \sim 1$  &  (\ref{eq:straightflowfield}) & $ \sim 1$  & $\approx \frac{1.96\, \ImpPe^{1/2}}{(1+1.18\, \ImpPe^{1/6})}$\\ 

\textit{Truncated wedge} &   &  &  & \\
Thin boundary layer regime &   &  &  & \\
$\delta \ll 1$ (slice-wise \Leveque regime) & $\ImpPe^{1/3}\gg 1$  & $\gamma(z) r+O(\delta^2)$ & $ \sim \ImpPe^{-1/3}$  & $\frac{3^{4/3} \zavg{\gamma^{1/3}} }{2\Upgamma(1/3)}  \ImpPe^{1/3} + f(\gamma,\beta) \beta$\\
$\delta \sim H \ll 1$ (confined) & $\ImpPe^{1/3}\gg 1$  & $\sim H^2$ & $ \sim H$  & $\sim H^3  \ImpPe$\\
Thick boundary layer regime &   &  &  & \\
$\beta\ll \beta\delta \ll 1 \ll \delta \ll H$, sub-regime (i) & $\beta \ll \ImpPe^{1/2}\ll 1$  & $\frac{3}{2}(1-4\frac{\theta^2}{\beta^2})+O(\delta^{-2},\beta,(\beta\delta)^{2})$ & $\sim \ImpPe^{-1/2}$  & $\frac{2}{\sqrt{\pi}}  \ImpPe^{1/2}+\beta/2$\\
$\beta\ll \beta\delta \ll 1 \ll \delta \sim H$ (confined) & $\beta \ll \ImpPe^{1/2}\ll 1$  & $\sim 1$ & $ \sim H$  & $\sim H  \ImpPe$\\

$\beta\ll \beta\delta \sim 1 \ll \delta \ll H$, sub-regime (ii) & $\beta \sim \ImpPe^{1/2}\ll 1$  & $ \frac{3}{2}(1-4\frac{\theta^2}{\beta^2})(1+\beta r)^2+O(\beta^2)$ & $\sim \ImpPe^{-1/2} \sim \beta^{-1}$  & $O(\ImpPe^{1/2})+O(\beta)$\\
$\beta\ll \beta\delta \sim 1 \ll \delta \sim H$ (confined) & $\beta \sim \ImpPe^{1/2}\ll 1$  & $\sim 1$ & $ \sim H$  & $\sim H  \ImpPe$\\

$\beta\ll 1 \ll \beta\delta \ll \delta \ll H$, sub-regime (iii) & $ \ImpPe^{1/2}\ll \beta \ll 1$  & $\frac{3}{2}(1-4\frac{\theta^2}{\beta^2})(\beta^2 r^2+2\beta r)+O(\delta^2\beta^4,1,\delta\beta^2)$ & $\sim \ImpPe_{\beta}^{-1/4}= \beta^{-1/2}\ImpPe^{-1/4}$  & $\sim \beta^{3/4}\ImpPe_\beta^{1/16}$\textdagger \\
$\beta\ll 1 \ll  \beta\delta \ll \delta \sim H$ (confined) & $\ImpPe^{1/2} \ll \beta \ll 1$  & $\sim \beta^2 H^2 +O(\beta H)$ & $ \sim H$  & $\sim H^3  \beta^2 \ImpPe$\\

Transition &   &  &  & \\
$ \delta \sim 1 \ll H$ & $\ImpPe \sim 1$  &  (\ref{eq:angledwallsflowfield}) & $ \sim 1 $  &  -- \\ 

    \end{tabular}}
    }
  \end{minipage}
\caption{Summary of all the asymptotic and scaling (indicated by $\sim$) predictions results  and fits (indicated by $\approx$) for the characteristic velocity in the diffusive boundary layer $U_\delta$, the characteristic diffusive boundary layer thickness $\delta$ and the Sherwood number $\Sh$ (see (\ref{eq:slicewiseflux})) in parallel  channels and truncated wedges. The results depend on the  \Peclet number $\ImpPe=\hat{U}_0\hat{w}^2/(\hat{L}\hat{D})$ or the curvature-rescaled \Peclet number $\ImpPe_\beta=\beta^2\ImpPe$, the opening angle $\beta$ and the channel height $H$. The function $f\sim O(1)$ must be computed numerically (see \S\ref{sec:angledwallsthin}). \textdagger Empirical scaling not confirmed analytically.
}
\noindent\rule{\textwidth}{0.4pt}
\label{tab:expcharacpara}
\end{table}
\end{landscape}


\begin{itemize}
\item In the thin boundary layer regime, $\ImpPe\gg 1$, both geometries follow the classical \Leveque regime with $\Sh\sim \ImpPe^{1/3}$. Since $\delta\ll 1$, the influence of the geometry, whether curved or not, and the effect of lateral confinement are negligible, except for $\beta\sim 1$ where $\Sh\sim \ImpPe^{1/3}+\beta/2$. We find that the effect of  the diffusive boundary layers due to the no-flux side walls is small, decreasing with increasing \Peclet number, which explains why  two-dimensional approximate models provide accurate results in this regime.

\item In the thick boundary layer limit, $\ImpPe\ll 1$, cross-channel diffusion is dominant and the concentration is uniform across the channel. For parallel wall channels, $U_\delta$ is asymptotically constant with distance from the  base of the channel. The resulting Sherwood number follows $\Sh\sim \ImpPe^{1/2}$. In parallel channels, we find a smooth transition between the thick and thin boundary layer regimes  at intermediate \Peclet numbers, $\ImpPe\sim 1$ that is empirically described by the Pad\'{e} approximant $\Sh_{approx}\approx 1.96 \ImpPe^{1/2}/(1+1.18\ImpPe^{1/6})$. In contrast, the Sherwood number in the truncated wedge geometry  follows a more complex behaviour across the transition regime and the thick boundary layer regime, depending on the opening angle $\beta$ and $\ImpPe$. As $\beta$ increases at a fixed $\ImpPe\lesssim 1$, we find that $\Sh$  has a linear dependence with increasing $\beta$. For $\beta \sim 1$, we find that the dependence of $\Sh$ with $\ImpPe$ decreases. This is due to curvature effects. As $\beta$ increases, the influence of the no-flux boundary condition at the side walls is lessened, which tends to enhance curvature-induced diffusion.

\item We also found another effect of increasing the opening angle in which the velocity field is less constrained by the side walls and increases quadratically with $\beta$ and $r$. Thick boundary layers then experience an increasing velocity with distance from the area of release: $U_{\delta} \sim (\beta r)^2$. In this regime, sub-regime (iii), we showed that the diffusive boundary layer thickness scales with a curvature-rescaled \Peclet number $\ImpPe_\beta=\beta^2\ImpPe$, such that $\delta\sim \ImpPe_{\beta}^{-1/4}$ in the limit $\ImpPe_{\beta}^{1/4}\ll \beta \ll 1$,  provided $\delta\ll H$ so that the flow is not constrained radially. However, we find that the Sherwood number $\Sh$ does not follow the intuitive scaling $\Sh\sim 1/\delta \sim \ImpPe_{\beta}^{1/4}$,  but instead appears to follow the empirical scaling $\Sh\sim \beta^{3/4} \ImpPe_\beta^{1/16}$. This enhanced flux is due  to increased diffusion from curvature effect. The curvature term in the cross-channel averaged advection--diffusion equation in cylindrical coordinates is a leading order term in this limit. Therefore, to predict accurately its effect on the resulting mass transfer, numerical computation is necessary.

\item We have also shown that it is not necessary to solve the full three-dimensional advection--diffusion equation to obtain an accurate estimate of $Sh$.
The results for $Sh$ predicted by solutions to the simplified two-dimensional cross-channel averaged advection--diffusion equations (\ref{eq:parallel2davgproblem}) and (\ref{eq:2dpolarproblem}), for parallel wall channels and truncated wedges respectively,  agree with the full three-dimensional numerical solution to better than \SI{5}{\percent} for $\ImpPe \leq 100$ for all curves calculated. We have also shown that this can be extended towards larger $\ImpPe$ by improving the numerical resolution. 
These simplified equations neglect  the contribution to the overall flux of the three-dimensional fluctuation flux $\zavg{u'c'}$, which can be responsible for up to \SI{50}{\percent} of the total streamwise flux at some height for large $\ImpPe$ and $\beta\sim 1$, but is negligible for low $\ImpPe$ as shown by our asymptotic analysis and numerical results.
It is therefore somewhat surprising that the significant non-zero contribution from $\zavg{u'c'}$, which varies with the height above the base, leads to only a small net contribution to the three-dimensional Sherwood number after integration over the depth of the flow. 
Nevertheless, our findings  reveal that the net effect of this $y$-varying $\zavg{u'c'}$ contribution, integrated over the depth of the flow, leads to only a small deviation on the two-dimensional Sherwood number for all \Peclet numbers and opening angles. 
The asymptotic analysis shows that in the limit of large $\ImpPe$ a two-dimensional computation over-predicts the Sherwood number by only \SI{2.4}{\percent} for channels with parallel walls and by less than \SI{4.5}{\percent} for truncated wedges with $\beta\leq \pi/2$. 
Therefore, the two-dimensional equations (\ref{eq:parallel2davgproblem}) (parallel channels) and (\ref{eq:2dpolarproblem}) (wedges) will be useful to many applications where improving computational speed is  critical. 


\item In addition, we have demonstrated that vertical (parallel channels) and radial (wedges) confinement  lead to $\Sh\sim U_{\delta} H \ImpPe$ when $\delta\to H$ for all the geometries studied and at any \Peclet numbers. Effectively, vertical or radial confinement has the strongest impact on reducing the flux. This is due to significant reduction in the  gradient of the concentration field normal to the area of release. This is an important finding for applications optimising  convective fluxes in narrow spaces.

\end{itemize}

The two geometries studied provide insight as to the impact of having other opening channel geometries such as convex or concave side walls of complex profiles. The impact occurs mainly in the thick boundary layer regime where the velocity field  varies depending on lateral confinement. From  scaling analysis, we find that $\delta\sim (U_{\delta}\ImpPe)^{-1/2}$, for $ \delta \ll H$. As the channel width increases, the resulting  $\Sh$ increases owing to a combination of enhanced streamwise advection and enhanced diffusion through curvature effects. An accurate dependence of $\Sh$ with flow and geometrical properties can be computed numerically using  the simplified two-dimensional equations (\ref{eq:parallel2davgproblem}) or (\ref{eq:2dpolarproblem}). Nevertheless, we note that  the dependence with the \Peclet number will be of the form $\Sh\sim \ImpPe^b$, where the exponent $b\leq 1/2$ is a function of $\ImpPe$, provided $U_{\delta}$ increases with $\delta$. We predict that this result is valid for all $\ImpPe\ll 1$ and $\delta \ll H$. For $\delta \sim H$, our previous result $\Sh\sim U_{\delta} H \ImpPe$ should hold, being valid for any geometry.

The low dependence of the average flux on cross-channel variations found for all  geometries and across all \Peclet numbers is an important result. Specifically, by neglecting  three-dimensional effects, the broad range of advanced analytical  techniques for two-dimensional problems can be exploited to obtain further result. For instance, conformal mapping  and potential flow techniques could explore how the Sherwood number depends on more complex geometry and flow profile in the $(x,y)$ or $(x,r)$ plane \citep[][]{bazant04,choi05}. Our results  strongly suggest that small variations to the geometry and flow profile in the cross-sectional  $(x,z)$ plane, beyond the geometries studied here, are unlikely to be important. However,  we leave the case of  more complex cross-sectional variations  for future studies.




\appendix


\setcounter{figure}{0}


\section{Additional material related to the flow field}\label{apdx:AddMat}

\begin{figure}
\centering
\includegraphics[keepaspectratio=true,width=\textwidth]{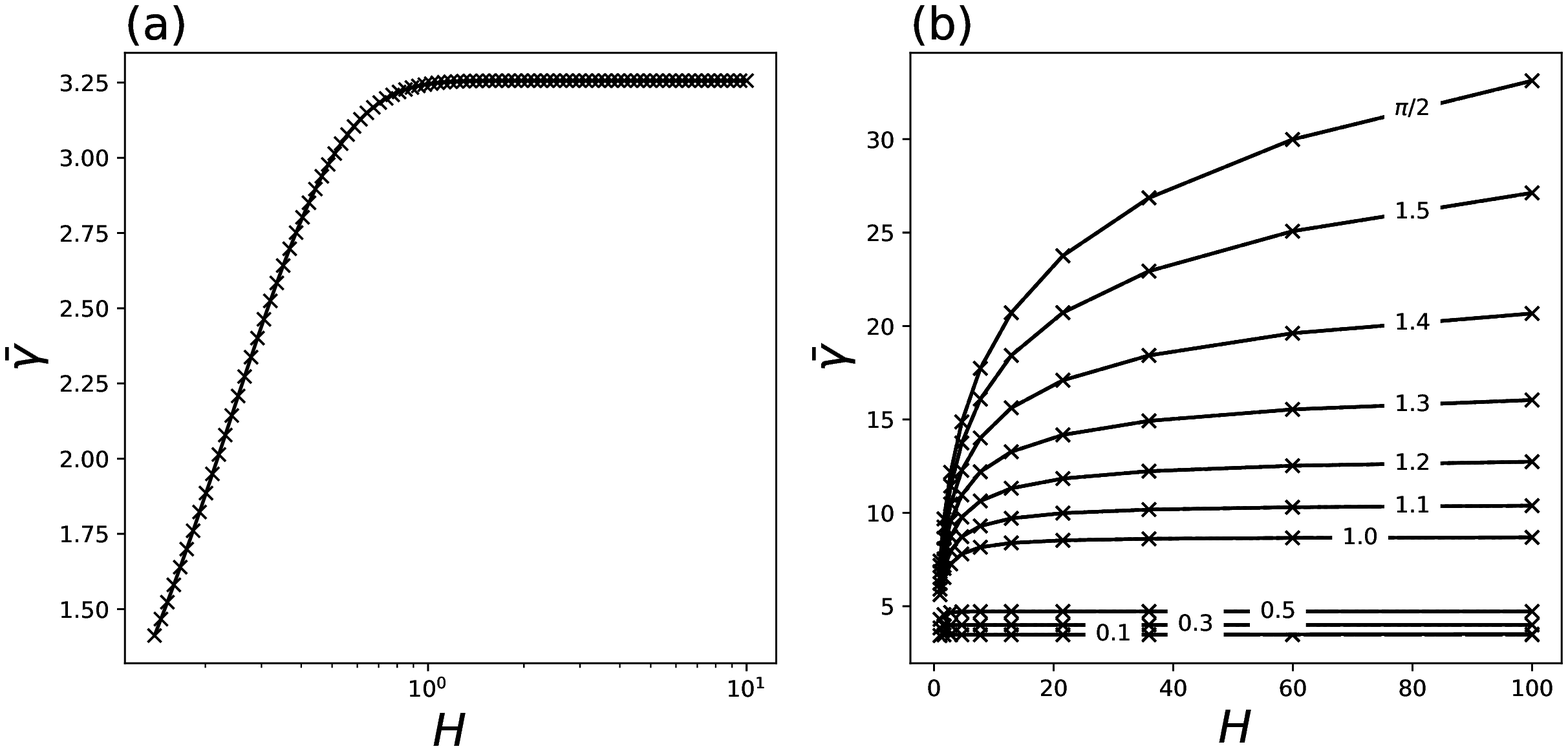}
\caption{ Variation $\zavg{\gamma}$ with $H$ for (\textit{a}) parallel channels, computed with (\ref{eq:shearUchannel}), and (\textit{b}) wedges with various $\beta$ (indicated on each curve), computed with (\ref{eq:wedgegamma}) (see table~\ref{tab:numdet}, appendix~\ref{apdx:NumSimDet}, for numerical details). In (\textit{a}), for $H>1.4$, the solution is truncated at $n=20000$ eigenpairs, for smaller channel heights fewer eigenpairs are used: $0.71<H\leq 1.4$, $n\leq 5000$; $0.27<H\leq 0.71$,  $n\leq 2000$; and $H\leq 0.27$, $n\leq 1000$ to avoid numerical overflows in the calculation of the series.}
\label{fig:GammaVSHandBeta}
\end{figure}

Figure \ref{fig:GammaVSHandBeta}(\textit{a}) shows the dependence of the cross-channel averaged bottom shear rate $\zavg{\gamma}$ with  $H$  for parallel channels. We find that $\zavg{\gamma}$ rapidly approaches an asymptotic value of approximately 3.26 for $H>1$. Figure \ref{fig:GammaVSHandBeta}(\textit{b}) presents similar results as in figure \ref{fig:GammaVSHandBeta}(\textit{a}) for truncated wedges of various  angles $\beta$. The mean shear rate $\zavg{\gamma}$ increases with increasing $\beta$. The channel height required for $\zavg{\gamma}$ to reach an asymptotic plateau increases with $\beta$. 

In figure \ref{fig:gammabarrationwedge}(\textit{a}) we show the dependence of the negative deviation $(\overline{\gamma^{1/3}}-\bar{\gamma}^{1/3})/\overline{\gamma^{1/3}}$ with $H$ and $\beta$. This  corresponds  to the deviation between the slice-wise and the two-dimensional solution of $\Sh$ in the thin boundary layer regime, see equation (\ref{eq:slicewiseflux}) for parallel-wall channels and (\ref{eq:truncatedwegeleveque}) for truncated wedges. The numerical results are computed using $n_{max}= 5000$ eigenpairs for $\tavg{\gamma}$ based on (\ref{eq:wedgegamma}).
For $\beta\lessapprox 1$, the deviation appears to reach an asymptotic plateau for  $H\lessapprox 100$. 
We see that for $\beta\to 0$ the magnitude of the deviation approaches the value for channels with parallel walls, approximately \SI{2.39}{\percent}. At $\beta\sim 1$, the deviation remains small, slightly larger than \SI{4}{\percent} for $H=100$ and $\beta=\pi/2$.

\section{Numerical computations}\label{apdx:NumSimDet}
\subsection{Verification of our numerical schemes}\label{app:verification}


We verified our numerical schemes for the computation of the three-dimensional and two-dimensional problems (see \S\ref{sec:straightwallstnumerics} and \S\ref{sec:angledwallstransition}) by recovering the well-known  \Leveque solutions for heat transfer in the thin boundary layer limit, which solves the scalar transport equation
\begin{equation}
    y\frac{\partial c}{\partial x} = \frac{\partial^2 c}{\partial y^2}, \label{eq:levequeproblem}
\end{equation}
for $0< x < \infty$ and $0<y<\infty$ with boundary conditions
\begin{equation}
    c(x=0,y)=0,\: c(0<x<L,0)=1,\: c(x,y\rightarrow \infty) \rightarrow 0. \label{eq:levequebc}
\end{equation}
The solution for this problem is given in classical texts such as \cite{Bejan}. In appropriate dimensionless form it is
\begin{equation}
    \breve{Sh} = \frac{3^{4/3}}{2\Gamma(1/3)}\breve{Pe}_L^{1/3}, \label{eq:stdleveque}
\end{equation}
 where $\breve{Sh}$ is the Sherwood number (defined as $\breve{Sh}=1/L\int_0^L\partial c/\partial y|_{y=0}\mathrm{d}x$) and the appropriate \Peclet number ($\breve{Pe}_L=1/L$) \citep{Bejan}. 

The two-dimensional numerical scheme 
used in this paper solves (\ref{eq:parallel2davgproblem}) with boundary conditions (\ref{eq:crosschannelavg1BC}a,b,d), on a finite domain with $0<y<H$. Since the governing equation (\ref{eq:levequeproblem}) with boundary conditions (\ref{eq:levequebc}) is defined for an unconfined flow ($0<y<\infty$), we expect the numerical solution to diverge from the analytical solution for small
$\breve{Pe}_L$ as the boundary layer assumption made in deriving (\ref{eq:levequeproblem}) is no longer valid (see also \citep{Bejan}) when the boundary layer thickness appropaches $H$. 

The three-dimensional numerical scheme  used in this paper solves (\ref{eq:cartesianfullproblem_a}) with boundary conditions (\ref{eq:cartesianfullproblem_aBC}a,b,d-f). It 
is expected to recover (\ref{eq:stdleveque}) if $u(y,z)=y$. Note that the velocity field in this model problem is inconsistent with the lateral boundary conditions at $z=\pm1/2$ in (\ref{eq:bcwithparallelwallsForAppendix}) and does not depend on $z$. Therefore, the three-dimensional code also solves the two-dimensional problem (\ref{eq:stdleveque}) provided, as for the two-dimensional case, that the boundary layer thickness is much smaller than the channel height.

\begin{figure}
    \centering
    \includegraphics[width=\textwidth]{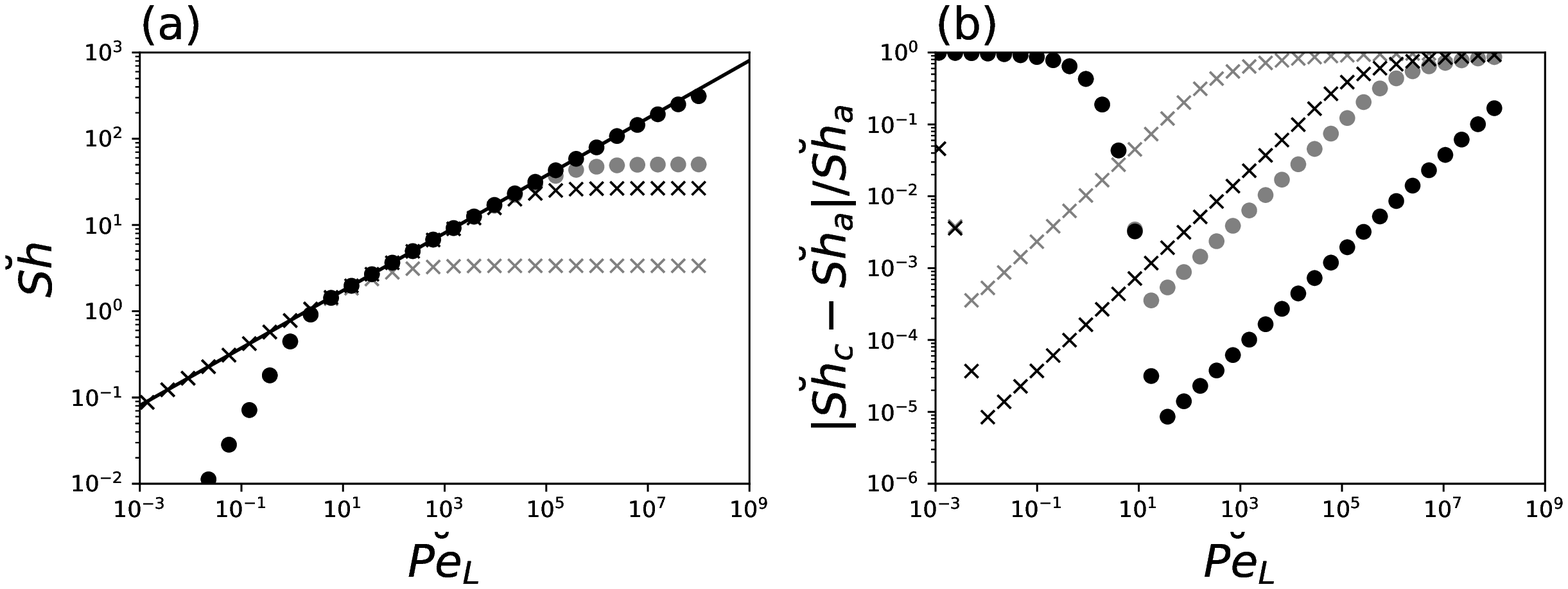}
    \caption{ Direct comparison between the three-dimensional numerical results for the Sherwood number (symbols) obtained with our three-dimensional numerical scheme for the classical \Leveque problem (\ref{eq:levequeproblem} with (\ref{eq:levequebc})) for two different domain heights and resolutions and  the analytical solution (\ref{eq:stdleveque}) (solid line). Crosses: $H=15$, $(N_y,N_z)=(50,5)$ (grey) and $(400,40)$ (black); circles: $H=1$, $(N_y,N_z)=(50,5)$ (grey) and $(400,40)$ (black). (b) Relative error for the data shown in (a).}
    \label{fig:verification3d}

\end{figure}

 Figure \ref{fig:verification3d} compares the three-dimensional numerical solutions for two different domain heights and resolutions (crosses: $H=15$, $(N_y,N_z)=(50,5)$ (grey) and $(400,40)$ (black); circles: $H=1$, $(N_y,N_z)=(50,5)$ (grey) and $(400,40)$ (black)) against (\ref{eq:stdleveque}).
In figure \ref{fig:verification3d}(a) the numerical solutions are plotted together with the analytical  solution   (\ref{eq:stdleveque}) (solid line). The error between the numerical and the analytical  solution is plotted in figure \ref{fig:verification3d}(b). A minimum error exists for both $H$: for $H=1$, the minimum is at $\breve{Pe}_L\approx 10^{1}$; 
for $H=15$ at $\breve{Pe}_L \approx 10^{-2}$.  The boundary layer thickness scales as $\delta \sim \breve{Pe}_L^{-1/3}$ \citep{Bejan}. Therefore, $\breve{Pe}_L\approx 10$ corresponds to $\delta \approx 0.5$ and $\breve{Pe}_L\approx \num{e-2}$ to $\delta \approx 5$, thus, the effect of vertical confinement on the boundary layer becomes non-negligible as $\delta \sim H$, as anticipated above.

Figure \ref{fig:verification3d}(b) shows that for $\breve{Pe}_L\gtrapprox 10$ ($H=1$) and $\breve{Pe}_L\gtrapprox 10^{-2}$ the error is less than \SI{1}{\percent} for about 2.5 orders of magnitude in $\breve{Pe}_L$ for the lower resolution calculations and about 4.5 orders of magnitude for the higher resolution calculations. The error increases again with increasing $\breve{Pe}_L$ as the boundary layer thickness decreases and approaches the numerical resolution. For  example, for $H=15$ resolved as $(400,40)$ (black crosses), $\breve{Pe}_L\approx \num{500}$ for an error of \SI{1}{\percent}. This corresponds to a boundary layer thickness of about \num{0.13}, which approaches the grid size (0.0375). At this resolution, the boundary layer can only be resolved with reduced accuracy for $\breve{Pe}_L > \num{500}$.

\begin{figure}
    \centering
    \includegraphics[width=\textwidth]{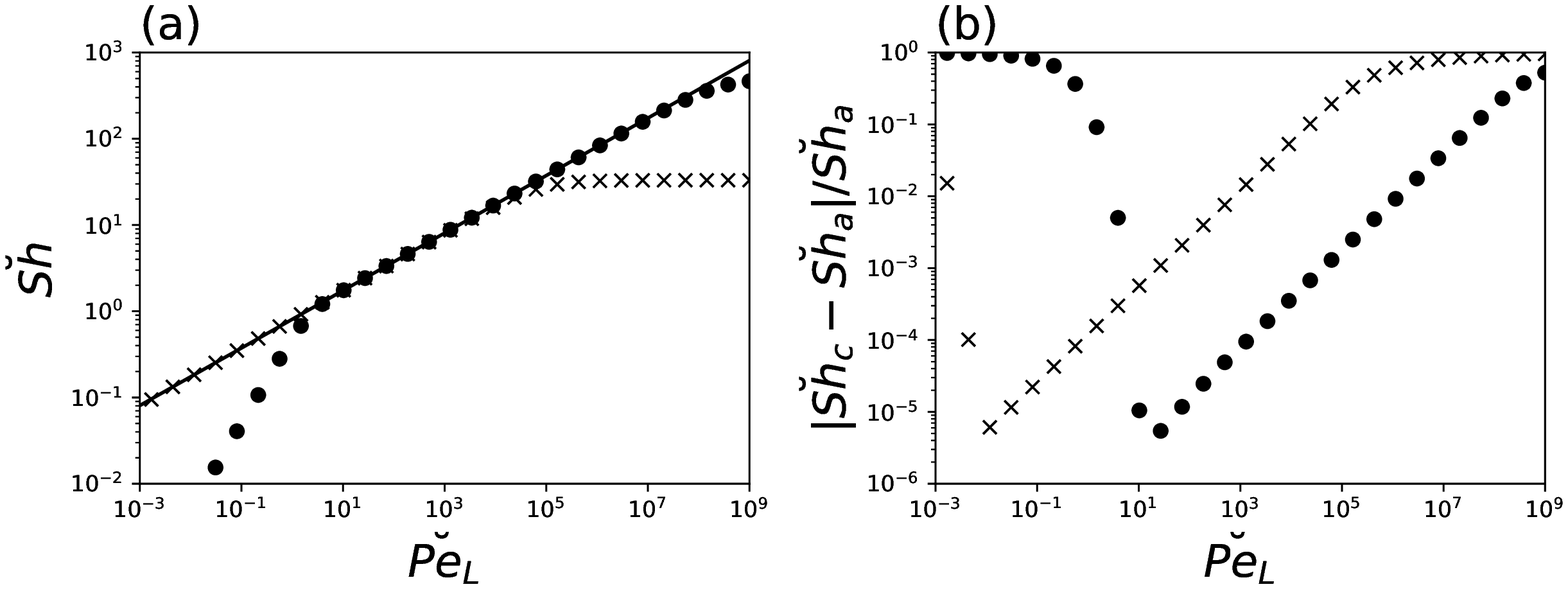}
    \caption{
    (a) Direct comparison between two-dimensional numerical results for the Sherwood number (symbols) obtained with our two-dimensional scheme for the classical \Leveque problem  and the analytical solution ((\ref{eq:stdleveque}) (solid line). Crosses: $H=15$, $N_y=25000$; circles: $H=1.25$, $N_y=25000$. (b) Relative error for the data shown in (a).
    \label{fig:verification2d}}
\end{figure}

We proceed analogously for the validation of the two-dimensional scheme.  The results are compared directly in figure \ref{fig:verification2d}(a). Figure \ref{fig:verification2d}(b) shows the relative error.
Similar to before, circles indicate solutions with $H=1.25$ and crosses with $H=15$. Since we were able to choose a large resolution on a one-dimensional grid ($N_y=25000$, consistent with other computations in this paper), we only investigate one resolution. Since the shooting method computes the eigenvalues serially \citep{Berry:1952} and the computational effort increases for larger eigenvalues, the truncation of the series is required nevertheless.
Consistent with the solutions obtained from the three-dimensional code, we find a minimum value of $\breve{Pe}_L$ at which the two solutions agree, which is consistent between both codes since both codes solve effectively the same problem. As can be seen,  an excellent agreement, to within (\SI{1}{\percent}), between both solutions is obtained for both channel heights for approximately four orders of magnitude. Computation of additional eigenpairs will extend the agreement between both solutions towards smaller $\breve{Pe}_L$ until the resolution limit is reached.

We note that our code verification spans multiple orders of magnitude in $\breve{Pe}_L$ and therefore in $\breve{Sh}$. Any computation spanning such a range of values is highly challenging and expensive and is eventually limited by numerical resolution. This is one of the reasons why we supplement the numerical solutions with asymptotic solutions throughout this paper.

\subsection{Numerical details}
Table  \ref{tab:numdet} provides details about the numerical computations presented in this study. The numerical calculations of $u$  for parallel channels (\ref{eq:straightflowfield}) and  for  wedges (\ref{eq:angledwallsflowfield}) were affected by numerical overflows at small  $H$ and large  $\beta$, even with 128 bit floating point precision.  As an error criterion we used that in the interior of the channel ($y=[0.001,0.999H],z=[-0.495,0.495]$, analogous for wedges) the relative error of the flow field (defined as the difference to a computation made with 1.5 times as many eigenpairs) should be less than \SI{1}{\percent}. For rectangular channels, velocity fields were computed with  at least 5000 eigenpairs and for wedge-shaped channels the velocity fields were computed with at least 2000 eigenpairs.


Semi-infinite domains (figures \ref{fig:contourcart}, \ref{fig:fluxcontributions}, \ref{fig:polarcontourplots}, \ref{fig:fluxcontributionswedge} and \ref{fig:polar2}) were approximated by solving the flow field in a channel of sufficient height (designated $H_f$) to ensure the flow field is independent with height. The height of the domain for solving the advection--diffusion problem (designated $H_D$) was then chosen such that $H_D> 2\delta$. We note that when $H_D=H_f$, the channel height is simply designated as $H$.


\subsection{Effect of truncation error}\label{sec:appconv}

\begin{figure}
    \centering
    \begin{tikzpicture}
    \node[inner sep=0pt] (left) at (0,0){
    \includegraphics[width=0.49\textwidth,keepaspectratio=true]{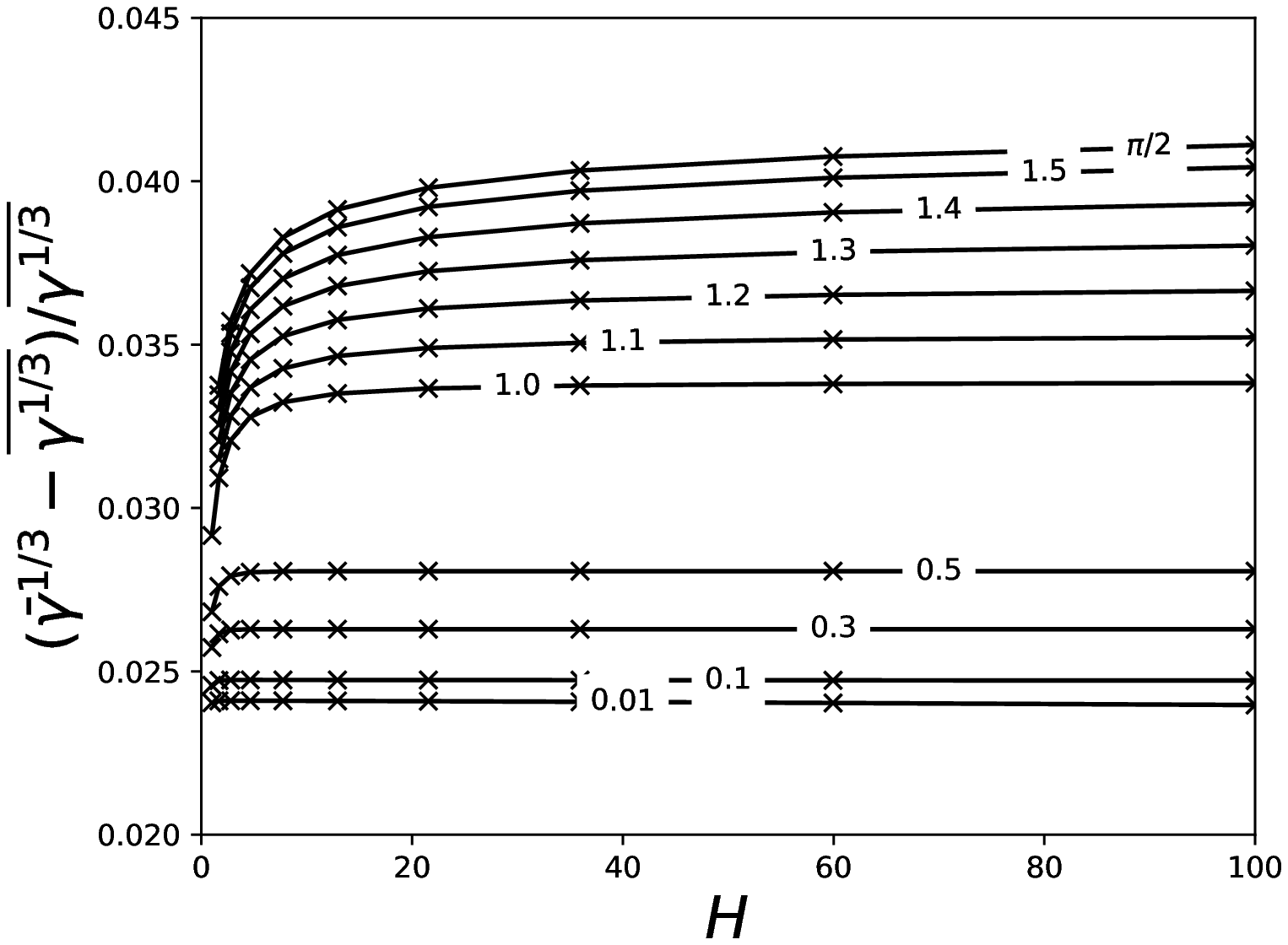}};
    \node[inner sep=0pt] (right) at (7,0){
    \includegraphics[width=0.49\textwidth,keepaspectratio=true]{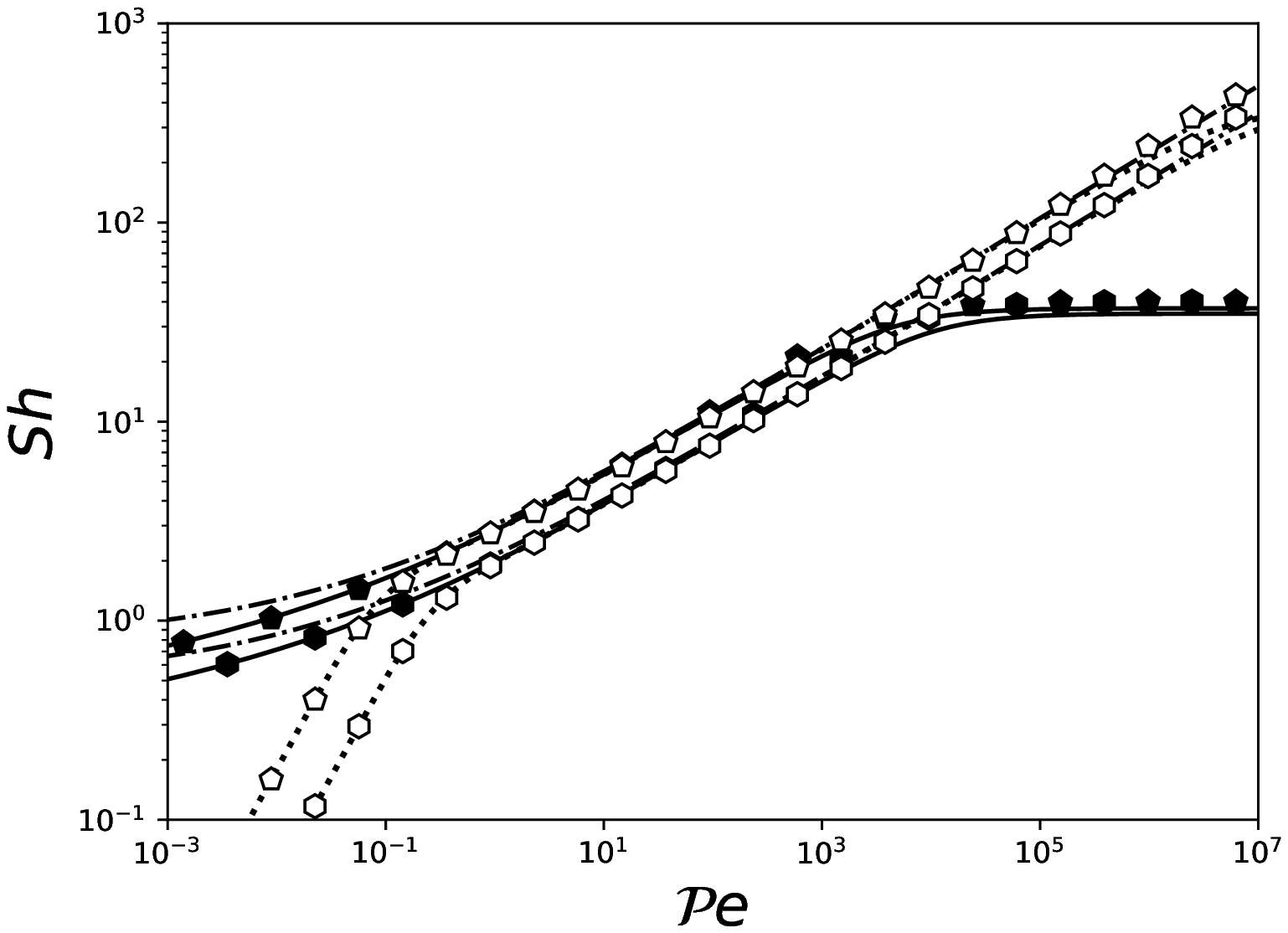}};
    \node[] at (-1.95,2.5) {(a)};
    \node[] at (4.9,2.5) {(b)};
    \end{tikzpicture}
    
    \caption{(\textit{a}) Deviation between the slice-wise   and  two-dimensional solutions of the averaged Sherwood number in  wedges, $(\bar{\gamma}^{1/3}-\overline{\gamma^{1/3}})/\overline{\gamma^{1/3}}$, versus $H$ and for various $\beta$ (indicated on each curve). See also equation (\ref{eq:truncatedwegeleveque}) and the discussion below for more detail. 
    (\textit{b}) The three-dimensional (black symbols) and two-dimensional (solid lines) numerical results shown in figure~\ref{fig:polar1} for $\beta=1$ and $\pi/2$ are reproduced here. The open symbols and the dotted lines show the three-dimensional and two-dimensional numerical results with increased resolution, at the same angles $\beta$, but for $H_D=1$ instead of $H_D=15$ in figure~\ref{fig:polar1}.
    }
    \label{fig:gammabarrationwedge}
\end{figure}

 Figure~\ref{fig:gammabarrationwedge}(\textit{b}) reproduces results from figure~\ref{fig:polar1}{,} for $\beta=1$ and $\pi/2$ (solid lines and black symbols), for which we noted a larger deviation from the asymptotic prediction (\ref{eq:polarthinexpansionflux}) (dash-dotted lines) for  $\ImpPe\gtrsim 10^3$.  Here the results are supplemented by  three-dimensional (open symbols) and two-dimensional (dotted lines) computations at the same $\beta$ but with a reduced domain height  for the advection--diffusion problem: from $H_D=15$ (figure~\ref{fig:polar1}) to $H_D=1$ (figure~\ref{fig:gammabarrationwedge}(\textit{b})) ($H_f=15$). The numerical results now agree with the asymptotic prediction (\ref{eq:polarthinexpansionflux}) for an extended range: up to  $\ImpPe\sim 10^6$. This is due to the increased resolution obtained when reducing the domain height, since the number of grid points is maintained fixed in all our computational domains (see table~\ref{tab:numdet},   appendix~\ref{apdx:NumSimDet}). The drawback of reducing $H_D$ is that the transition to the radially confined regime occurs at higher values of $\ImpPe$.


\clearpage 

\begin{table}
\centering
\scalebox{0.88}{
 \centering   
\begin{tabular}{lllllllll}

\hline
Figure       &          &  \multicolumn{2}{c}{Resolution}                  &                            &     &         &  EP     &Notes\\
 & $\ImpPe$     &  $N_y$ or $N_r$   & $N_z$ or $N_\theta$& $H_D$ & $\beta$ & $H_f$ & $n_{max}$ &    \\
 \hline
\ref{fig:contourcart}(\textit{a}) & $10^6$& 400 & 40 & 0.1 & - & 30 & 16000& 3D LP\\
\ref{fig:contourcart}(\textit{b}) & $10^5$ & 400 & 40 & 0.2 & - & 30 & 16000 &3D LP\\
\ref{fig:contourcart}(\textit{c,d})& $10^4, 10^3$ & 400 & 40 & 1 & - & 30 & 16000 & 3D LP\\
\ref{fig:contourcart}(\textit{e,f}) & $10^2, 10^1$ & 400 & 40 & 5 & - & 30 & 16000 &3D LP\\
\ref{fig:contourcart}(\textit{g}) &$1$  & 400 & 40 & 15 & - & 30 & 16000 &3D LP\\
\ref{fig:contourcart}(\textit{h}) & $10^{-1}$ & 400 & 40 & 5 & - & 30 & 16000 &3D LP\\
\hline
\ref{fig:fluxcontributions} & $10^6$& 400 & 40 & 0.1 & - & 30 & 16000 & 3D LP\\
\ref{fig:fluxcontributions} & $10^5$ & 400 & 40 & 0.2 & - & 30 & 16000 &3D LP\\
\ref{fig:fluxcontributions} & $10^4, 10^3$ & 400 & 40 & 1 & - & 30 & 16000 & 3D LP\\
\ref{fig:fluxcontributions} & $10^2, 10^1$ & 400 & 40 & 5 & - & 30 & 16000 &3D LP\\
\ref{fig:fluxcontributions} &$1$  & 400 & 40 & 15 & - & 30 & 16000 &3D LP\\
\ref{fig:fluxcontributions} & $10^{-1}$ & 400 & 40 & 5 & - & 30 & 16000 &3D LP\\
\hline
\ref{fig:cartfull} & $10^{-3}$--$10^{4}$ & 400 & 40 & 1.25,5,15 & - & 1.25,5,15  & 16000 & 3D LP\\
\ref{fig:cartfull} & $10^{-3}$--$10^4$ & 25000 & -  & 15 & - & 15  & 5000 & 2D SM \\
\hline
\ref{fig:polarcontourplots}(\textit{a,b}) & $10^4$, $10^3$ & 400 & 40 & 1 & 0.3 & 200 & 16000 & 3D LP \\
\ref{fig:polarcontourplots}(\textit{c}) & $10^2$ & 400 & 40 & 2 & 0.3 & 200 & 16000 & 3D LP \\
\ref{fig:polarcontourplots}(\textit{d,e}) & $10^1$, $10^0$ & 400 & 40 & 10 & 0.3 & 200 & 16000 &3D LP \\
\ref{fig:polarcontourplots}(\textit{f,g,h}) & $10^{-1}$, $10^{-2}$, $10^{-4}$ & 400 & 40 & 100 & 0.3 & 200 & 16000 &3D LP \\
\hline
\ref{fig:fluxcontributionswedge} & $10^6$ & 400 & 40 & 0.1 & 0.1-$\pi/2$ & 200 &16000& 3D LP\\
\ref{fig:fluxcontributionswedge} & $10^5$ & 400 & 40 & 1 & 0.1--$\pi/2$ & 200 &16000 & 3D LP\\
\ref{fig:fluxcontributionswedge} & $10^4$, $10^3$, $10^2$ & 400 & 40 & 2 & 0.1--$\pi/2$ & 200 & 16000& 3D LP\\
\ref{fig:fluxcontributionswedge} & $10$, $1$ & 400 & 40 & 10 & 0.1--$\pi/2$ & 200 &16000& 3D LP\\
\ref{fig:fluxcontributionswedge} & $10^{-1}$, $10^{-2}$, $10^{-3}$ & 400 & 40 & 100 & 0.1--$\pi/2$ & 200 & 16000&3D LP\\
\hline
\ref{fig:polar1} & $10^{-6}$--$10^{4}$ & 400 & 40 & 15 & 0.01-$\pi$ & 15 & 16000 & 3D LP \\
\ref{fig:polar1} & $10^{-6}$--$10^4$ & 25000 & - & 15 & 0.01-$\pi$ & 15 & 5000 & 2D SM\\
\hline
\ref{fig:polar2} & all shown & 25000 & - & 1000 & 0.01 - $\pi$/2  & 1000 & 5000 &  2D SM \\
\hline
\ref{fig:gammabarrationwedge}(\textit{b})  & $10^{-3}$--$10^{7}$ & 12500 & - & 1 & 1, $\pi$ & 15 & 3000 & 2D SM \\
\ref{fig:gammabarrationwedge}(\textit{b})  & $10^{-3}$--$10^{7}$ & 400 & 40 & 1 & 1, $\pi$ & 15 & 16000 & 3D LP \\
\ref{fig:gammabarrationwedge}(\textit{b})  & $10^{-3}$--$10^{7}$ & 25000 & - & 15 & 1, $\pi$ & 15 & 5000 & 2D SM\\
\ref{fig:gammabarrationwedge}(\textit{b})  & $10^{-3}$--$10^{7}$ & 400 & 40 & 15 & 1, $\pi$ & 15 & 16000 & 3D LP \\
				 
 \end{tabular}
}
 
\caption{
Details about the numerical calculations of  (\ref{eq:straightevproblem})--(\ref{eq:straightevproblemBC}) for parallel  channels and (\ref{eq:wedgeevproblem})--(\ref{eq:wedgeevproblemBC}) for wedges. Abreviations: EP - eigenpairs,  LP - Lapack solver, SM - Shooting method; $H_D$ is the domain height  for the advection--diffusion problem; $N_y$ and $N_r$ are the number of grid points in the $y$- and $r$-directions, and  $N_z$ and $N_\theta$ in the $z$- and $\theta$-directions; $H_f$ is the channel height for the calculation of the velocity; and $n_{max}$ is the number eigenpairs used. }
\label{tab:numdet}
\noindent\rule{\textwidth}{0.4pt}
\end{table}

\section{First-order correction in channels with a truncated wedge geometry}

\subsection{Thin boundary layer regime}\label{sec:appthinboundarylayerregime}


In equation (\ref{eq:polarlevequebase}) we change the variables from $(x, r )$ to $(\xi,\eta)$, with $\xi = x^{1/3}/\beta^{-1}$, which represents the ratio of $ \delta\sim x^{1/3} $ and $r_i=\beta^{-1}$, and $\eta =  r /x^{1/3}$ the similarity variable for the advection--diffusion equation at leading order. We obtain
\begin{equation}
\frac{\partial^2  c }{\partial\eta^2}+\gamma\frac{\eta^2}{3}\frac{\partial  c }{\partial\eta} + \xi \left( \eta \frac{\partial^2  c }{\partial\eta^2} + \frac{\partial  c }{\partial\eta}\left(1+ \gamma\frac{\eta^3}{3} \right) - \gamma\frac{\eta}{3} \frac{\partial  c }{\partial\xi} \right) - \xi^2 \gamma \frac{\eta^2}{3} \frac{\partial  c }{\partial\xi}=0.
\end{equation}
Substituting a Poincar\'e expansion: $ c (\xi,\eta)= c _0(\eta) +\xi  c _1(\eta)  + \ldots$, we find at order $\xi^0$
\begin{equation}\label{eq:c0thinBLwedge}
\fd{ c _0}{\eta}{2} + \gamma \frac{\eta^2}{3} \fd{ c _0}{\eta}{}=0,
\end{equation}
 which, as expected, leads to our modified \Leveque solution (\ref{eq:levequestraightslicewiseconcentrationfield}) (substituting $(y,z)$ by $(r,\theta)$ and with $\gamma(\theta)$ following (\ref{eq:wedgegamma})) and the flux (\ref{eq:truncatedwegeleveque}). For $n\geq 1$, we find
\begin{equation}\label{eq:cnthinBLwedge}
\fd{ c _n}{\eta}{2} + \gamma \frac{\eta^2}{3} \fd{ c _n}{\eta}{} - n \gamma \frac{\eta}{3}  c _n = \sum_{i=0}^{n-1} (-1)^{(n-i)} \eta^{(n-1-i)} \fd{ c _i}{\eta}{}.
\end{equation}
The boundary conditions for (\ref{eq:c0thinBLwedge}) and (\ref{eq:cnthinBLwedge}) are
\begin{equation}
 c _0(\eta=0)=1,\  c _n(\eta=0)= 0,\ \forall n\geq 1,\ \textrm{and} \  c _n(\eta\to +\infty)\to 0, \ \  \forall n\geq0.
\end{equation}
The next term, at order $\xi^1$, is
\begin{equation}
c_1(x,r,\theta) = -\frac{r}{2x^{1/3}} \frac{\Upgamma (1/3, \gamma(\theta) r^3/(9x))}{\Upgamma\left(1/3\right)}.
\end{equation}

\subsection{Thick boundary layer regime, sub-regime (i)}\label{sec:appthickboundarylayerregime}

\textit{[1-8]} 
Substituting  $\eta=r/x^{1/2}$ and $\epsilon=1/x^{1/2}$ and using $\tavg{u}=1+O \left(\delta^{-2},\beta,(\beta\delta)^2 \right)$,  (\ref{eq:crosschannelavg3}) becomes, at leading order,
\begin{equation}
\left( 1+\eta\frac{\beta}{\epsilon} \right) \frac{1}{2} \left( \eta \fp{\tavg{c}}{\eta}{} +\epsilon\fp{\tavg{c}}{\epsilon}{} \right) +  \left( 1+ \eta\frac{\beta}{\epsilon} \right) \fp{\tavg{c}}{\eta}{2} + \frac{\beta}{\epsilon} \fp{\tavg{c}}{\eta}{} = 0,
\end{equation}
with $\eta = O(1)$, $\epsilon= O(1/\delta) \ll 1$ and $\beta/\epsilon= O(\beta\delta) \ll 1$. Using a two-parameter  expansion: $\tavg{c}(\eta,\epsilon)=\tavg{c}_0(\eta)+\epsilon\tavg{c}_{11}(\eta)+(\beta/\epsilon)\tavg{c}_{12}(\eta)+O\left(\delta^{-2},\beta,(\beta\delta)^2 \right)$, we find at leading order  $\tavg{c}_0=\textrm{Erfc}(\eta/2)$, similar to (\ref{eq:straightwallthickblconcentration}) in  parallel channels as expected intuitively, and which satisfies the boundary conditions $\tavg{c}_0(0)=1$ and $\tavg{c}_0(\eta\to +\infty)=0$. At the next order in $O(\epsilon)$,
\begin{equation}
\fd{\tavg{c}_{11}}{\eta}{2} + \frac{\eta}{2} \fd{\tavg{c}_{11}}{\eta}{} +\frac{1}{2}\tavg{c}_{11}=0.
\end{equation}
The solution is
\begin{equation}
\tavg{c}_{11} = K_{11}e^{-\eta^2/4}\textrm{Erfi}\left(\frac{\eta}{2}\right),
\end{equation}
with $\textrm{Erfi}(\boldsymbol\cdot)$  the imaginary error function, and $K_{11}$ a  constant of order $O(1)$, which we determine below. The solution satisfies the  boundary conditions $\tavg{c}_{11}(0)=0$ and $\tavg{c}_{11}(\eta\to +\infty)=0$. Similarly, at order $O(\beta/\epsilon)$, $ \tavg{c}_{12}$ satisfies
\begin{equation}
\fd{\tavg{c}_{12}}{\eta}{2} + \frac{\eta}{2} \fd{\tavg{c}_{12}}{\eta}{} -\frac{1}{2}\tavg{c}_{12}=-\fd{\tavg{c}_{0}}{\eta}{}.
\end{equation}
The solution is
\begin{equation}\tag{\ref{eq:thickBLwedge1stordercorrectioni}}
\tavg{c}_{12} = -\frac{\eta}{2} \textrm{Erfc}\left(\frac{\eta}{2}\right),
\end{equation}
which satisfies the  boundary conditions $\tavg{c}_{12}(0)=0$ and $\tavg{c}_{12}(\eta\to +\infty)=0$. We can now compute the Sherwood number including the corrections at order $O(\epsilon,\beta/\epsilon)$,
\begin{equation}
    \Sh = \frac{2}{\sqrt{\pi}}\ImpPe^{1/2} - \frac{K_{11}}{\sqrt{\pi}}\ImpPe \ln(\ImpPe^{-1}) + \frac{\beta}{2},
\end{equation}
As $\beta\to 0$, we must recover the result (\ref{eq:straightwallfarfieldflux}) in parallel channels. Hence, $K_{11}=0$ and $\tavg{c}_{11}=0$.

\bibliographystyle{elsarticle-num-names}
\bibliography{BLPaperBibtex}

\end{document}